# On Direct Integration for Mirror Curves of Genus Two and an Almost Meromorphic Siegel Modular Form

Albrecht Klemm, Maximilian Poretschkin, Thorsten Schimannek, and Martin Westerholt-Raum

**Abstract**

This work considers aspects of almost holomorphic and meromorphic Siegel modular forms from the perspective of physics and mathematics. The first part is concerned with (refined) topological string theory and the direct integration of the holomorphic anomaly equations. Here, a central object to compute higher genus amplitudes, which serve as the generating functions of various enumerative invariants, is provided by the so-called propagator. We derive a universal expression for the propagator for geometries that have mirror curves of genus two which is given by the derivative of the logarithm of Igusa's cusp form $\chi_{10}$. In addition, we illustrate our findings by solving the refined topological string on the resolutions of the three toric orbifolds $\mathbb{C}^3/\mathbb{Z}_3$, $\mathbb{C}^3/\mathbb{Z}_5$ and $\mathbb{C}^3/\mathbb{Z}_6$.

In the second part, we give explicit expressions for lowering and raising operators on Siegel modular forms, and define almost holomorphic Siegel modular forms based on them. Extending the theory of Fourier-Jacobi expansions to almost holomorphic Siegel modular forms and building up on recent work by Pitale, Saha, and Schmidt, we can show that there is no analogue of the almost holomorphic elliptic Eisenstein series $\widetilde{E}_2$. In the case of genus 2, we provide an almost meromorphic substitute for it. This, in particular, leads us to a generalization of Ramanujan's differential equation for $\widetilde{E}_2$.

The two parts are intertwined by the observation that the meromorphic analogue of $\widetilde{E}_2$ coincides with the physical propagator. In addition, the generalized Ramanujan identities match precisely the physical consistency conditions that need to be imposed on the propagator.

## Contents

# 1 | Introduction and Summary

The topological string theory on non-compact Calabi-Yau geometries $M$ provides exact moduli dependent amplitudes in the $N = 2$ low energy effective action of type II string theory on these backgrounds. Expanded in flat coordinates for the moduli, these exact amplitudes are generating functions of related symplectic invariants, namely the Gromov-Witten-, Donaldson-Thomas- or Pandharipande-Thomas invariants. From the physics point of view the most natural invariants are the actual numbers $N^{\beta}_{j_L,j_R} \in \mathbb{N}$ of refined BPS states with a given representation $(j_L, j_R)$ of the 5d little group and the charge $\beta \in H_2(M,\mathbb{Z})$. Non-compact Calabi-Yau geometries exhibit typically an $U(1)_{\mathscr{R}}$ isometry, which is most obvious in the toric cases. This $U(1)_{\mathscr{R}}$ symmetry allows a motivic decomposition of the Pandharipande-Thomas moduli space of stable pairs and a mathematical definition of the $N^{\beta}_{j_L,j_R}$ [CKK14].

Topological string theory on non-compact Calabi-Yau spaces exhibits mirror symmetry and is related by string-gauge theory duality to three dimensional Chern-Simons theory [Wit95] and matrix models [AKMV04]. This makes A-model localization techniques [KZ01], B-model approaches, see e.g. [HKR08], and large N-duality methods available for its solutions. The large N Chern-Simons theory leads to the topological vertex [AKMV05]. The Chern-Simons theory large N expansion can be seen as mirror dual to a matrix model large N expansion [AKMV04]. The Ward identities of the matrix models leads to an iterative solution of the topological string in the genus [Mar08],[BKMP09]. Recently possible non-perturbative completions of the topological string were studied using a free Fermi-gas approach [MP12] and quantum mechanical systems [GHM14]. This led to a profound connection of topological string theory to operator analysis [KM15].

The full moduli dependence of the closed amplitudes $F^{(g)}$ is most efficiently studied in the B-model using modularity and the holomorphic anomaly equation of BCOV [BCOV94]. For topological strings the $F^{(g)}$ can be compared at special points in the moduli space with the A-model localization computation e.g. of Gromov-Witten invariants. A first check of mirror symmetry at higher genus was performed in [KZ01] for the non-compact toric $\mathcal{O}(-3)\to\mathbb{P}^2$ Calabi-Yau space. The BCOV equations can be generalized to the refined holomorphic anomaly equations (2.6) [HK12][KW11][1], which can be used to compute the refined amplitudes $F^{(n,g)}$. The corresponding check of refined mirror symmetry for topological strings was done in [CKK14] on the same geometry $\mathcal{O}(-3)\to\mathbb{P}^2$.

The B-model of a wide class of local Calabi-Yau geometries is defined by a mirror geometry, whose data are a family of Riemann surfaces $\mathcal{C}_{\mathfrak{g}}$ of genus $\mathfrak{g}$ and a meromorphic differential $\lambda$. These data $(\mathcal{C}_{\mathfrak{g}},\lambda)$ define also $N=2$ Seiberg-Witten theories, where $\mathcal{C}_{\mathfrak{g}}$ is known as Seiberg-Witten curve and $\lambda$ as Seiberg-Witten differential [SW94], and matrix models [Mar08],[BKMP09] with algebraic spectral curve $\mathcal{C}_{\mathfrak{g}}$. In this case $\lambda$ defines the filling fractions. So far the B-model approach using modularity properties has been developed for $\mathfrak{g}=1$ starting with [HK07]. Here we will extend the formalism to $\mathfrak{g}=2$. This extension applies to the calculation of closed amplitudes in all the settings mentioned above.

The occurrence of elliptic quasi-modular- and almost holomorphic modular forms [Shi86] in the solution of the holomorphic anomaly equation was noticed in [HK07]. In particular the identification of the propagator

$$S^{tt}=c^t_\tau\frac{1}{2\pi i}\frac{\partial}{\partial\tau}F^{(0,1)}=\frac{1}{2\pi i}\frac{\partial}{\partial\tau}\,c^t_\tau\left(\log(y)+4\log(|\eta||f_1|)\right)=\frac{c^t_\tau}{12}\widetilde{E}_2+G_2 \qquad (1.1)$$

with the almost-holomorphic Eisenstein series $\widetilde{E}_2=E_2-\frac{3}{\pi y}$ up to a meromorphic ambiguous piece $G_2$ was found [2]. This was in the context of $N=2$ Seiberg Witten theories of rank one on $\mathbb{R}^4$, which do emerge in the field theory limit of Type II theories on non-compact Calabi-Yau spaces, whose mirror geometry is a family of elliptic curves $\mathcal{C}_{\mathfrak{g}=1}$. Here $\eta$ is the Dedekind eta-function, $\tau\in\mathbb{H}^{(\mathfrak{g}=1)}$ is in the Siegel (Poincaré) upper half space (plane) and we denote[3] $\tau=x+iy$. $c^t_\tau$ is a constant whose geometrical significance is explained following eq. (1.3). Further $f_1(z)$, a rational function of the absolute modular invariant $z$[4] of the family $\mathcal{C}_1$, is the holomorphic ambiguity at genus one. $G_2$ is hence a meromorphic weight 2 form of the monodromy group $\Gamma$ of $\mathcal{C}_{\mathfrak{g}=1}$. Here $\Gamma\subset\mathrm{SL}(2,\mathbb{Z})$ is a congruent subgroup and often a meromorphic weight 2 form of $\Gamma$ does not exits, which fixes $f_1=1$. In any case we may set $f_1=1$ and hence $G_2$ to zero in the definition of the propagator. This merely modifies the form of the higher genus holomorphic or modular ambiguity. It was argued in [HK07] that the latter can be fixed by regularity and the *conifold gap* for the unrefined theory. In [HKR08] this fact has been more carefully established.

[1]The one of [KW11] is correct after a shift.

[2]It is also implicit in the invariant $N=4$ sector of the $\mathbb{Z}_3\otimes\mathbb{Z}_3$ orbifold example in [BCOV94].

[3]Due to a conflict with other parameters we use in the physical part also explicitly $\mathrm{Re}(\tau)$ and $\mathrm{Im}(\tau)$ for $x$ and $y$.

[4]It is traditionally called $u=\frac{1}{2}\mathrm{Tr}\Phi^2$ for Seiberg Witten curves.

Historically almost holomorphic (elliptic) modular forms were defined by Shimura [Shi86], who called them "nearly holomorphic". The occurrence of the regularized quasi modular Eisenstein series $E_2$ in topological string theory can be traced back to the world-sheet contact terms in the derivation of the holomorphic anomaly equation [BCOV94]. Related occurrences of $E_2$ due to space contact terms in Seiberg-Witten theory on compact target spaces, where one integrates over the phase space, were noted in [MW97] and in 2d QCD on the torus in [Dij95]. The occurrence of $E_2$ in the latter case was related systematically to almost holomorphic forms in [KZ95], where the isomorphism of the ring of quasi modular and almost holomorphic modular forms, the closures of these rings under the Serre- and Maass derivative respectively and the relation to weak Jacobi-forms[5] was explained. Quasi modular forms have Fourier expansions. Fourier coefficients attached to quasi modular forms are not only counting refined BPS states, but they have been connected to a vast variety of combinatorial and geometric quantities – see, for example, [BO00] and papers citing it. In many cases no recursive holomorphic anomaly is known. E.g. for the 2d QCD example it was argued in [Rud94] that such a recursion does not exist.

The general structure on almost holomorphic modular forms, which is particularly useful when holomorphic anomaly equations govern the theory, is as follows: The non-holomorphic Maass derivative acts as raising operator R on weight $k$ modular forms as $\partial_\tau - \frac{ki}{2}y^{-1}$. In addition, there is the lowering operator $\mathrm{L} = y^2\partial_{\bar\tau}$, which acts trivially on quasi modular forms, but non-trivially on almost holomorphic ones. The commutation relation $[\mathrm{L},\mathrm{R}] = \frac{-k}{4}$ is crucial when proving that $\widetilde{E}_2$ is essentially the only proper quasi modular form that does not arise from differentiating other ones.

This structure allows for the *direct integration* of the holomorphic anomaly equations (2.6) and an effective way to determine the $F^{(g,n)}$ for mirror curves $\mathscr{C}_{\mathfrak{g}=1}$ of genus one [ABK08][GKMW07]. The covariant derivative in the holomorphic anomaly equation (2.6) becomes the Maass derivative after changing from the flat coordinate $t$ to the $\tau$ coordinate using $\frac{\mathrm{d}t}{\mathrm{d}\tau} = C^{-1}_{ttt}$. The fact that the Maass derivative closes on the generators of the ring of holomorphic forms by the Ramanujan identities, e.g. if the modular group of the family is $\Gamma_0 = \mathrm{PSL}(2,\mathbb{Z})$ they read

$$\frac{1}{2\pi i}\partial_\tau \widetilde{E}_2 = \frac{1}{12}\left(\widetilde{E}_2^2 - E_4\right), \quad \frac{1}{2\pi i}\partial_\tau E_4 = \frac{1}{3}\left(E_2E_4 - E_6\right), \quad \frac{1}{2\pi i}\partial_\tau E_6 = \frac{1}{2}\left(E_2E_6 - E_4^2\right), \tag{1.2}$$

implies that the r.h.s. of (2.6) is a polynomial in $\widetilde{E}_2$, with coefficients that are rational functions in the absolute modular invariant $z$. The uniqueness of the an-holomorphic $\widetilde{E}_2$ implies further that the an-holomorphic derivative on the l.h.s of the holomorphic anomaly equation (2.6) can be replaced by a derivative by $\widetilde{E}_2$. Hence the recursive equations (2.6) can be solved by integration w.r.t. to $\widetilde{E}_2$ up to an integration constant. This constant in $\widetilde{E}_2$ is another rational function in $z$ called the holomorphic or modular ambiguity. In the refined case it can also be completly determined from regularity requirements and a refined gap condition [KW11] [HK12][HKPK13]. Note that $F^{(g)}$ or $F^{(g,n)}$ are meromorphic functions and in fact it is convenient to define a meromorphic almost holomorphic generator (4.5).

The purpose of this paper is to extend this analysis from families of elliptic mirror curves $\mathscr{C}_{\mathfrak{g}=1}$ to families of mirror curves of genus two $\mathscr{C}_{\mathfrak{g}=2}$. A key observation is that in the generalization of (1.1) the Dedekind $\eta$-function is replaced by the Igusa cusp form [6] $\chi_{10}$ of weight 10, so that the propagators becomes

$$\mathrm{S}^{ij} = C^i_p C^j_q \frac{1}{2\pi i}\frac{\partial}{\partial \tau_{pq}} F^{(0,1)}_{mod} = \frac{1}{2\pi i}\frac{1}{10}\left(\partial_{\tau_{pq}} \log\left(\chi_{10}\right) + \frac{5}{i}\left(\mathrm{Im}\tau\right)^{-1}_{pq}\right) C^i_p C^j_q . \tag{1.3}$$

Here we set in $F^{(0,1)}_{mod}$ the holomorphic ambiguity $f_1 = 1$. Of course this corresponds just to choice of the ambiguity in the propagator. In this universal relation the indices $i, j$ refer to the flat coor-dinates as in (1.1) and the relative coefficients of the terms on the r.h.s. are fixed by the weight. The matrix $C^i_p$ is the intersection matrix of the homology basis of a genus two surface, which is in general different from the identity matrix. In fact, for toric geometries, the non-standard intersection form $C^i_p$ is determined by the intersection pairing in the even-dimensional homology of the A-model geometry due to mirror symmetry. In addition, we show that the physical consistency conditions on the propagator coincide with generalized Ramanujan identities which are derived in the mathematical half of this paper. In the genus 2 case one has due to the Riemann Roch theorem a normal form with a corresponding theory of classifying invariants, the Igusa invariants, which replace the j-function of the elliptic case. The Fourier expansion of these invariants allows to extract the period matrix $\tau$ given the algebraic normal form of the Riemann surface, without any further calculation.

[5] Based on [KMW12] quite spectacular all genus results have been obtained in [HKK15] for compact elliptical fibered Calabi-Yau 3 folds using this relation to weak Jacobi-forms.

[6] There are two classical normalizations of the weight 10 genus two cusp form. The Maass lift $\phi^{(2)}_{10}$ used in section nine is related to it by $4\chi_{10} = \phi^{(2)}_{10}$.

Even though Shimura studied nearly holomorphic modular forms in full generality the theory of quasi Siegel modular forms has not yet been developed. We fill this gap in this paper, and reveal several interesting, possibly unexpected phenomena. The theory of Siegel differential operators is well-known, but it has not yet been made explicit. Section 6 contains expressions for raising (6.13) and lowering (6.6) operators, as well as their commutation relations (Lemma 6.11 and 6.12).

Fourier Jacobi expansions are important tools, when investigating Siegel modular forms. The theory of Jacobi forms which arise from vector valued Siegel modular forms was initiated recently in [IK11], and we extend it to the case of genus 2 almost holomorphic Siegel modular forms. Furthermore, we set up a framework for Jacobi forms that arise from vector valued Siegel modular forms of arbitrary genus. We do not characterize them as precisely as in the genus 2 case; rather this will be the theme of a sequel to this paper.

In Section 8, we classify almost holomorphic Siegel modular forms of genus 2, and find that every proper quasi Siegel modular form arises from differentiating other ones. We thus complete the independent classification by Pitale, Saha, and Schmidt [PSS15], which exhausted all almost holomorphic modular forms except for those of weight $\det^3 \mathrm{sym}^l$. As opposed to the case of elliptic modular forms constant modular forms do not occur in the image of the lowering operator. This insight naturally leads us to Section 9, in which we study the genus 2 case in detail. We introduce almost meromorphic Siegel modular forms with determined singular locus, which comprise the inverse of Igusa's $\chi_{10}$ [Igu62]. All of a sudden, we observe behavior very similar to the genus 1 case: There is an almost meromorphic Siegel modular form $S^{(2)}$ whose image under the lowering operator is 1. It naturally does not occur in the range of the raising operator; That is, the associated meromorphic quasi Siegel modular forms cannot be obtained as a derivative of any meromorphic Siegel modular form: It is analogue to $\widetilde{E}_2$. In Proposition 9.3 we provide an analogue of the Ramanujan differential equation for $S^{(2)}$.

Increasing the genus of the mirror curve is e.g. very natural for the mirror curves of resolved orbifolds for the $A$-model geometry. Consider e.g. the $\mathbb{C}^3/\mathbb{Z}_3$ orbifold. Resolving the singularity leads to $\mathcal{O}(-3) \to \mathbb{P}^2$ geometry. The Kähler parameter of the $\mathbb{P}^2$ is mapped to the one modulus of the genus one mirror geometry. If one considers the $\mathbb{C}^3/\mathbb{Z}_5$ orbifold the mirror will have $\mathfrak{g} = 2$. More precisely, it is a special two parameter family of genus two curves, where the two parameters correspond to the two blow up divisors of the $\mathbb{C}^3/\mathbb{Z}_5$ blowup geometry. These blow ups have toric descriptions and the genus $\mathfrak{g}$ of the mirror curve is given by the interior points of the 2d toric diagram, which represents the trace of the non-compact 3d cone that represents $M$. It is not directly related to the number of (Kähler) parameters $n_k$, which is given by the total number of points in the 2d toric diagram minus 3. The inner points correspond to dynamical fields of the 4d gauge theory, while the rest of $n_k$ parameters correspond to mass or non-dynamical parameters. The latter are given by the non-vanishing residua of the meromorphic form $\lambda$. Their number is not directly related to $\mathfrak{g}$, even though by the Riemann Roch theorem the latter determines an upper bound on the number of non dynamical parameters. An easy example, where such non-dynamical fields arise in the orbifold construction with a genus two mirror curve is the $\mathbb{C}^3/\mathbb{Z}_6$ orbifold. Our main computational contribution is to solve the $\mathbb{C}^3/\mathbb{Z}_5$ and the $\mathbb{C}^3/\mathbb{Z}_6$ cases to $n + g = 3$ and $n + g = 1$ respectively, checking the general formalism in great detail.

As reviewed in [BGHZ08], Siegel modular forms can be constructed from products of $\theta$ functions

$$\theta[\epsilon] = \sum_{m\in\mathbb{Z}^{\mathfrak{g}}} e^{2\pi i\left[(m+\epsilon'/2)^t\tau(m+\epsilon'/2)+(m+\epsilon'/2)\epsilon''/2\right]}, \tag{1.4}$$

where $\epsilon = \begin{pmatrix}\epsilon'\\ \epsilon''\end{pmatrix}$ with $\epsilon',\epsilon'' \in \{0,1\}^{\mathfrak{g}}$ encoding the spin structure. The simplest way is to raise the product of the $2^{\mathfrak{g}-1}(2^{\mathfrak{g}-1}+1)$ $\theta$-functions with *even spin structure* to a power so that no phase transformation occurs. E.g. for $\mathfrak{g}=1$: $2^8\eta^{24} = \left(\theta\begin{bmatrix}0\\0\end{bmatrix}\theta\begin{bmatrix}0\\1\end{bmatrix}\theta\begin{bmatrix}1\\0\end{bmatrix}\right)^8$, for $\mathfrak{g}=2$: $2^{14}\chi_{10} = -\prod_{\epsilon \text{ even}}\theta[\epsilon]^2$, while for $\mathfrak{g}=3$ the product of the 36 even theta functions defines a Siegel cusp form of weight 18 under $\mathrm{Sp}_6(\mathbb{Z})$. It is therefore natural to speculate that (1.1,1.3) get replaced, at least for genus 3, using the corresponding canonical cusp form and adapting the coefficients to the corresponding weight.

**Acknowledgments** It is a pleasure to thank Mirjam Cvetic, Hans Jockers, Ameya Pitale, and Don Zagier for discussions. AK thanks for support by KL 2271/1-1 and NSF DMS-11-01089. The work of MP has been supported by a scholarship of the Deutsche Telekom Stiftung and by the grant DE-SC0007901. He also thanks the BCTP for hospitality. TS is supported by a scholarship by the graduate school BCGS. MWR thanks the Max Planck Institute for Mathematics for their hospitality.

# 2 | Calculating Refined BPS Invariants

In this section we provide some necessary background material. We start with a review of the notion of refined BPS numbers in paragraph 2.1. Afterwards, we present the mirror symmetry for toric A-model geometries. We move on with a review of (refined) B-model computational techniques in 2.2. In particular, we rigorously discuss how the non-symplectic basis of the mirror curve modifies the well-known relations that apply to the case of a standard homology basis. Finally, we end by pointing out the quasi- respectively almost modular structure of the generating functions of the (un-)refined BPS invariants in paragraph 2.4.

**§2.1 The refined A-model.** In this subsection we provide some necessary background material for the discussion of refined BPS invariants. These can be meaningfully defined within five-dimensional theories with eight supercharges and an additional $U(1)_{\mathscr{R}}$ symmetry. Such theories are naturally engineered by M-theory compactifications on toric Calabi-Yau threefolds $X$. Physically, these refined BPS invariants count particles arising from M2-branes that wrap holomorphic curves within $X$. Besides the homology class $\beta \in H_2(X,\mathbb{Z})$, these particles are in addition classified by their spins $(j_L, j_R)$ under the five-dimensional little group $SU(2)_L \times SU(2)_R$. The multiplicities $N^{\beta}_{j_L,j_R}$ are called the refined BPS invariants. These are counted by a five-dimensional index which reads

$$Z_{BPS}(\epsilon_1,\epsilon_2) = \mathrm{Tr}(-1)^{2(J_L+J_R)}\exp\left[-\left((\epsilon_1-\epsilon_2)J_L^3+(\epsilon_1+\epsilon_2)J_R^3+(\epsilon_1+\epsilon_2)J_{\mathscr{R}}^3+\beta H\right)\right]. \tag{2.1}$$

Here we have denoted the respective Cartan generators by $J_*$. It is crucial to note that the invariance under deformations follows from the twist of the generator $J_R$ by $J_{\mathscr{R}}$. It turns out to be convenient to re-organize the refined BPS invariants in the so-called refined free energy that takes the form

$$\mathscr{F}(\epsilon_1,\epsilon_2,t) = \sum_{\substack{j_L,j_R=0\\k=1}}^{\infty}(-1)^{2(j_L+j_R)}\frac{N^{\beta}_{j_L,j_R}}{k}\frac{\sum_{m_L=-j_L}^{j_L} q_L^{km_L}}{2\sinh\left(\frac{k\epsilon_1}{2}\right)}\frac{\sum_{m_R=-j_R}^{j_R} q_R^{km_R}}{2\sinh\left(\frac{k\epsilon_2}{2}\right)}Q^{k\beta}, \qquad Q^{\beta}=e^{-\beta\cdot t}. \tag{2.2}$$

Here $t_i$ denote the Kähler moduli and $q_{L/R} = e^{\epsilon_1\pm\epsilon_2}$. There is a second set of invariants, denoted by $n^{\beta}_{g_L,g_R}$ which are related to the refined BPS numbers by a change of spin basis given as

$$I_*^n = \left(2\,[0]_* + \left[\frac{1}{2}\right]_*\right)^{\otimes n}, \qquad * \in \{L,R\}, \tag{2.3}$$

which is easily recognized as the spin content of the Lefshetz decomposition of an $n$-torus. In terms of this new basis, the $n^{\beta}_{g_L,g_R}$ are explicitly given as

$$\sum_{g_L,g_R} n^{\beta}_{g_L,g_R} I_L^{g_L} \otimes I_R^{g_R} = \sum_{j_L,j_R} N^{\beta}_{j_L j_R} \left[\frac{j_L}{2}\right] \otimes \left[\frac{j_R}{2}\right]. \tag{2.4}$$

While both sets of invariants take values in the integers, the refined BPS invariants $N^{\beta}_{j_L,j_R}$ are in addition non-negative. It is also useful to rewrite the refined free energy as

$$\mathcal{F}(\epsilon_1,\epsilon_2,t) = \sum_{n,g=0}^{+\infty} (\epsilon_1+\epsilon_2)^{2n}(\epsilon_1\epsilon_2)^{g-1}\mathcal{F}^{(n,g)}(t). \tag{2.5}$$

One recognizes the genus expansion of the unrefined topological string in the limit $\epsilon_1 = -\epsilon_2$. The $\mathcal{F}^{(n,g)}(t)$ can be recursively determined using the refined holomorphic anomaly equations

$$\bar{\partial}_{\bar{i}} F^{(n,g)} = \frac{1}{2}\bar{C}_{\bar{i}}^{jk}\big(D_j D_k F^{(n,g-1)} + \sum_{m,h}{}' D_j F^{(m,h)} D_k F^{(n-m,g-h)}\big), \quad n+g>1, \tag{2.6}$$

which are discussed in more detail in section 2.3 and taking the holomorphic limit

$$\mathcal{F}^{(n,g)}(t) = \lim_{\bar{t}\to i\infty} F^{(n,g)}(t,\bar{t}). \tag{2.7}$$

Mathematically, the formulation of refined BPS invariants is based on the notion of stable pairs. These are defined as a pure sheaf $\mathcal{F}$ of complex dimension one with

$$\mathrm{ch}_2(\mathcal{F}) = \beta, \qquad \chi(\mathcal{F}) = n, \tag{2.8}$$

where $\beta \in H_2(X,\mathbb{Z})$ gets identified with the D2-brane charge and $n$ determines the number of D0-branes. In addition, one requires a section $s \in H^0(\mathcal{F})$ that generates $\mathcal{F}$ outside a finite set of points $Q$. All this information can be organized within an exact sequence

$$0 \longrightarrow I_C \longrightarrow \mathcal{O}_X \xrightarrow{s} \mathcal{F} \longrightarrow Q \longrightarrow 0,$$

where $I_C$ denotes the annihilation ideal of a curve $C$ representing $\beta$. The moduli space of stable pairs with charges $\beta, n$ is denoted $P_n(X,\beta)$ and carries a symmetric and perfect obstruction theory. This implies the existence of a virtual fundamental class $[P_n(X,\beta)]^{\mathrm{vir}}$ which can be integrated to a number, the Pandharipande Thomas invariant

$$P_{n,\beta} = \int_{P_n(X,\beta)} [P_n(X,\beta)]^{\mathrm{vir}}. \tag{2.9}$$

These are organized in the following generating function

$$Z_{PT} = \sum_{n,\beta} P_{n,\beta} q^n Q^{\beta}. \tag{2.10}$$

The latter is conjectured, and in the toric case proven, to coincide with the generating function of disconnected Gromov-Witten invariants

$$Z_{GW} = \exp(\mathcal{F}_{GW}(\lambda,Q)), \qquad \mathcal{F}_{GW}(\lambda,Q) = \sum_{\beta\neq 0}\sum_{g} \lambda^{2g-2}Q^{\beta}, \tag{2.11}$$

provided one identifies $q = -e^{i\lambda}$.

Only recently the notion of refinement on the level of PT-invariants has been established in [CKK14]. It requires an $\mathbb{C}^*$-action on $P_n(X,\beta)$ with finitely many isolated fixed points $p$ and gives rise to a virtual Bialynicki-Birula cell decomposition. More precisely, for any fixed point $p$ one considers the decomposition of the tangent space into eigenspaces of positive and negative characters $\chi$ of the $\mathbb{C}^*$-action

$$T_pX = \bigoplus_{\chi\in X(\mathbb{C}^*)} T_p^\chi X = \underbrace{\bigoplus_{\chi>0} T_p^\chi X}_{T_p^+ X} \oplus \underbrace{\bigoplus_{\chi<0} T_p^\chi X}_{T_p^- X}, \tag{2.12}$$

and defines $d_p^\pm = \dim\left(T_p^\pm X\right)$. This construction allows to define the virtual motive

$$\left[P_n(X,\beta)\right]^{\text{vir}} = \sum_{p\in P_n(X,\beta)^{\mathbb{C}^*}} \left(-\mathbb{L}^{-\frac{1}{2}}\right)^{d_p^+-d_p^-}, \tag{2.13}$$

where $\mathbb{L}$ denotes the absolute motive of $\mathbb{C}$. Interpreting the right-hand side as a direct sum of irreducible SU(2) characters, one ends up with the following decomposition into refined PT invariants

$$[P_{1-p_a}(X,\beta)]^{vir} = \sum_{j_R}(-1)^{2j_R} N^\beta_{p_a/2,j_R}\left[j_R\right] \tag{2.14}$$

where $\left[j_R\right]$ denotes the representations associated to the respective characters. In particular the refined PT partition function

$$Z_{PT}^{ref} = \prod_{\beta,j_L,j_R}\prod_{m_{L/R}=-j_{L/R}}^{j_{L/R}}\prod_{m=1}^{\infty}\prod_{j=0}^{m-1}\left(1-\mathbb{L}^{-m/2+1/2+j-m_R}(-q)^{m-2m_L}Q^\beta\right)^{(-1)^{2(j_L+j_R)}N^\beta_{j_L,j_R}}, \tag{2.15}$$

coincides with the refined string partition function

$$Z^{ref} = \exp\left[\mathscr{F}(\epsilon_1,\epsilon_2,t)\right], \tag{2.16}$$

where $\mathscr{F}(\epsilon_1,\epsilon_2,t)$ denotes the refined free energy defined in (2.2).

**§2.1.1 Local A-model geometries.** A toric (i.e. in particular non-compact) Calabi-Yau three-fold $M$ is specified as the quotient of some open subset of $\mathbb{C}^{k+3}$ by a group $G = \left(\mathbb{C}^*\right)^k \times C$, where the last factor denotes a possible discrete part

$$M = \left(\mathbb{C}^{k+3} - Z\right)/G. \tag{2.17}$$

The continuous part of the group action is specified by $k$ charge vectors $Q^\alpha \in \mathbb{Z}^{k+3}$ satisfying the Calabi-Yau condition

$$\sum_{i=1}^{k+3} Q_i^\alpha = 0, \tag{2.18}$$

that act on the open subset as

$$x_i \mapsto \mu_\alpha^{Q_i^\alpha} x_i. \tag{2.19}$$

Finally, $Z$ denotes the Stanley-Reisner ideal which is the fix point set of the group action (2.19) and needs to be substracted in order to make $M$ being well-defined as a variety. All this information is encoded in the so-called toric diagram.

$$\left(\begin{array}{ccc|cccc} a_1 & b_1 & 1 & Q_1^1 & \dots & Q_1^n & \leftarrow D_1 \\ & \vdots & & & \vdots & & \\ a_k & b_k & 1 & Q_k^1 & \dots & Q_k^n & \leftarrow D_k \\ & & & \uparrow & & \uparrow & \\ & & & C^1 & & C^n & \end{array}\right) \tag{2.20}$$

Here $D_k = \{x_k = 0\}$ denotes the divisor associated to the coordinate $x_k$ corresponding to the ray $e_k = (a_k\ b_k\ 1)$ within the toric diagram. The $C^\alpha$ denote a basis of the Mori cone and constitute a distinguished basis of $H_2(X,\mathbb{Z})$. In particular, the intersection number between the divisor $D_i$ and the curve $C^\alpha$ is given by

$$Q_i^\alpha = C^\alpha \cdot D_i\,. \tag{2.21}$$

From a physics point of view, the toric variety (2.17) has an interpretation as the vacuum field configuration of a two-dimensional Sigma-model. Here the coordinates $x_i$ get identified as the vacuum expectation values of the scalars of the chiral multiplets having charge vectors $Q_i$ under the gauge group $U(1)^k$. The vacuum field configuration arises from the D-term constraints

$$D^\alpha = \sum_{i=1}^{k+3} Q_i^\alpha |x_i|^2 = r^\alpha. \tag{2.22}$$

The inequivalent vacua are obtained by dividing out the group $U(1)^k$. In addition, $r^\alpha = \int_{C^\alpha} J + iB$ correspond to the complexified Kähler parameters, where $J$ denotes the Kähler class and $B$ is the Kalb Ramond field.

**§2.2 The refined B-model.**

**§2.2.1 Local B-model geometries.** The local B-model geometry can also be constructed in terms of the toric charge vectors using the Hori-Vafa method. For this one introduces two $\mathbb{C}$-valued coordinates $w^+, w^-$ as well as the homogeneous coordinates $x_i := e^{y_i}$ which are subject to the rescaling

$$x_i \longmapsto \lambda x_i, \qquad \lambda \in \mathbb{C}^*\,, \tag{2.23}$$

and are further constrained by

$$(-1)^{Q_0^\alpha} \prod_{i=1}^{k+3} x_i^{Q_i^\alpha} = z_\alpha\,. \tag{2.24}$$

The local mirror geometry is then defined by

$$w^+ w^- = H = \sum_{i=1}^{k+3} x_i\,, \tag{2.25}$$

where the constraints (2.24) as well as the rescaling relation (2.23) can be used to eliminate all but two $(x,y)$ of the homogeneous coordinates such that the geometry takes the form of a conic bundle over a family of Riemann surfaces $H(x,y,z_\alpha)$ parametrized by the complex structure moduli $z_\alpha$. The local B-model geometry (2.25) naturally inherits a holomorphic three-form from the ambient space which is given as

$$\Omega = \frac{dH dx dy}{Hxy}. \tag{2.26}$$

After integrating out the non-compact directions this gets reduced to a meromorphic one-form defined on $H$

$$\lambda = \frac{\log(y)dx}{x}. \tag{2.27}$$

As the Riemann surfaces arising in this way are generically non-compact, one does not expect to be able to find a normalizable symplectic basis in general. Instead the most standard form of a basis $A^i, B_j$ of $H_1(H,\mathbb{Z})$ is subject to the intersections

$$A^i \cap A^j = 0, \qquad B_i \cap B_j = 0, \qquad A^i \cap B_j = n^i_j, \quad n^i_j \in \mathbb{Z}. \tag{2.28}$$

In fact, these intersections can be determined from toric diagram of the A-model geometry, see also the discussion in section 2.2.3. For the following discussion we always make use of a normalizable basis and devote the section 2.2.2 to analyse the modifications in the case of a non-normalizable basis.

One refers to the respective period integrals of $\lambda$ along this basis as the A- and B-periods respectively, denoting them by $t_i$ and $t_D^i$ respectively. These periods are annihilated by a set of linear differential operators, called the Picard-Fuchs equations

$$D_\alpha = \prod_{Q_i^\alpha>0} \partial_{x_i}^{Q_i^\alpha} - \prod_{Q_i^\alpha<0} \partial_{x_i}^{-Q_i^\alpha} . \tag{2.29}$$

The solutions are constructed by the Frobenius method. Starting with the fundamental period [CKYZ99]

$$w_0(\vec{z},\vec{\rho}) = \sum_{\vec{n}} \frac{1}{\Gamma\left(Q_i^\alpha(n^\alpha+\rho^\alpha)+1\right)}\left((-1)^{Q_0^\alpha} z^\alpha\right)^{n^\alpha+\rho^\alpha} , \tag{2.30}$$

one obtains a constant solution, as well as a number of linear logarithmic solutions.

$$X^0 = w_0(\vec{z},0) = 1, \qquad t_i = \frac{1}{2\pi i}\frac{\partial}{\partial\rho^i} w_0(\vec{z},\vec{\rho})|_{\rho=0} . \tag{2.31}$$

Second order derivatives with respect to $\vec{\rho}$ are not solutions to (2.29), but suitable linear combinations [7] of them constitute the B-periods $t_D^i$. The number of these solutions is combinatorically determined by the toric diagram as well. In fact, any inner point gives rise to a single as well as a doubly logarithmic solution (that are dual to each other), while the other points give rise to further linear logarithmic solutions. The first class of solutions corresponds to true complex structure moduli of the Riemann surface, while the second class of solutions is identified with additional residues $m_k$ of the meromorphic one-form that correspond to non-normalizable directions in the complex structure moduli space and therefore are merely parameters of the geometry rather than real moduli. As the residues of the meromorphic one-form give rise to masses of fundamental matter in the Seiberg-Witten theory engineered by this Riemann surface, these parameters are also referred to as mass parameters. From a homological point of view, these moduli correspond to curves that have no compact dual within the geometry, compare also the discussion in 2.2.3. Yet another interpretation is given as isomonodromic deformations. In fact, the periods may undergo non-trivial monodromies when encircling singularities in the complex structure moduli space. As these monodromies have to respect the symplectic form, they have to constitute a subgroup $\Gamma \subset \mathrm{Sp}_{2\mathfrak{g}}(\mathbb{Z})$. This action takes explicitly the form

$$\tilde{t}_i = a_i^j t_j + b_{ij} t_D^j + E_i^k m_k \tag{2.32}$$

$$\tilde{t}_D^i = c^{ij} t_j + c_j^i t_D^j + F^{ik} m_k \tag{2.33}$$

$$\tilde{m}_k = m_k \qquad \begin{pmatrix} a & b \\ c & d \end{pmatrix} \in \mathrm{Sp}_{2\mathfrak{g}}(\mathbb{Z}) . \tag{2.34}$$

The pairs of true periods and their duals are related to the prepotential $F^{(0,0)}$ as

$$t_D^i = \frac{\partial F^{(0,0)}}{\partial t_j} . \tag{2.35}$$

The period matrix of the mirror geometry $\tau = \left\{\tau_{ij}\right\}_{(i,j=1,\ldots,g)}$ is related to the prepotential as

$$\tau_{ij} = -\frac{\partial^2}{\partial t_k \partial t_l} F^{(0,0)} . \tag{2.36}$$

[7] Clearly, one may also add logarithmic solutions of order zero and one to the double-logarithmic solutions. The latter do not affect the instanton part of the prepotential, but only its classical contribution. Nevertheless, they are crucial in order to obtain a set of solutions that forms an integral basis for the monodromies. See [Hos06] for a conjecture how this basis is obtained. Recently, the choice of basis has been related to the $\hat{\Gamma}$-classes and the hemisphere partition function, see [CBKL14] for an application to fourfolds and more detailed references.

The third derivatives of the prepotential $F^{(0,0)}$

$$C_{ijk} = \partial_{t_i}\partial_{t_j}\partial_{t_k}F^{(0,0)}\,, \tag{2.37}$$

are called the Yukawa couplings. Finally there is the Kähler potential

$$K = \frac{1}{2i}\left(t_i \bar{t}^i_D - \bar{t}_i t^i_D\right). \tag{2.38}$$

Note that this form only holds true in the local case, in contrast to the compact case where the right hand side of (2.38) gets dressed by a logarithm. From this one deduces the Kähler metric

$$g_{t_i\bar{t}_{\bar{j}}} = \frac{\partial^2 K}{\partial t_i \partial \bar{t}_{\bar{j}}} = \mathrm{Im}\tau\,. \tag{2.39}$$

**§2.2.2 Modifications for non-compact Riemann surfaces.** In this subsection we discuss the modifications of the results obtained in the previous section for a canonically normalized basis of the Riemann surface.

We start with a basis $(A_i, B^j)$ of $A-$ and $B-$cycles of a Riemann surface of genus $\mathfrak{g}$. Their intersection matrix defines the symplectic form

$$J_{\mathfrak{g}} = \begin{pmatrix} & \mathbb{1}_{\mathfrak{g}} \\ -\mathbb{1}_{\mathfrak{g}} & \end{pmatrix}, \tag{2.40}$$

and with respect to this one defines the pair of dual cycles

$$t_i = \int_{A_i}\lambda, \qquad t^j_D = \int_{B^j}\lambda, \tag{2.41}$$

where $\lambda$ is the meromorphic one-form that is inherited from the Calabi-Yau three-form. As already discussed, $t_i$ and $t^j_D$ provide an integral basis for the monodromy group and are related via

$$t^i_D = \frac{\partial F^{(0,0)}}{\partial t_i}\,. \tag{2.42}$$

In contrast, if there is no normalizable basis, i.e. the $A-$ and $B-$ cycles intersect as

$$J^{\text{non-norm.}}_{\mathfrak{g}} = \begin{pmatrix} & C \\ -C & \end{pmatrix}, \tag{2.43}$$

the dual periods are given as non-trivial linear combinations of derivatives of the prepotential [Hos06]

$$t^i_D = C^i_j \frac{\partial F^{(0,0)}}{\partial t_j}\,, \tag{2.44}$$

and it is the pair $(t_i, t^j_D)$ that provides an integral basis for the monodromy group.

In order to derive the modification of the period matrix, we note the following behavior under the symplectic transformation

$$\begin{pmatrix} A \\ B \end{pmatrix} \mapsto \begin{pmatrix} C & 0 \\ 0 & C^{-1} \end{pmatrix}\begin{pmatrix} A \\ B \end{pmatrix} = \begin{pmatrix} CA \\ C^{-1}B \end{pmatrix}, \qquad \tau \mapsto C\tau C \quad \text{for} \quad \begin{pmatrix} C & 0 \\ 0 & C^{-1} \end{pmatrix} \in \mathrm{Sp}_{2\mathfrak{g}}(\mathbb{Z})\,. \tag{2.45}$$

Instead of a change from $(A, B)$ to $(CA, C^{-1}B)$, we can interpret (2.45) also as transforming into the basis $(A, C^{-1}B)$ while changing $J_{\mathfrak{g}}$ into $J^{\text{non-norm.}}_{\mathfrak{g}}$ at the same time. I.e. (2.45) changes the canonical symplectic form into a non-canonical form which is characterized by the intersection matrix $C$. Accordingly, the period matrix is connected to the prepotential as

$$\tau_{ij} = -C^k_i C^l_j \frac{\partial^2 F^{(0,0)}}{\partial t_k \partial t_l}\,. \tag{2.46}$$

**§2.2.3 Mirror symmetry and the mirror map.** Mirror symmetry relates the B-model on a three-fold $X$ to the A-model on the mirror three-fold $Y$. Mathematically, it is a conjectured equivalence between the bounded derived category of coherent sheaves on $X$ and the the derived Fukaya category $DFuk(Y,J)$ with the Kähler form $J$ viewed as a symplectic form. This equivalence gives in particular rise to a symplectomorphism

$$mir: K^c(X) \longrightarrow H_3(Y,\mathbb{Z})\,. \tag{2.47}$$

Here, $K^c(X)$ denotes the Grothendieck or K-theory group of coherent sheaves on $X$. As our main interest is in non-compact three-folds, we have to demand that the latter are compactly supported. As a consequence of the Grothendieck Riemann Roch theorem the Chern character map gives rise to a further symplectomorphism

$$K^c(X)\otimes\mathbb{Q} \cong A^c_{\text{even}}(X)\otimes\mathbb{Q}\,, \tag{2.48}$$

that identifies the pairing on $K^c(X)$

$$\chi(R,S) = \int \text{ch}(R^*)\text{ch}(S)\text{td}(X)\,, \tag{2.49}$$

with the intersection pairing on $A^c_{\text{even}}(X)$. The latter denotes the compactly supported Chow group which coincides with the homology group in case of crepant resolutions of $\mathbb{C}^3/G$ where $G$ is an abelian subgroup of $\text{SL}(3,\mathbb{Z})$ which is the concern of the present publication. For this class of geometries the generators of as well as the pairing in the compact K-theory group can be calculated group-theoretically using the McKay correspondence[8].

Roughly speaking, $mir$ identifies A-cycles in the mirror geometry with curves in $X$ and B-cycles with divisors such that the natural intersection pairing between curves and divisors on the left side gets identified with that of A- and B-cycles on the right side. To be more precise, we have the following identification, as compactly summarized in table 1. For every generator $C^\alpha$ of the Mori cone there is a dual element of the Kähler cone $J_k$. Also we have a basis of $H_4(X,\mathbb{Z})$ given by divisors $D_k$. To all these elements we associate the following sheaves [Hos06; OFS02]

$$C_\alpha \quad\longrightarrow\quad \mathcal{O}_{C_\alpha}(-J_\alpha) := \mathcal{O}_C\otimes\mathcal{O}(-J_\alpha), \tag{2.50}$$

$$D_k \quad\longrightarrow\quad \mathcal{O}_{D_k}(\mathcal{L}_k) := \mathcal{O}_D\otimes\mathcal{L}_k, \quad \mathcal{L}_k = \mathcal{O}_X(-a_{k,1}J_1\ldots - a_{k,l}J_l)\,. \tag{2.51}$$

Here, $\mathcal{O}_{C_\alpha}$ and $\mathcal{O}_{D_k}$ are supported[9] on $C_\alpha$ and $D_k$. The coefficients in (2.51) are chosen such that

$$\chi\left(\mathcal{O}_{D_j}(\mathcal{L}_j),\mathcal{O}_{D_k}(\mathcal{L}_k)\right) = 0\,. \tag{2.52}$$

Finally there is the D0-brane charge which is supported on a skyscraper sheaf $\mathcal{O}_p$ and corresponds to the trivial solution of the Picard-Fuchs equations. By construction, the only non-trivial intersection between the K-theory generators is given by

$$C_{\alpha\beta} = \chi\left(\mathcal{O}_{C_\alpha}(-J_\alpha),\mathcal{O}_{D_\beta}(\mathcal{L}_\beta)\right)\,, \tag{2.53}$$

which corresponds to the intersection between divisors and curves and can directly be read off from (2.20). We end by noting that the matrix $C$ defined by (2.53) is in general not regular. This can be traced back to the fact that there are curves which have no compact dual in $H_4(X,\mathbb{Q})$. These are precisely given by the mass parameters discussed in section 2.25.

[8]Strictly speaking, this is only guaranteed for the resolution given by $G$-Hilb.

[9]One way to construct these sheaves is to consider a coherent sheaf $S_V$ on a compact sub-manifold $\iota: V \hookrightarrow X$ and to compute the compact Chern character $\text{ch}^c(S) := \text{ch}(\iota_* S_V)$ by the Grothendieck Riemann Roch theorem $\iota_*(\text{ch}(S_V)\text{td}(V)) = \text{ch}(\iota_* S_V)\text{td}(X)$.

| Solution of the PF eqn. | Charge | $H_{\text{even}}(X,\mathbb{Z})$ | $K^c(X)$ | comp. dual |
|---|---|---|---|---|
| 1 | D0 | $H_0(X,\mathbb{Z})$ | $\mathcal{O}_p$ | no |
| transcendental single log. | D2 | $H_2(X,\mathbb{Z})$ | $\mathcal{O}_{C^\alpha}(-J_\alpha)$ | yes |
| rational single log. | D2 | $H_2(X,\mathbb{Z})$ | $\mathcal{O}_{C^\alpha}(-J_\alpha)$ | no |
| double log. | D4 | $H_4(X,\mathbb{Z})$ | $\mathcal{O}_{D_j}(\mathcal{L}_j)$ | yes |

Table 1: The elements of the compact K-theory group together with their associated D-brane charges and mirror maps.

In the second part of this subsection we aim to discuss a a second - obviously not unrelated - notion of mirror map which provides an identification of the Kähler moduli in the A-model on $Y$ in terms of the complex structure moduli of the B-model on $X$. Concretely, one identifies for every normalizable complex structure parameter $z_i$ the logarithmic solution given as

$$t_i = \frac{1}{2\pi i}\log(z_i) + \Sigma(z_j)\,, \tag{2.54}$$

where $\Sigma(z_j)$ denotes a power series with the Kähler modulus $t_i$. The relation with the homological mirror map is obtained by noting that the complex structure moduli are parameterized by $H^{(0,1)}(X, T^{(1,0)}X) \cong H^{(2,1)}(X)$ and the Kähler moduli by $H^{(1,1)}(X)$ and passing to the respective Poincaré dual homology. The transcendental mirror map is finally given by inverting (2.54) in favor of the monodromy invariant variables $Q_i = e^{2\pi i t_i}$. In contrast, moduli corresponding to non-normalizable directions have rational mirror maps.

**§2.3 The refined holomorphic anomaly equations.** Refined holomorphic anomaly equations have been proposed in [HK12; HKPK13]. They read

$$\bar\partial_{\bar i} F^{(n,g)} = \frac{1}{2}\bar C_{\bar i}^{jk}\Big(D_j D_k F^{(n,g-1)} + \sum_{m,h}{}' D_j F^{(m,h)} D_k F^{(n-m,g-h)}\Big)\,, \quad n+g>1\,. \tag{2.55}$$

Here the prime signals the omission of $(m,h)=(0,0)$ and $(n,g)$. The covariant derivatives are with regard to the bundle $T^{(1,0)}X$ and its symmetric powers. More precisely, for a correlation function of $k$ operators they take the form

$$D_t = \partial_t - k\Gamma_t\,, \tag{2.56}$$

where $\Gamma$ denotes the Christoffel symbol of the Kähler metric (2.39). In general, the free energies at genus $g$ transform as a section of a line bundle $\mathcal{L}^{2g-2}$ that gives an additional contribution to the connection in (2.56). However, in the local limit, $\mathcal{L}$ becomes trivial and (2.56) is exact.

While the unrefined holomorphic anomaly equations can be rigorously derived, a worldsheet interpretation of the refined holomorphic anomaly equations is so far missing. However, it was conjectured in [HKPK13] that the second deformation parameter corresponds to the insertion of $n$ times an operator $\mathcal{O}$ into the worldsheet of genus $g$. Accordingly, the refined free energies take now the form

$$F^{(n,g)} = \int_{\bar{\mathcal{M}}_g} \langle \mathcal{O}^n \prod_{k=1}^{3g-3} \beta^k \bar\beta^k\rangle_g dm\wedge d\bar m\,. \tag{2.57}$$

Thus, the second sum in (2.55) gets an interpretation as the distribution of the $n$ points among the degenerated components of the Riemann surface. In particular, it was conjectured that

$$F^{(n+1,0)} = \langle \phi^{(0)}(0)\phi^{(0)}(1)\phi^{(0)}(\infty)\mathcal{O}^n\rangle_{g=0}\,. \tag{2.58}$$

This implies in particular that $F^{(1,0)}$ is purely holomorphic such that its most general ansatz reads

$$F^{(1,0)} = \frac{1}{24}\log\left(\Delta \prod_i u_i^{a_i}\prod_j m_j^{b_j}\right)\,, \tag{2.59}$$

where $\Delta$ denotes the discriminant locus of the mirror curve and the constants $a_i, b_j$ are fixed by requiring regularity at infinity and the knowledge of a few BPS numbers. In contrast, $F^{(0,1)}$ corresponds to the unrefined free energy and obeys a holomorphic anomaly equation as well. The most general ansatz for $F^{(0,1)}$ is given in the local limit by

$$F^{(0,1)} = \frac{1}{2}\log\left(\Delta^a \prod_i u_i^{a_i} m_j^{b_j} |g^{-1}_{i\bar{j}}|\right). \tag{2.60}$$

Here $g_{z_i\bar{z}_j} = \partial_{z_i}\bar{\partial}_{\bar{z}_j}K$ refers to the Kähler metric (2.39) transformed into the coordinates $z_i$. This can be re-expressed as

$$g_{z_i\bar{z}_j} = \left(\frac{\partial^2 K}{\partial t_k \partial \bar{t}_{\bar{l}}}\right)\frac{\partial t_k}{\partial z_i}\frac{\partial \bar{t}_{\bar{l}}}{\partial \bar{z}_{\bar{j}}} = (\mathrm{Im}\,\tau_{kl})\, G_i^k \bar{G}_{\bar{j}}^{\bar{l}}. \tag{2.61}$$

The quantity

$$G_i^j = \frac{\partial t_i}{\partial z_j}, \tag{2.62}$$

is often referred to as the topological metric in the local and holomorphic limit. Clearly, the factor $|\mathrm{Im}\tau_{ij}|$ is not holomorphic. Accordingly, $F^{(0,1)}$ can be split as

$$\begin{aligned} F^{(0,1)} &= \frac{1}{2}\log\left(|\mathrm{Im}\tau_{ij}|^{-1}\right) + \frac{1}{2}\log\left(|(\bar{G}^{-1})_{\bar{i}}^{\bar{j}}|\right) + \frac{1}{2}\log\left(\Delta^a \prod_i u_i^{a_i} m_j^{b_j} |(G^{-1})_i^j|\right) \\ &= \frac{1}{2}\log\left(|\mathrm{Im}\tau_{ij}|^{-1}\right) + \frac{1}{2}\log\left(|(\bar{G}^{-1})_{\bar{i}}^{\bar{j}}|\right) + \mathscr{F}^{(0,1)}. \end{aligned} \tag{2.63}$$

As discussed in detail in section 4 for genus one and two, $F^{(0,1)}$ is invariant under modular transformations, but not holomorphic. In contrast, $F^{(0,1)}$ alone is not modular invariant.

Having discussed the antagonistic behavior under modularity respectively holomorphicity of $F^{(0,1)}$, we remark that the ambiguity can be fixed with the help of a few known BPS numbers, analogously to the case of $F^{(1,0)}$. Alternatively, they can also be determined from the behavior of $\mathscr{F}^{(0,1)}$ near singularities, e.g.

$$\lim_{z_i\to 0} \mathscr{F}^{(0,1)} = -\frac{1}{24}\sum_{j=1}^{h^{2,1}} t_j \int_X c_2 J_j. \tag{2.64}$$

Besides the prepotential $F^{(0,0)}$, the knowledge of the free energies at genus one is sufficient to determine the higher genus free energies $F^{(n,g)}$ using the refined holomorphic anomaly equations up to an integration constant, called the ambiguity. The direct integration method proceeds by calculating the so-called propagators $S^{ij}$. They can be used to re-express the anti-holomorphic derivative of $F^{(n,g)}$ as

$$\bar{\partial}_{\bar{i}} F^{(n,g)} = C_{\bar{i}}^{jk}\frac{\partial F^{(n,g)}}{\partial S^{jk}}, \tag{2.65}$$

which implies that $F^{(n,g)}$ is a polynomial in $S^{ij}$ of degree $3(g+n)-3$. The propagators are determined from a - in the multi-moduli case overdetermined - set of equations that reads for local Calabi-Yau manifolds as follows

$$D_i S^{kl} = -C_{imn}S^{km}S^{lm} + f_i^{kl}, \tag{2.66}$$

$$\Gamma_{ij}^k = -C_{ijl}S^{kl} + \tilde{f}_i^{kl}, \tag{2.67}$$

$$\partial_i F^{(0,1)} = \frac{1}{2}C_{ijk}S^{jk} + A_i. \tag{2.68}$$

In these equations the ambiguities $f_i^{kl}, \tilde{f}_i^{kl}, A_i$ can be solved from the following ansätze

$$\begin{aligned} f_i^{kl} &= \frac{h(z_i)}{\prod_i z_i^{m_i}\Delta^p}, \\ A_i &= \partial_i\left(\tilde{a}_j \log\Delta_j + \tilde{b}_j \log z_j\right). \end{aligned}$$

The first formula applies to both, $f_i^{kl}$ and $\tilde{f}_i^{kl}$ and $p$ is either zero or one while $h(z_i)$ is a polynomial.

Finally, the integration constant in (2.55) can be fixed by matching the refined free energies with a very particular behaviour of the refined free energies at the conifold locus known as the gap condition. It arises from integrating out a particle becoming massless by performing a Schwinger Loop computation, where the coordinate $t$ denotes the flat coordinate normal to the singularity

$$\begin{aligned}\mathscr{F}(s,g_s,t) &= \int_0^\infty \frac{ds}{s}\frac{\exp(-st)}{4\sinh(s\epsilon_1/2)\sinh(s\epsilon_2/2)}+\mathcal{O}(t^0) \qquad (2.69)\\ &= \Big[-\frac{1}{12}+\frac{1}{24}(\epsilon_1+\epsilon_2)^2(\epsilon_1\epsilon_2)^{-1}\Big]\log(t)\\ &\quad+\frac{1}{\epsilon_1\epsilon_2}\sum_{g=0}^{\infty}\frac{(2g-3)!}{t^{2g-2}}\sum_{m=0}^{g}\hat{B}_{2g}\hat{B}_{2g-2m}\epsilon_1^{2g-2m}\epsilon_2^{2m}+\dots\\ &\quad=\Big[-\frac{1}{12}+\frac{1}{24}sg_s^{-2}\Big]\log(t)+\Big[-\frac{1}{240}g_s^2+\frac{7}{1440}s-\frac{7}{5760}s^2g_s^{-2}\Big]\frac{1}{t^2}\\ &\quad+\Big[\frac{1}{1008}g_s^4-\frac{41}{20160}sg_s^2+\frac{31}{26880}s^2-\frac{31}{161280}s^3g_s^{-2}\Big]\frac{1}{t^4}+\mathcal{O}(t^0)\\ &\quad+\text{contributions to } 2(g+n)-2>4\,. \qquad (2.70)\end{aligned}$$

Here, $\hat{B}_m=\left(\frac{1}{2^{m-1}}-1\right)\frac{B_m}{m!}$ with $B_m$ denoting the Bernoulli numbers.

**§2.4 Background independence of the topological string partition function and modularity.** Besides the (refined) holomorphic anomaly equations, there is a second way to compute the higher genus free energies which (so far) only applies to the unrefined case. This method is based on an analysis of the monodromy group $\Gamma$ that strongly constrains the form of the free energies. Namely it requires them to transform as almost/quasi modular forms of $\Gamma$, as explained in the following. In this case the topological string partition function is interpreted as a state $|Z\rangle$ in the Hilbert space obtained by quantizing $H^3(M,\mathbb{Z})$ where $g_s$ plays the role of $\hbar$. While $|Z\rangle$ as an abstract state is invariant under the monodromy action, this does not neccessarily have to hold for its wavefunction (i.e. the projection of $|Z\rangle$ onto a particular set of coordinates). There are two important choices of coordinates. The first set $\{z_i,\bar{z}_i\}$ refers to picking a complex structure (holomorphic polarization), while the second set $\{x_i,p_i\}$ refers to a background symplectic structure (real polarization).

The results of this analysis, performed in [ABK08] are as follows:

1. In the holomorphic polarization the free energies are almost modular forms of the monodromy group $\Gamma$ (i.e. modular invariant but not holomorphic) having an expansion

$$\widetilde{F}^{(g)}=F_0^{(g)}+\sum_{i=1}^{3g-3}h_g^i(\tau)(\mathrm{Im}\tau)^{-i}$$

with $h^i(\tau)$ being holomorphic functions. From a physics point of view these are determined by the holomorphic anomaly equation up to an integration constant.

2. In the real polarization the free energies are quasi modular forms of the monodromy group $\Gamma$ (i.e. holomorphic but not modular invariant) $F^{(g)}$ and in fact $F^{(g)}\equiv F_0^{(g)}$.

3. The holomorphic anomaly equation can be derived from transforming from the real polarization to the complex polarization.

4. The propagator can be shown to be a modular form of weight two

$$\widetilde{E}^{IJ}(\tau,\bar{\tau})=E^{IJ}(\tau)+\left((\mathrm{Im}\tau)^{-1}\right)^{IJ},$$

transforming as

$$\widetilde{E}^{IJ}(\tau,\bar{\tau}) \mapsto (c\tau+d)^I_K (c\tau+d)^J_L \widetilde{E}^{KL}(\tilde{\tau},\bar{\tilde{\tau}}), \tag{2.71}$$

under

$$\tau \longmapsto \gamma\tau = (a\tau+b)(c\tau+d)^{-1}, \qquad \gamma = \begin{pmatrix} a & b \\ c & d \end{pmatrix} \in \mathrm{Sp}_{2\mathfrak{g}}(\mathbb{Z}). \tag{2.72}$$

Here $E$ takes values in $\mathbb{C}^{\mathfrak{g}\times\mathfrak{g}}$. In fact, any free energy $\widetilde{F}^{(g)}$ in holomorphic polarization can be written as a polynomial in $\widetilde{E}^{IJ}(\tau,\bar{\tau})$ where the coefficients are given by meromophic modular forms.

# 3 | Invariants of Riemann surfaces and Their Modular Properties

This section is devoted to discuss the invariants of elliptic and genus two curves respectively. We mainly focus on the Igusa invariants that classify the latter. In particular, their Fourier expansion is used in the sections 4 and 5 in order to extract the period matrix of the mirror curve.

**§3.1 Invariants of genus one curves.** Any genus one curve can be represented in Weierstrass normal form

$$y^2 = 4x^3 - g_2(u,m_i)x - g_3(u,m_i). \tag{3.1}$$

Here $u$ denotes the the true modulus of the curve that corresponds to the complex structure modulus $\tau$, while the parameters $m_i$ denote possible isomonodromic deformations as discussed in section 2.2.1. The coefficients $g_2, g_3$ are not true invariants, but enjoy a rescaling symmetry

$$g_2 \mapsto r^4 g_2, \qquad g_3 \mapsto r^6 g_3. \tag{3.2}$$

In particular there is an $r$ such that

$$E_4 = 12r^4 g_2, \qquad E_6 = 216 r^6 g_3, \qquad \Delta_{\mathrm{mod}} = r^{12}\Delta_{\mathrm{dis}}. \tag{3.3}$$

Here, we denote the discriminant locus respectively the modular discriminant by

$$\Delta_{\mathrm{mod}} = \frac{1}{1728}\left(E_4^3(\tau) - E_6^2(\tau)\right), \qquad \Delta_{\mathrm{dis}} = g_2^3(u,m_i) - 27g_3^2(u,m_i). \tag{3.4}$$

In contrast to $g_2, g_3$, the associated $j$−function

$$j = \frac{g_2^3(u,m_i)}{\Delta(u,m_i)} = \frac{E_4^3(\tau)}{E_4^3(\tau)-E_6^2(\tau)} = \frac{1}{q} + 744 + \ldots, \qquad q = e^{2\pi i\tau}, \tag{3.5}$$

is a true invariant.

**§3.2 Invariants of genus two curves.** Any Riemann surface $R$ of genus $g = 2$ can be represented as hyperelliptic curve

$$\begin{aligned} y^2 &= v_0x^6 - v_1x^5 + v_2x^4 - v_3x^3 + v_4x^2 - v_5x + v_6, \\ &= \prod_{i=1}^{6}(x-\lambda_i) \quad v_i, \lambda_i \in \mathbb{C}. \end{aligned} \tag{3.6}$$

To see this one notices that a hyperelliptic curve is a double cover of $\mathbb{P}^1$ which is equivalent to a function of degree two on $R$. The existence of the latter is guaranteed by the Riemann Roch theorem.

Two such hyperelliptic curves are conformally equivalent and define therefore the same Riemann surface if their branch points $\lambda_i$ differ by a fractional linear transformation. Thus the moduli space of the hyperelliptic curves is locally parametrized by the $\lambda_i$ and has dimension

$$\dim \mathcal{M}_g^{\mathrm{hyp}} = 2g + 2 - \dim \mathrm{PSL}_2(\mathbb{Z}) = 2g - 1. \tag{3.7}$$

Note that this coincides for $g = 2$ with dimension of the moduli space of Riemann surfaces

$$\dim \mathcal{M}_g = 3g - 3. \tag{3.8}$$

**§3.2.1 The Igusa invariants.** In this subsection we discuss the Igusa invariants which provide a natural generalization of $g_2, g_3$ which occur in the Weierstrass normal form and the absolute $j$-invariant.

Given a genus two curve in the form (3.6), the Igusa invariants are defined as

$$A(v_i) = v_0^2 \sum_{15} (12)^2(34)^2(56)^2, \quad (3.9)$$

$$B(v_i) = v_0^4 \sum_{10} (12)^2(23)^2(31)^2(45)^2(56)^2(64)^2, \quad (3.10)$$

$$C(v_i) = v_0^6 \sum_{60} (12)^2(23)^2(31)^2(45)^2(56)^2(64)^2(14)^2(25)^2(36)^2, \quad (3.11)$$

$$D(v_i) = v_0^{10} \prod_{i<j} (\lambda_i - \lambda_j). \quad (3.12)$$

In these expressions we have summed over all permutations $\sigma \in \mathfrak{S}_6$ and used the abbreviation $(ij) = (\lambda_{\sigma(i)} - \lambda_{\sigma(j)})$. The invariants (3.9), (3.10), (3.11), (3.12) for a sextic of the form (3.6) are given by

$$\begin{aligned}
A =& 6v_3^2 - 16v_2v_4 + 40v_1v_5 - 240v_0v_6, \\
B =& 48v_6v_2^3 + 4v_4^2v_2^2 - 12v_3v_5v_2^2 + 300v_0v_5^2v_2 + 4v_1v_4v_5v_2 - 180v_1v_3v_6v_2 - 504v_0v_4v_6v_2 + 48v_0v_4^3 \\
& - 12v_1v_3v_4^2 - 80v_1^2v_5^2 + 1620v_0^2v_6^2 + 36v_1v_3^2v_5 - 180v_0v_3v_4v_5 + 324v_0v_3^2v_6 + 300v_1^2v_4v_6 \\
& - 540v_0v_1v_5v_6, \\
C =& -36v_5^2v_2^4 - 160v_4v_6v_2^4 - 24v_4^3v_2^3 - 96v_0v_6^2v_2^3 + 76v_3v_4v_5v_2^3 + 60v_3^2v_6v_2^3 + 616v_1v_5v_6v_2^3 \\
& + 8v_3^2v_4^2v_2^2 + 26v_1v_3v_5^2v_2^2 - 640v_0v_4v_5^2v_2^2 - 900v_1^2v_6^2v_2^2 - 24v_3^3v_5v_2^2 + 28v_1v_4^2v_5v_2^2 \\
& + 424v_0v_4^2v_6v_2^2 + 492v_1v_3v_4v_6v_2^2 - 876v_0v_3v_5v_6v_2^2 - 160v_0v_4^4v_2 + 76v_1v_3v_4^3v_2 \\
& + 1600v_0v_1v_5^3v_2 + 330v_0v_3^2v_5^2v_2 + 64v_1^2v_4v_5^2v_2 + 3060v_0v_1v_3v_6^2v_2 + 20664v_0^2v_4v_6^2v_2 \\
& + 492v_0v_3v_4^2v_5v_2 - 238v_1v_3^2v_4v_5v_2 - 198v_1v_3^3v_6v_2 - 640v_1^2v_4^2v_6v_2 - 18600v_0^2v_5^2v_6v_2 \\
& - 468v_0v_3^2v_4v_6v_2 - 1860v_1^2v_3v_5v_6v_2 + 3472v_0v_1v_4v_5v_6v_2 - 36v_1^2v_4^4 + 60v_0v_3^2v_4^3 - 320v_1^3v_5^3 \\
& + 2250v_0^2v_3v_5^3 - 119880v_0^3v_6^3 - 24v_1v_3^3v_4^2 + 176v_1^2v_3^2v_5^2 - 900v_0^2v_4^2v_5^2 - 1860v_0v_1v_3v_4v_5^2 \\
& - 10044v_0^2v_3^2v_6^2 + 2250v_1^3v_3v_6^2 - 18600v_0v_1^2v_4v_6^2 + 59940v_0^2v_1v_5v_6^2 + 72v_1v_3^4v_5 + 616v_0v_1v_4^3v_5 \\
& + 26v_1^2v_3v_4^2v_5 - 198v_0v_3^3v_4v_5 + 162v_0v_3^4v_6 - 96v_0^2v_4^3v_6 - 876v_0v_1v_3v_4^2v_6 - 2240v_0v_1^2v_5^2v_6 \\
& + 330v_1^2v_3^2v_4v_6 + 1818v_0v_1v_3^2v_5v_6 + 1600v_1^3v_4v_5v_6 + 3060v_0^2v_3v_4v_5v_6, \\
D =& v_0^2\Delta.
\end{aligned} \quad (3.13)$$

The invariants $A, B, C, D$ are the genus two analogues of $g_2$ and $g_3$ in the Weierstrass form. I.e. the rescaled coordinates

$$r^2A, \qquad r^4B, \qquad r^6C, \qquad r^{10}D, \qquad r \in \mathbb{C}, \quad (3.14)$$

define the same curve. In particular, one can find a proportionality constant such that one has [10]

$$A = -24r^2\frac{\chi_{12}}{\chi_{10}}, \qquad B = 4r^4E_4, \qquad C' = \frac{1}{2}(AB - 3C) = 4r^6E_6, \quad D = -2^{14}r^{10}\chi_{10}. \quad (3.15)$$

Altogether these invariants define a point in $\mathbb{P}^{1,2,3,5}$ and the complement of the divisor $D = 0$ describes the moduli space of genus two curves. However, if three roots in (3.6) coincide, the genus two curve degenerates into a product of elliptic curves. At the same time, the three invariants $B, C, D$ vanish simultaneously such that the two-dimensional subspace parametrizing the two factors gets mapped onto the point $[1:0:0:0]$. One can show [Igu62; Igu67] that the following absolute invariants

$$x_1 = 2^43^2\frac{B}{A^2}, \qquad x_2 = -2^63^3\frac{AB-3C}{A^3}, \qquad x_3 = 2\cdot 3^5\frac{D}{A^5}, \quad (3.16)$$

[10]The Maass lift $\phi_{10}^{(2)}$ used in section 9 is related to Igusa's cusp form by $4\chi_{10} = \phi_{10}^{(2)}$.

serve as good uniformizing parameters around $[1:0:0:0]$. In particular, by introducing the following quantities

$$y_1 = \frac{x_1^3}{x_3}, \qquad y_2 = \frac{x_2^2}{x_3}, \qquad y_3 = \frac{x_1^2 x_2}{x_3}. \tag{3.17}$$

one obtains the relations between the invariants of the degenerated genus two curve and its respective elliptic factors as

$$y_1 = j(\tau_1)j(\tau_2), \qquad y_2 = \left(j(\tau_1) - 2^6 3^3\right)\left(j(\tau_2) - 2^6 3^3\right). \tag{3.18}$$

In this work we will eventually be interested to use the Fourier expansion of (3.16) in order to extract the complex structure parameters of a given genus two curve. We review some basic properties of the ring of Siegel modular forms and the computation of their Fourier expansions in the appendix C.

# 4 | Direct Integration Algorithms based on Normal Forms of Mirror Curves

The methods presented in section 2.3 are perfectly sufficient in order to compute the (refined) free energies up to arbitrary worldsheet genus. However, for mirror curves of genus one and two, the fact that these possess a normal form allows to directly extract many quantities from the toric diagram. The key observation is that all quantities of the associated topological field theory can be expressed in terms of invariants of the curves and their period matrices. The latter are obtained by inverting the Fourier expansion of the associated absolute invariants which are constituted by the j-function respectively the Igusa invariants. In contrast, there is no way to circumvent the direct integration procedure to obtain the higher genus invariants in general. However there is one universal object, the propagator, which in fact can be written down once the period matrices are known. It is given by the second Eisenstein series $E_2$ in the case of genus one and by the derivative of $\log\left(\chi_{10}\right)$ in the case of genus two.

Mathematically, for the case of genus one this is reflected in the fact that it is precisely the second Eisenstein series whose adjunction turns the ring of modular forms into the ring of almost respectively quasi modular forms. We show in section 9 that a similar statement can be made for genus two. The complete modular structure which is ensured to hold at least in the unrefined case is encoded in the monodromy group of the individual mirror curves and needs to be determined case by case.

In the following two subsections we discuss the algorithmic integration procedures for genus one and two separately.

**§4.1 Direct Integration for genus one mirror curves.** We summarize the integration procedure in figure 1 and discuss in addition some salient points of the algorithm.

**§4.1.1 Obtaining the periods.** The relation between the A-period $t$ of the meromorphic differential and the period of the mirror curve at a cusp point in the moduli space is given by

$$\frac{dt}{du} = \sqrt{\frac{E_4(\tau) g_3(u, m_i)}{E_6(\tau) g_2(u, m_i)}} \overset{!}{=} \frac{1}{2\pi i}\int_\mu \frac{dx}{y}. \tag{4.1}$$

It is obvious that this expression is not invariant under the rescalings (3.2). This ambiguity is fixed by demanding that it coincides with the explicit integral over the vanishing cycle $\mu$ which implies the leading behaviour $t = \frac{1}{2\pi i}\log(u) + \ldots$. It is trivial, but also important to note that (4.1) is also the holomorphic limit of the topological metric and is to be identified[11] with the proportionality constant $r$ in (3.3).

[11] Up to a numerical factor of 18.

**§4.1.2 Obtaining the propagator.** It is also possible to derive a closed expression for the propagator. One notices from (2.60) and (2.68) that one can choose $A_t$ in such a way that one obtains [12]

$$\frac{1}{2}C_{ttt}\mathscr{S}^{tt} = -\frac{1}{24}\partial_t \log(\Delta_{\mathrm{mod}}) \,. \tag{4.2}$$

From this one easily obtains

$$\mathscr{S}^{tt} = \frac{c_0}{12}E_2 \,. \tag{4.3}$$

We denote by $c_0$ the intersection number of the A-cycle with the B-cycle. Here we made use of (2.46) and the well-known relation

$$E_2(\tau) = \frac{1}{2\pi i}\frac{d}{d\tau}\log(\Delta_{\mathrm{mod}}). \tag{4.4}$$

After tensor transforming to the $u$-coordinates, one obtains

$$\mathscr{S}^{uu} = \frac{c_0}{12}\frac{g_3(u,m_i)}{g_2(u,m_i)}\frac{E_2(\tau)E_4(\tau)}{E_6(\tau)}. \tag{4.5}$$

In this derivation we have used the generic form (2.63) of $\mathscr{F}^{(0,1)}$. Taking into account the modular complement in equation (2.61), the right hand side of (4.2) picks up an additional summand

$$-\frac{1}{2}\log(\mathrm{Im}\tau) \,, \tag{4.6}$$

which precisely cancels the modular factor of the discriminant[13]. The relation (4.3) gets accordingly replaced by

$$S^{tt} = \frac{c_0}{12}\left(E_2 - \frac{3}{\pi \mathrm{Im}\tau}\right). \tag{4.7}$$

In addition, one has to check in general that the relations (2.66) and (2.67) are satisfied as well. Here we note that for genus two curves the relation (2.66) becomes just the Ramanujan identity

$$\partial_\tau E_2 = \frac{1}{12}\left(E_2^2 - E_4\right). \tag{4.8}$$

This follows from (4.3) and (2.46). In particular, one finds that

$$f_t^{tt} = \frac{c_0^2}{144}C_{ttt}E_4(q)\,. \tag{4.9}$$

After transforming to $u$-coordinates one obtains

$$f_z^{zz} = \frac{c_0^2}{12}g_2 C_{zzz}\,. \tag{4.10}$$

[12] In the following, we use $\mathscr{S}$ to denote the holomorphic part of the propagator.

[13] Actually this is not completely true. Under a modular transformation $\tau \mapsto \frac{a\tau+b}{c\tau+d}$, $\frac{1}{\mathrm{Im}\tau}$ transforms into $\frac{|c\tau+d|^2}{\mathrm{Im}\tau}$. But (2.61) also contains a term $\bar\partial_{\bar u}\bar t$ which is not taken into account in (4.2). Altogether these three contributions are modular invariant.

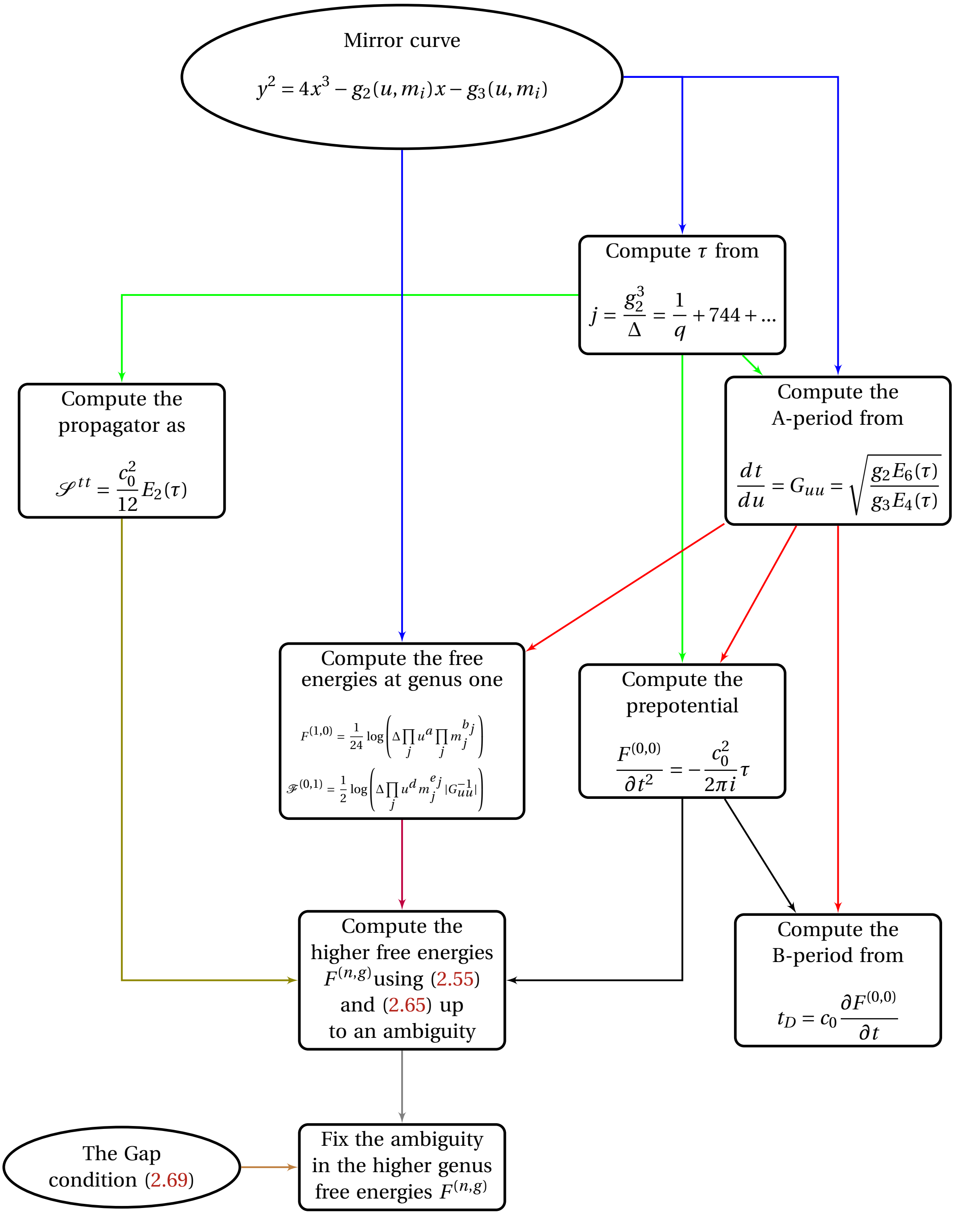

Figure 1: This diagram shows the algorithm based on genus one mirror curves to compute the free energies via direct integration. Boxes show quantities that can derived purely from the mirror curve while clouds denote information that is needed in addition.

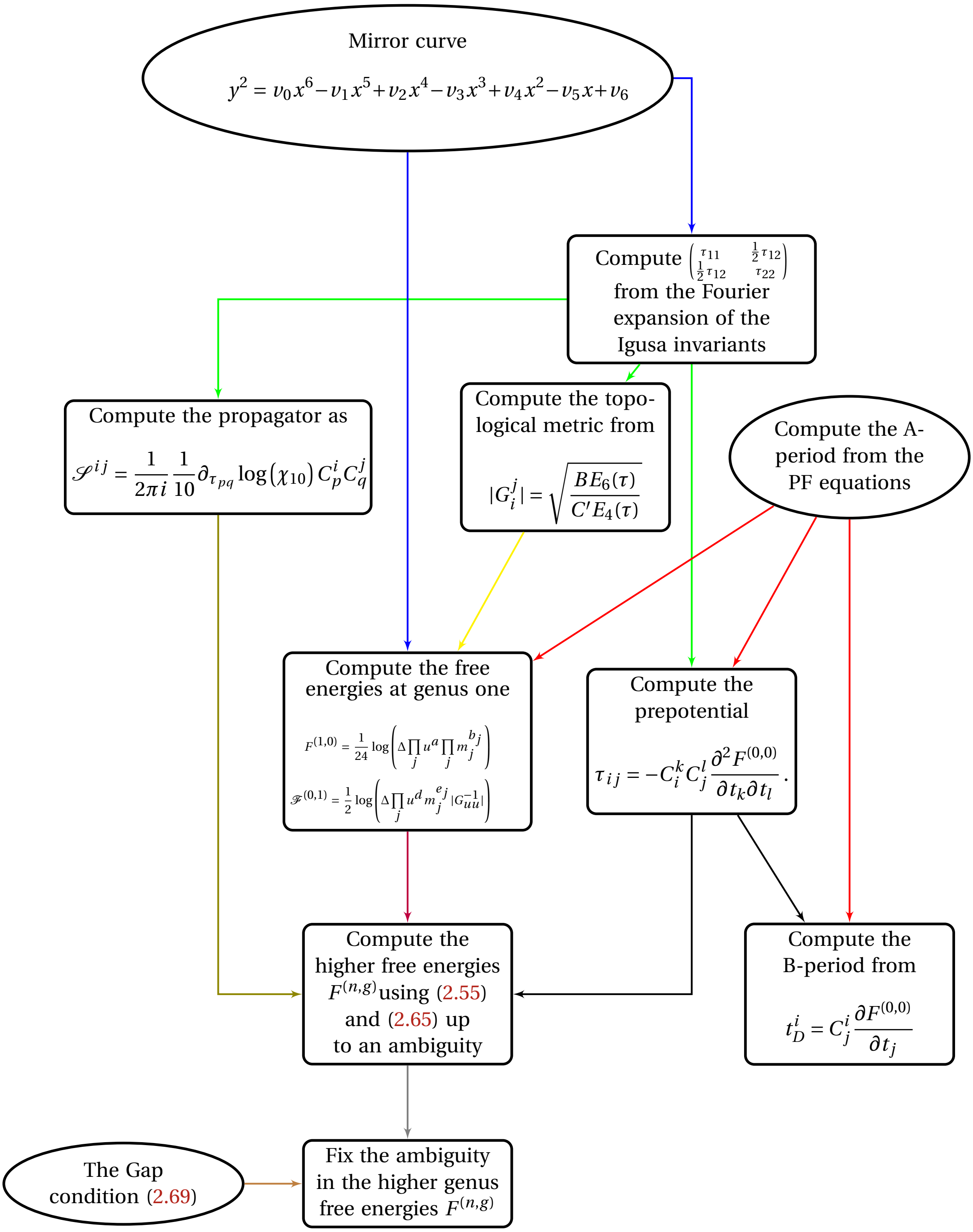

Figure 2: This diagram shows the algorithm based on genus two mirror curves to compute the free energies via direct integration. Boxes show quantities that can derived purely from the mirror curve while clouds denote information that is needed in addition. In contrast to the genus one case, also the knowledge of the A-periods is required.

### §4.2 Direct integration for genus two mirror curves.

**§4.2.1 Obtaining the periods and the topological metric.** Unfortunately there seems to be no known generalization of (4.1) in order to obtain the A-periods of the meromorphic differential directly. Therefore these have to be computed by making use of the Picard-Fuchs equations in the case of genus two. However we conjecture that (4.1), interpreted as the topological metric, generalizes as follows to the case of genus two

$$|G_{ij}| = \mathrm{Det}\left(\partial_{u_i} t_j\right) = \sqrt{\frac{E_6(\tau)B(u_i,m_j)}{E_4(\tau)C'(u_i,m_j)}}. \tag{4.11}$$

Clearly, also this expression is not invariant under rescalings and we further conjecture that this freedom can be fixed by requiring a leading term of $\prod_i \frac{1}{u_i}$. We have successfully checked this conjecture for the examples of the resolutions of $\mathbb{C}^3/\mathbb{Z}_5$ and $\mathbb{C}^3/\mathbb{Z}_6$, see also the discussion in sections 5.2 and 5.3. Finally we note that (4.11) provides up to a numerical factor the proportionality constant $r$ in (3.14).

**§4.2.2 Obtaining the propagator.** The same derivation as above applies to the case of genus two. Analogously to the previous discussion, one chooses the ambiguity in (2.68), such that one obtains

$$\frac{1}{2}C_{ijk}S^{jk} = -\frac{1}{20}\partial_i \log\left(\chi_{10}\right) - \frac{1}{2}\partial_i \log\left(\det\mathrm{Im}\tau\right). \tag{4.12}$$

Here, we abbreviate $t_i$ by $i$ and have used (3.14).

In this case one finds the following equality for the propagator

$$S^{ij} = \frac{1}{2\pi i}\frac{1}{10}\left(\partial_{\tau_{pq}} \log\left(\chi_{10}\right) + \frac{5}{i}\left(\mathrm{Im}\tau\right)^{-1}_{pq}\right) C_i^p C_j^q. \tag{4.13}$$

As for the well-known case of genus one, it is interesting to wonder whether the ambiguity (2.66) is also fixed by a generalized Ramanujan identity. In fact, we derive such an identity for the almost meromorphic Siegel modular form $\partial_\tau \log\left(\chi_{10}\right)$ in section 9, see (9.3) and (9.5) for an explicit realization.

We recall the constraint (2.66) for the propagator which reads in the flat coordinates $t_i$

$$\partial_i \mathscr{S}^{kl} = -C_{imn}\mathscr{S}^{km}\mathscr{S}^{lm} + f_i^{kl}. \tag{4.14}$$

Again, we can make use of the relation (2.46) to re-write this as

$$\left(R_{\mathrm{sym}^2}\mathscr{S}\right)^{rs,mn} = t(\mathscr{S}\otimes\mathscr{S})^{rs,mn} + \frac{\partial t^i}{\partial \tau_{rs}}\left(C^{-1}\right)^m_k \left(C^{-1}\right)^m_l f_i^{kl}. \tag{4.15}$$

By abuse of notation, $\mathscr{S}^{mn}$ denotes here only $\partial_{\tau_{mn}} \log\left(\chi_{10}\right)$ and the definition of the operators $R_{\mathrm{sym}^2}$ and $t(S\otimes S)$ is provided in section 6.1. Comparing this expression to (9.3), it follows that the ambiguities $f_i^{mn}$ are universally determined and can be expressed in terms of the meromorphic Siegel modular form $f_{\mathrm{RS}}$ of weight $\mathrm{sym}^2\otimes\mathrm{sym}^2$ as

$$f_i^{mn} = -C_{irs}C^r_k C^s_l C^m_o C^n_p (f_{\mathrm{RS}})^{kl,op}. \tag{4.16}$$

# 5 | Examples

In this section we exemplify the theory which has been developed in sections 2, 3 and 4 by solving the topological string on three different toric geometries. The first one is given by local $\mathbb{P}^2$. Besides being the standard example of a geometry with genus one mirror curve it also serves as an interesting limiting case of the resolution of $\mathbb{C}^3/\mathbb{Z}_5$. The latter geometry has a genus two mirror curve with two real moduli and no mass parameter deformations. Finally, we consider an example with a mass parameter which is provided by the resolution of $\mathbb{C}^3/\mathbb{Z}_6$. In all calculations we find our formalism to be perfectly confirmed.

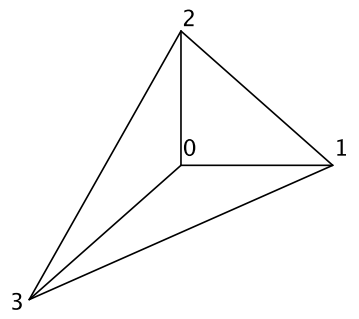

$$\left(\begin{array}{ccc|c} 0 & 0 & 1 & -3 \\ 1 & 0 & 1 & 1 \\ 0 & 1 & 1 & 1 \\ -1 & -1 & 1 & 1 \end{array}\right) \tag{5.1}$$

## §5.1 Solving the Topological String on local $\mathbb{P}^2$.

**§5.1.1 The geometry and its mirror.** First we will solve the refined topological string on the anti-canonical bundle $\mathbb{K}_{\mathbb{P}^2} = \mathcal{O}(-3) \to \mathbb{P}^2$. The solution to this problem is well known and will serve as a reference for the latter cases. Here the mirror curve is of genus one and all physical quantities can be expressed through modular or quasi-modular forms. We read off from the toric diagram that the A-cycle intersects the B-cycle $-3$ times. In addition, we find that the coordinates

$$X_0 = uxyw, \qquad X_1 = x^2 y, \qquad X_2 = wy^2, \qquad X_3 = w^2 x, \tag{5.2}$$

automatically fulfill the constraint

$$z = \frac{X_1 X_2 X_3}{X_0^3}, \qquad u^{-3} = z, \tag{5.3}$$

that can be read off from the toric diagram. Accordingly, the mirror curve $\Sigma(z)$ is given by

$$x^2 y + w^2 y + wx^2 + uxyw = 0\,. \tag{5.4}$$

Here $z$ denotes the complex structure modulus that takes values in the punctured sphere $\mathcal{M}(\Sigma(z)) = \mathbb{P}^1 \setminus \{z = 0, z = -\frac{1}{27}, \frac{1}{z} = 0\}$. Using birational transformations as e.g. provided by Nagell's algorithm this can be brought into Weierstrass form

$$y^2 = 4x^3 - \frac{1}{12}(1+24z)\,x - \frac{1}{216}(1+36z+216z^2)\,. \tag{5.5}$$

As a next step, one inverts the $j$-function in order to evaluate $\tau$ as a function of $z$ and obtains

$$q(z) = -z^3 + 45z^4 - 1512z^5 + 45672z^6 + \mathcal{O}(z^7)\,, \qquad q = e^{2\pi i\tau}\,. \tag{5.6}$$

We can also use the knowledge about $q$ in order to evaluate the A-period according to (4.1) as

$$t = \log(z) - 6z + 45z^2 - 560z^3 + \frac{17325}{2} z^4 + \mathcal{O}(z^5)\,. \tag{5.7}$$

As a consistency check one notices that the Picard-Fuchs operator

$$\mathcal{D} = \Theta^2 + z(3\Theta+2)(3\Theta+1)3\Theta = \mathcal{L}\Theta, \qquad \Theta = z\partial_z\,, \tag{5.8}$$

annihilates the A-period, as expected. Having computed the period we can exploit (2.36) to determine the prepotential

$$F^{(0,0)} = -\frac{1}{18}X_A^3 + 3Q - \frac{45}{8}Q^2 + \frac{244}{9}Q^3 - \frac{12333}{64}Q^4 + \frac{211878}{125}Q^5 + \mathcal{O}(Q^6)\,. \tag{5.9}$$

Here, we have introduced the monodromy invariant variable $Q = e^t$. Also, we note that the correctly normalized B-period is given as

$$t_D = -\frac{1}{6}\log(z)^2 + \frac{1}{3}X_A\log(z) - 3z + \frac{141}{4}z^2 + \frac{1486}{3}z^3 - \frac{129805}{16}z^4 + \mathcal{O}(z^5)\,. \tag{5.10}$$

Finally, we find that the Yukawa couplings are given as

$$C_{zzz} = -\frac{1}{3}\frac{1}{z^3(1+27z)}\,. \tag{5.11}$$

The genus one free energies are given by

$$\mathscr{F}^{(0,1)} = -\frac{1}{12}\log(z^7(1+27z)) - \frac{1}{2}\log(\frac{\partial X_A}{\partial z})\,, \tag{5.12}$$

and

$$F^{(1,0)} = \frac{1}{24}\log(\frac{1+27z}{z}). \tag{5.13}$$

**§5.1.2 Direct integration.** In this subsection we discuss the direct integration procedure applied to the example of local $\mathbb{P}^2$. First of all, one notices that

$$z_c = 1+27z\,, \tag{5.14}$$

is a good coordinate to perform computations around the conifold locus. Indeed, using (4.1) one directly obtains the A-period around the conifold locus as

$$t_c = \alpha\left(z_c + \frac{11}{18}z_c^2 + \frac{109}{243}z_c^3 + \frac{9389}{26244}z_c^4 + \frac{88351}{295245}z_c^5 + \mathscr{O}(z_c^6)\right) \tag{5.15}$$

Here, the proportionality constant $\alpha$ parametrizes the matching of the conifold period with the mass of D2-branes wrapping the vanishing cycle. In practice, it gets fixed by the gap condition (2.69). As a next step, we use the results of section 4.1 in order to determine the propagator,

$$S^{zz} = \frac{3}{4}z^2 + 9z^3 - 54z^4 + 756z^5 - 13284z^6 + 260496z^7 - 5451624z^8 + 119116656z^9 + \mathscr{O}\left(z^{10}\right). \tag{5.16}$$

Now the holomorphic anomaly equation (2.55) can be solved recursively and the holomorphic ambiguity at each step can be fixed by implementing the gap condition (2.70).

### §5.2 Solving the Topological String on $\mathbb{C}^3/\mathbb{Z}_5$.

**§5.2.1 The geometry and its mirror.** We now apply the direct integration procedure to the topological string on a mirror-pair where the A-model geometry is the resolution of the orbifold $\mathbb{C}^3/\mathbb{Z}_5$. The toric data is given in (5.17) and one can see that the Kaehler-structure is parametrised by two moduli, both corresponding to normalisable directions in the moduli-space.

$$\left(\begin{array}{ccc|cc|c} 0 & 0 & 1 & -3 & 1 & x_0 \\ 1 & 0 & 1 & 1 & -2 & x_1 \\ 2 & 0 & 1 & 0 & 1 & x_2 \\ 0 & 1 & 1 & 1 & 0 & x_3 \\ -1 & -1 & 1 & 1 & 0 & x_4 \\ & & & \uparrow & \uparrow & \\ & & & \mathbb{P}^1_b & \mathbb{P}^1_f & \end{array}\right) \tag{5.17}$$

By investigating the scaling relations, one easily sees that the divisor obtained by setting $x_0 = 0$ is given by a $\mathbb{P}^2$ while one obtains a $\mathbb{F}_3$ from $x_1 = 0$. In addition there are two generators of the Mori cone. They correspond to the hyperplane class of $\mathbb{P}^2$, which can also be identified with the base $\mathbb{P}^1_b$ of $\mathbb{F}_3$ while the curve $\mathbb{P}^1_f$ gets identified with the fiber of the latter. There are no mass parameters present and the mirror curve is now given by a Riemann surface of genus two. From the two generators of the Mori cone we construct the mirror curve and read off the moduli

$$z_1 = \frac{X_1X_3X_4}{X_0^3}\,, \qquad z_2 = \frac{X_0X_2}{X_1^2}\,. \tag{5.18}$$

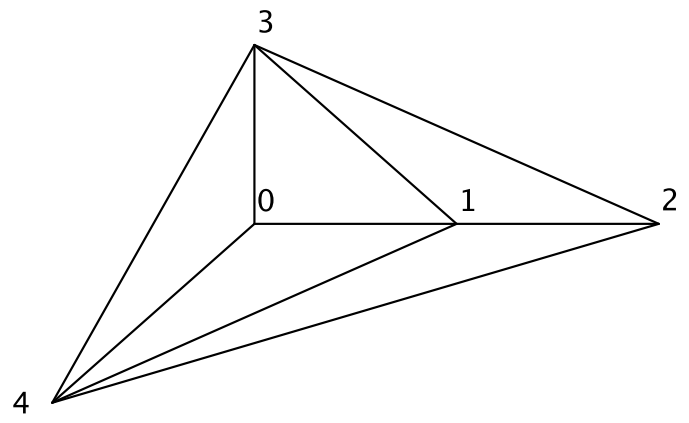

Figure 3: Toric diagram of the A-model geometry, which is the resolution of $\mathbb{C}^3/\mathbb{Z}_5$.

The constraints can be easily solved using the following choice of coordinates

$$X_0 = u\,xyz, \quad X_1 = v\,x^2 y, \quad X_2 = \frac{x^3 y}{z}, \quad X_3 = y^2 z, \quad X_4 = xz^2, \tag{5.19}$$

where the relation between $u$, $v$ and $z_1$, $z_2$ is given as

$$u = z_1^{-\frac{2}{5}} z_2^{-\frac{1}{5}}, \qquad v = z_1^{-\frac{1}{5}} z_2^{-\frac{3}{5}}. \tag{5.20}$$

Note that the coordinates $u$, $v$ are the natural coordinates from a diagrammatic point of view and signal two normalizable complex structure moduli. In contrast, $z_1$, $z_2$ are the physical complex structure moduli that appear in the mirror map. Having clarified the choice of coordinates, the mirror curve takes the form

$$y^2 = -4x^5 + z_1^{-\frac{4}{5}} z_2^{-\frac{2}{5}} x^4 + 2z_1^{-\frac{3}{5}} z_2^{-\frac{4}{5}} x^3 + (1+2z_2) z_1^{-\frac{2}{5}} z_2^{-\frac{6}{5}} x^2 + 2z_1^{-\frac{1}{5}} z_2^{-\frac{3}{5}} x + 1. \tag{5.21}$$

The discriminant locus[14] of the curve is given by

$$\Delta = 1 + 27z_1 + 3125 z_1^2 z_2^3 + 4z_2^2(4 + 125z_1) - z_2(8 + 225z_1). \tag{5.22}$$

In addition, one finds three Picard Fuchs operators

$$\begin{aligned}
\mathscr{L}_1 &= (\Theta_1 - 2\Theta_2)\Theta_1^2 - z_1\left[(\Theta_2 - 3\Theta_1)^3 - 3(\Theta_2 - 3\Theta_1)^2 + 2\Theta_2 - 6\Theta_1\right], \\
\mathscr{L}_2 &= (\Theta_2 - 3\Theta_1)\Theta_2 - z_2\left[(2\Theta_2 - \Theta_1)^2 + 2\Theta_2 - \Theta_1\right], \\
\mathscr{L}_3 &= \Theta_2\Theta_1^2 + z_1 z_2 (\Theta_2 - 3\Theta_1)(2\Theta_2 - \Theta_1)(\Theta_2 - 3\Theta_1 - 1).
\end{aligned} \tag{5.23}$$

Here we have used the common abbreviation for logarithmic derivatives $\Theta_i = z_i \frac{d}{dz_i}$. Note that linear dependent relations of the generators in the toric diagram can lead to linear independent Picard-Fuchs operators which is what happens here. In particular, the third differential operator is needed in order to exclude a triple logarithmic solution which is allowed by the first two operators. If one does not take into account the third operator, the solutions to the first two operators take the form $(1, t^i, \partial_i F, 2F - t^i \partial_i F)$ which is familiar from the compact case but not expected for the non-compact case.

After having excluded the triple-logarithmic solution, one then finds five solutions that are anni-hilated by the Picard-Fuchs operators $\mathscr{L}_i$,

$$\begin{aligned}
\sigma_0 &= 1\\
\sigma_1 &= \log(z_1)-6z_1-z_2+45z_1^2-\frac{3}{2}z_2^2-560z_1^3-18z_1^2z_2-\frac{10}{3}z_2^3+\mathcal{O}(z^4)\\
\sigma_2 &= \log(z_2)+2z_1+2z_2-15z_1^2+3z_2^2+\frac{560}{3}z_1^3+6z_1^2z_2+\frac{20}{3}z_2^3+\mathcal{O}(z^4)\\
\sigma_3 &= 3z_1-z_2^2-z_1z_2-\frac{141}{4}z_1^2-z_2\log(z_2)-\frac{3}{2}z_2^2\log(z_2)-\frac{3}{10}\log(z_2)^2\\
&\quad -\frac{1}{5}\log(z_1)\log(z_2)+2z_1\log(z_1)-15z_1^2\log(z_1)-\frac{1}{5}\log(z_1)^2+\mathcal{O}(z^3),\\
\sigma_4 &= -2z_2-7z_2^2-z_2\log(z_2)-\frac{3}{2}z_2^2\log(z_1)-\frac{3}{5}\log(z_1)\log(z_2)-\frac{1}{10}\log(z_1)^2\\
&\quad -3z_2\log(z_2)-\frac{9}{2}z_2^2\log(z_2)-\frac{9}{10}\log(z_2)^2+\mathcal{O}(z^3),
\end{aligned} \tag{5.24}$$

and a basis for the periods takes the form [Hos06]

$$\Pi = \begin{pmatrix} \tilde{t}_1^D \\ \tilde{t}_2^D \\ t_1 \\ t_2 \\ 1 \end{pmatrix} = \begin{pmatrix} -3 & 1 & \lambda_{11} & \lambda_{12} & \rho_1 \\ 1 & -2 & \lambda_{21} & \lambda_{22} & \rho_2 \\ 0 & 0 & 1 & 0 & \rho_3 \\ 0 & 0 & 0 & 1 & \rho_4 \\ 0 & 0 & 0 & 0 & 1 \end{pmatrix} \begin{pmatrix} \partial_{t_1}F^{(0,0)} \\ \partial_{t_2}F^{(0,0)} \\ t_1 \\ t_2 \\ 1 \end{pmatrix}, \tag{5.25}$$

with

$$\left(\partial_{t_1}F^{(0,0)}, \partial_{t_2}F^{(0,0)}, t_1, t_2\right) = \left(\frac{1}{(2\pi i)^2}\sigma_3, \frac{1}{(2\pi i)^2}\sigma_4, \frac{1}{2\pi i}\sigma_1, \frac{1}{2\pi i}\sigma_2\right). \tag{5.26}$$

[14]We refer to the discriminant locus as the product of non-trivial irreducible components of the discriminant. The discriminant is up to a numerical factor given by the Igusa invariant $D$ in (5.33).

Here, the $\lambda_{kl}$ as well as $\rho_m$ parameterize the ambiguity to add logarithmic solutions of order zero and one to the double logarithmic solutions and do not influence the non-perturbative part of the prepotential but they are crucial in order to obtain an integral basis. However, as we do not refer to such a basis in the following we do not fix these ambiguities. By inverting the A-periods $\sigma_1, \sigma_2$ (5.24), we readily obtain the mirror map as

$$\begin{aligned}
z_1(Q_1,Q_2) &= Q_1+6Q_1^2+Q_1Q_2+9Q_1^3+10Q_1^2Q_2+\mathcal{O}(Q^4), &(5.27)\\
z_2(Q_1,Q_2) &= Q_2-2Q_1Q_2-2Q_2^2+5Q_1^2Q_2+6Q_1Q_2^2+3Q_2^3+\mathcal{O}(Q^4), &(5.28)
\end{aligned}$$

where we have introduced the monodromy invariant variables $Q_i = e^{\sigma_i}$.

**§ 5.2.2 Extracting the complex structure moduli from the mirror.** As a next step we extract the period matrix of the mirror curve that takes the explicit form

$$\tau = \begin{pmatrix} \tau_{11} & \frac{1}{2}\tau_{12} \\ \frac{1}{2}\tau_{12} & \tau_{22} \end{pmatrix}. \tag{5.29}$$

In contrast to the torus, where the number of normalizable complex structure moduli of the mirror curve matches the moduli of a generic elliptic curve, one finds in the case at hand that the family of mirror curves parametrized by $z_1, z_2$ only defines a hypersurface within the Siegel fundamental domain $\mathscr{F}_2 = \Gamma_2\backslash\mathbb{H}^{(2)}$. Using the formulae (3.13), one obtains for the mirror curve (5.21)

$$\begin{aligned}
A &= -8z_1^{\frac{2}{5}}z_2^{\frac{4}{5}}\left(-1+z_2\left(4+40z_1\right)\right), &(5.30)\\
B &= 4z_1^{\frac{4}{5}}z_2^{\frac{8}{5}}\left(1+24z_1+2400z_1^2z_2^3-8z_2\left(1+25z_1\right)+z_2^2\left(16+440z_1-80z_1^2\right)\right), &(5.31)\\
C &= -8z_1^{\frac{6}{5}}z_2^{\frac{12}{5}}\left(-1-20z_1+72z_1^2+8z_1^2z_2^4\left(1009+10900z_1\right)+4z_2\left(3+75z_1+92z_1^2\right)\right.\\
&\quad \left.-4z_2^2\left(12+365z_1+1652z_1^2\right)+16z_2^3\left(4+145z_1+948z_1^2+320z_1^3\right)\right), &(5.32)\\
D &= 4096z_1^6z_2^9\left(1+27z_1+3125z_1^2z_2^3+4z_2^2\left(4+125z_1\right)-z_2\left(8+225z_1\right)\right). &(5.33)
\end{aligned}$$

Computing the absolute invariants (3.16) and comparing them to their Fourier expansions (C.18) whose computation is presented in appendix C one obtains

$$\begin{aligned} q_1(z_1,z_2) &= -z_1^3+z_1^3z_2+45z_1^4-z_1^3z_2^2-60z_1^4z_2-1512z_1^5+\mathcal{O}(z^6)\,, \\ q_2(z_1,z_2) &= z_1^2z_2^5+16z_1^2z_2^6-5z_1^3z_2^5+\mathcal{O}(z^9)\,, \\ r(z_1,z_2) &= -z_1-3z_1z_2+15z_1^2-13z_1z_2^2+20z_1^2z_2-279z_1^3+\mathcal{O}(z^4)\,. \end{aligned} \tag{5.34}$$

Here, we have denoted $q_1 = e^{2\pi i\tau_{11}}$, $q_2 = e^{2\pi i\tau_{22}}$ and $r = e^{2\pi i\tau_{12}}$. In addition, we note that in the limit of restricting to the local $\mathbb{P}^2$ model, corresponding to $z_2 \to 0$ only $q_1$ survives and gets indeed identified with (5.6). It is also satisfying to check that $x_1$ and $x_2$ reduce to the Weierstrass coefficients $g_2, g_3$ in (3.1). Using the knowledge of the $\tau$-matrix as well as of the periods, we can determine the prepotential. For this purpose we make use of (2.46) noting that

$$C = \begin{pmatrix} -3 & 1 \\ 1 & -2 \end{pmatrix}, \tag{5.35}$$

and find that

$$\begin{aligned} F^{(0,0)} &= -\frac{1}{15}T_1^3-\frac{1}{10}T_1^2T_2-\frac{3}{10}T_1T_2^2-\frac{3}{10}T_2^3+3Q_1-2Q_2-\frac{45}{8}Q_1^2+4Q_1Q_2-\frac{1}{4}Q_2^2 \\ &\quad +\frac{244}{9}Q_1^3-10Q_1^2Q_2+3Q_1Q_2^2-\frac{2}{27}Q_2^3+\mathcal{O}(Q^4)\,. \end{aligned} \tag{5.36}$$

In particular, denoting $K_{ij} = \partial_{t_i^A}\partial_{t_j^A}F^{(0,0)}$, one finds the explicit relations[15]

$$\tau_{11} = -4K_{11}+4K_{12}-K_{22}\,, \tag{5.37}$$

$$\tau_{22} = -K_{11}+6K_{12}-9K_{22}\,, \tag{5.38}$$

$$\tau_{12} = -2K_{11}+7K_{12}-3K_{22}\,. \tag{5.39}$$

In addition, once the prepotential is known, we can directly write down the Yukawa couplings which are found to be

$$C_{z_1z_1z_1} = -\frac{2+9z_1-16z_2-95z_1z_2+32z_2^2+300z_1z_2^2}{5z_1^3\Delta}, \tag{5.40}$$

$$C_{z_1z_1z_2} = -\frac{1+27z_1-8z_2-210z_1z_2+16z_2^2+400z_1z_2^2}{5z_1^2z_2\Delta}, \tag{5.41}$$

$$C_{z_1z_2z_2} = -\frac{3+81z_1-14z_2-405z_1z_2+8z_2^2+325z_1z_2^2}{5z_1z_2^2\Delta}, \tag{5.42}$$

$$C_{z_2z_2z_2} = -\frac{9+243z_1-17z_2-540z_1z_2+4z_2^2+225z_1z_2^2}{5z_2^3\Delta}. \tag{5.43}$$

The knowledge of the A-periods enables us to check the conjecture (4.11) and we find in this case that

$$|G_{ij}| = \sqrt{\frac{E_6(\tau)B(z_i,m_j)}{z_1^{\frac{8}{5}}z_2^{\frac{6}{5}}E_4(\tau)C'(z_i,m_j)}}\,, \tag{5.44}$$

where we have fixed the rescaling ambiguity accordingly. Finally, with the help of a few known BPS numbers the unkowns in (2.59) and (2.60) can be determined and one finds the refined free energies at genus one

$$F^{(1,0)} = \frac{1}{24}\log\left(\Delta z_1^{-2}z_2^{-3}\right), \qquad \mathcal{F}^{(0,1)} = -\frac{1}{12}\log\left(\Delta z_1^{\frac{38}{5}}z_2^{\frac{39}{5}}\right)-\frac{1}{2}\log\left(|G_{ij}|\right),$$

[15] Clearly, the prepotential can also be integrated by determing the B-periods first. From this perspective, this is merely a check of (2.46).

$$\begin{aligned} F^{(1,0)} &= -\frac{1}{12}T_1-\frac{1}{8}T_2+\frac{7}{8}Q_1-\frac{1}{6}Q_2-\frac{129}{16}Q_1^2+\frac{5}{6}Q_1Q_2-\frac{1}{12}Q_2^2 \\ &\quad +\frac{589}{6}Q_1^3-\frac{65}{6}Q_1^2Q_2+\frac{7}{8}Q_1Q_2^2-\frac{1}{18}Q_2^3+\mathcal{O}(Q^4)\,, \end{aligned} \tag{5.45}$$

$$\begin{aligned} \mathscr{F}^{(0,1)} &= -\frac{2}{15}T_1-\frac{3}{20}T_2+\frac{1}{4}Q_1-\frac{1}{6}Q_2-\frac{3}{8}Q_1^2+\frac{1}{3}Q_1Q_2-\frac{1}{12}Q_2^2 \\ &\quad -\frac{23}{3}Q_1^3-\frac{5}{6}Q_1^2Q_2+\frac{1}{4}Q_1Q_2^2-\frac{1}{18}Q_2^3+\mathcal{O}(Q^4)\,. \end{aligned} \tag{5.46}$$

After this discussion, we are prepared to proceed to the direct integration which provides us with the higher genus free invariants.

**§5.2.3 Direct Integration.** In order to determine the higher genus free energies, we first have to calculate the propagator. Recall that the propagator is in the genus two case given by a matrix

$$\mathscr{S}=\begin{pmatrix}\mathscr{S}^{11}(z_1,z_2) & \mathscr{S}^{12}(z_1,z_2)\\ \mathscr{S}^{21}(z_1,z_2) & \mathscr{S}^{22}(z_1,z_2)\end{pmatrix}. \tag{5.47}$$

Instead of using equations (2.66), (2.67) and (2.68), we follow the procedure in section 4.2. Using the relation (5.39) one finds that the following ansatz for the components of (5.47) solves (2.68),

$$\mathscr{S}^{t_1t_1} = \frac{1}{10}\left(9\frac{\partial}{\partial\tau_{11}}\log(\chi_{10})+3\frac{\partial}{\partial\tau_{12}}\log(\chi_{10})+\frac{\partial}{\partial\tau_{22}}\log(\chi_{10})\right), \tag{5.48}$$

$$\mathscr{S}^{t_1t_2} = \frac{1}{10}\left(-3\frac{\partial}{\partial\tau_{11}}\log(\chi_{10})-\frac{7}{2}\frac{\partial}{\partial\tau_{12}}\log(\chi_{10})-2\frac{\partial}{\partial\tau_{22}}\log(\chi_{10})\right), \tag{5.49}$$

$$\mathscr{S}^{t_2t_2} = \frac{1}{10}\left(\frac{\partial}{\partial\tau_{11}}\log(\chi_{10})+2\frac{\partial}{\partial\tau_{12}}\log(\chi_{10})+4\frac{\partial}{\partial\tau_{22}}\log(\chi_{10})\right). \tag{5.50}$$

This expression can be directly written down without any further computation, once the toric diagram of the A-model geometry is known and the $\tau$-matrix has been extracted from the mirror curve by inverting the Fourier expansion of the Igusa invariants as performed in the previous section. After tensor transforming to the coordinates $z_i$ these read explicitly

$$\mathscr{S}^{z_1z_1} = \frac{7}{10}z_1^2+9z_1^3-\frac{3}{10}z_1^2z_2-54z_1^4-6z_1^3z_2+\mathcal{O}(z^5)\,, \tag{5.51}$$

$$\mathscr{S}^{z_1z_2} = -\frac{3}{20}z_1z_2-3z_1^2z_2+\frac{3}{5}z_1z_2^2+18z_1^3z_2+7z_1^2z_2^2+\mathcal{O}(z^5)\,, \tag{5.52}$$

$$\mathscr{S}^{z_2z_2} = \frac{3}{10}z_2^2+z_1z_2^2-\frac{6}{5}z_2^3-6z_1^2z_2^2-4z_1z_2^3+\mathcal{O}(z^5)\,. \tag{5.53}$$

In addition, we have checked that these propagators obey the over-determined system of equations provided by (2.67). Even more importantly, the ambiguities take precisely the form[16] (4.16). Let us emphasize once more that this reflects the fact that the ambiguity of the propagator is completely fixed by the generalization of the Ramanujan identities for genus two as presented in (9).

In order to fix the holomorphic ambiguities $f^{(g_L,g_R)}$ that arise after integration of the holomorphic anomaly equation (2.55), we have to impose the gap condition at the conifold. To do this we have to choose a rational point on the discriminant locus in the moduli space and repeat the previous analysis. Throughout the following discussion, we will work with the point

$$(z_1,\,z_2)=(3,\,-2/9)\,. \tag{5.54}$$

[16] For the ambitious reader who wants to re-check this calculation we note that our internal conventions for $f_{\mathrm{RS}}^{mn,kl}$ differ by a factor of $-1/2$ for the off-diagonal elements on each side.

The logarithmic singularities in the periods at the conifold are of the form $\log(\Delta)$. However since we will need to transform objects from the large radius point to the conifold it is desirable that the new set of variables $z_i^c(z)$ can easily be inverted. The most convenient ansatz is a linear transformation

$$z_i^c = b_i\left(z_1 - z_1^{c,0}\right) + c_i\left(z_2 - z_2^{c,0}\right). \tag{5.55}$$

For the expansion of the singular periods to be well behaved in these coordinates we choose $z_o^c$ orthogonal and $z_p^c$ parallel to the conifold locus, i.e. $\nabla_z z_o^c \sim \nabla_z \Delta$ and $\nabla_z z_p^c \cdot \nabla_z \Delta = 0$. Then one can expand the logarithm as

$$\begin{aligned} \log(\Delta) &= \log(z_o^c + \mathcal{O}\left[(z^c)^2\right]) = \log(z_o^c) + \log\left(1 + \frac{\mathcal{O}\left[(z^c)^2\right]}{z_o^c}\right) \\ &= \log(z_o^c) + \sum_n^\infty \frac{(-1)^{n+1}}{n}\left(\frac{\mathcal{O}\left[(z^c)^2\right]}{z_o^c}\right)^n . \end{aligned} \tag{5.56}$$

and observes that a period $\omega = \pi\log(\Delta) + f$ has the structure

$$\omega(z_o^c, z_p^c) = \pi(z_o^c, z_p^c)\log(z_o^c) + \pi(z_o^c, z_p^c)\sum_n^\infty \frac{(-1)^{n+1}}{n}\left(\frac{\mathcal{O}\left[(z^c)^2\right]}{z_o^c}\right)^n + f(z_o^c, z_p^c). \tag{5.57}$$

Using a linear ansatz we find the variables

$$\begin{aligned} z_o^c &= -9 + z_1 - 27 z_2 , \\ z_p^c &= -\frac{727}{9} + 27 z_1 + z_2 , \end{aligned} \tag{5.58}$$

at the point $(3, -2/9)$ which are orthogonal and tangential to the discriminant locus. After transforming the Picard-Fuchs operators into these new variables one obtains again five periods,

$$t_c^0 = 1, \tag{5.59}$$

$$\begin{aligned} t_c^{A,1} &= z_o^c - \frac{271476409}{2127869700}\left(z_o^c\right)^2 + \frac{116172}{16120225} z_o^c z_p^c - \frac{2187}{5861900}\left(z_p^c\right)^2 + \mathcal{O}(z^3), \\ t_c^{A,2} &= z_p^c - \frac{8420610821}{590880856630} z_o^c z_p^c - \frac{2408185701}{295440428315}\left(z_p^c\right)^2 + \mathcal{O}(z^3), \\ t_c^{B,1} &= \frac{271062557874}{35748291826115} z_o^c z_p^c - \frac{6636689883}{12999378845860}\left(z_p^c\right)^2 + z_o^c \log\left(z_o^c\right) - \frac{271476409}{2127869700}\left(z_o^c\right)^2 \log\left(z_o^c\right) \\ &\quad + \frac{116172}{16120225} z_o^c z_p^c \log\left(z_o^c\right) - \frac{2187}{5861900}\left(z_p^c\right)^2 \log\left(z_o^c\right) + \mathcal{O}(z^3), \\ t_c^{B,2} &= \left(z_o^c\right)^2 - \frac{1359666 z_o^c z_p^c}{809425831} - \frac{793881}{809425831}\left(z_p^c\right)^2 + \mathcal{O}(z^3). \end{aligned} \tag{5.60}$$

Note that these are not analytical continuations of the periods (5.24), (5.25) obtained at the large radius point. As there are besides the trivial solution three more solutions that take the form of a power series and only one logarithmic solution, a comment how the splitting into $A-$ and $B-$periods is performed is in order. One identifies the power series in front of the logarithm $t_c^{B,1}$ as the period $t_c^{A,1}$. $t_c^{A,2}$ is choosen so that it starts with $z_p^c$ and we found a possible addition of $X_c^{B,2}$ to cancel out in all physical quantities. Also note that $X_c^{B,1}$ contains terms with negative powers in $z_o^c$ at higher order. Finally, we find the following mirror map

$$\begin{aligned} z_o^c &= T_1^c + \frac{271476409}{2127869700}\left(T_1^c\right)^2 - \frac{116172}{16120225} T_1^c T_2^c + \frac{2187}{5861900}\left(T_2^c\right)^2 + \mathcal{O}(T^3), \\ z_p^c &= T_2^c + \frac{8420610821}{590880856630} T_1^c T_2^c + \frac{2408185701}{295440428315}\left(T_2^c\right)^2 + \mathcal{O}(T^3). \end{aligned} \tag{5.61}$$

At the large radius point the ambiguities $A_i, f_i^{mn}, \tilde{f}_{ij}^m$ and Yukawa couplings $C_{ijk}$ are rational expressions in the moduli $z_i$ and can be tensor transformed into the conifold coordinates $z^c$. Again the topological metric can be calculated from the A-periods via

$$G_{ij}^c = \frac{\partial X_c^{A,i}}{\partial z_j^c}, \tag{5.62}$$

and solving the over-constrained system of equations (2.66), (2.67) and (2.68) yields the propagator at the conifold. We have integrated the refined holomorphic anomaly equation (2.55) and implemented the gap-condition (2.69) up to genus three. The first two orders of the refined free energies at genus two are given by

$$\begin{aligned}
\mathcal{F}^{(2,0)} &= \frac{7}{46080} + \frac{47}{1280} z_1 - \frac{1}{240} z_2 - \frac{753}{640} z_1^2 - \frac{13}{3840} z_1 z_2 - \frac{1}{60} z_2^2 + \mathcal{O}(z^3),\\
\mathcal{F}^{(1,1)} &= -\frac{7}{11520} - \frac{73}{200} z_1 + \frac{19}{600} z_2 + \frac{2949}{200} z_1^2 + \frac{31}{300} z_1 z_2 + \frac{19}{150} z_2^2 + \mathcal{O}(z^3),\\
\mathcal{F}^{(0,2)} &= \frac{1}{1920} + \frac{109}{640} z_1 - \frac{11}{1200} z_2 - \frac{23331}{3200} z_1^2 - \frac{233}{9600} z_1 z_2 - \frac{11}{300} z_2^2 + \mathcal{O}(z^3),
\end{aligned} \tag{5.63}$$

and the leading terms for genus three are

$$\begin{aligned}
\mathcal{F}^{(3,0)} &= \frac{31}{77414400} - \frac{3229}{1152000} z_1 + \frac{83}{2304000} z_2 + \mathcal{O}(z^3),\\
\mathcal{F}^{(2,1)} &= -\frac{31}{12902400} + \frac{346669}{10240000} z_1 - \frac{1023}{1280000} z_2 + \mathcal{O}(z^3),\\
\mathcal{F}^{(1,2)} &= \frac{41}{9676800} - \frac{104421}{1280000} z_1 + \frac{13277}{2880000} z_2 + \mathcal{O}(z^3),\\
\mathcal{F}^{(0,3)} &= -\frac{1}{483840} + \frac{26454323}{460800000} z_1 - \frac{27563}{4800000} z_2 + \mathcal{O}(z^3).
\end{aligned} \tag{5.64}$$

The refined GV invariants are listed in Appendix D. In addition we find the following non-vanishing refined BPS invariants

$$N_{0,1}^{(1,0)} = N_{0,0}^{(1,1)} = N_{0,1}^{(1,1)} = N_{0,1}^{(1,2)} = 1\,. \tag{5.65}$$

**§5.3 Solving the Topological String on $\mathbb{C}^3/\mathbb{Z}_6$.** In the previous section we discussed local $\mathbb{C}^3/\mathbb{Z}_5$ which is the simplest geometry with mirror curve of genus two as it only has two true moduli. Now we present the discussion of the resolution[17] of the orbifold $\mathbb{C}^3/\mathbb{Z}_6$ which provides an example with three moduli out of which one is just a deformation (mass) parameter.

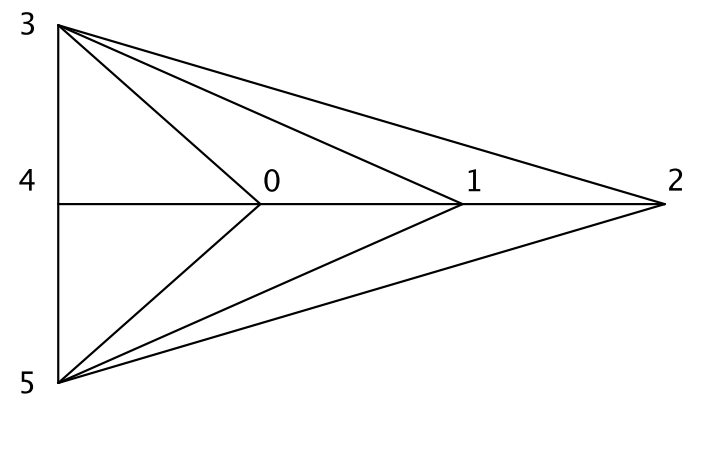

Figure 4: The A-model geometry of $\mathbb{C}^3/\mathbb{Z}_6$.

[17] Note that there are two ways to make $\mathbb{Z}_6$ act on $\mathbb{C}^3$. The second way leads to a geometry whose mirror curve has genus one.

The toric diagram is shown in figure 4 and the corresponding data is listed in (5.66).

$$\left(\begin{array}{ccc|ccc|c} 0 & 0 & 1 & -2 & 1 & 0 & x_0 \\ 0 & 1 & 1 & 1 & -2 & 0 & x_1 \\ 0 & 2 & 1 & 0 & 1 & 0 & x_2 \\ 1 & -1 & 1 & 0 & 0 & 1 & x_3 \\ 0 & -1 & 1 & 1 & 0 & -2 & x_4 \\ -1 & -1 & 1 & 0 & 0 & 1 & x_5 \end{array}\right) \tag{5.66}$$

We note that the orbifold is resolved by two Hirzebruch surfaces, $\mathbb{F}_2$ and $\mathbb{F}_4$ which correspond to the divisors $\{x_0 = 0\}$ and $\{x_1 = 0\}$ respectively. In addition there is the non-compact divisor $\{x_4 = 0\}$ that has the topology of $\mathbb{C} \times \mathbb{P}^1$ which resolves the singularity $\mathbb{C}^2/\mathbb{Z}_2$ corresponding to the point on the edge of the toric diagram.

By analogous considerations to those for the case $\mathbb{C}^3/\mathbb{Z}_5$ one finds the mirror curve by the Hori-Vafa method to be given as

$$y^2 = \left(z_3 z_1^2 - \frac{z_1^2}{4}\right)x^6 - \frac{z_1}{2}x^5 + \left(\frac{z_1}{2} + \frac{1}{4}\right)x^4 - \left(\frac{z_1 z_2}{2} + \frac{1}{2}\right)x^3 - \left(\frac{z_2}{2} + \frac{1}{4}\right)x^2 - \frac{z_2}{2}x - \frac{z_2^2}{4}. \tag{5.67}$$

Note that the variables $z_i$ are already those which correspond to the three Mori vectors of the toric Diagram which need to be contrasted to the two real moduli and the mass parameters appearing in the toric diagram. The discriminant locus of the mirror curve is found to consist of two irreducible components which are given as

$$\begin{aligned} \Delta_1 =& 1 - 8z_1 + 16z_1^2 - 64z_1^2 z_3 - 8z_2 + 68z_1 z_2 - 144z_1^2 z_2 + 576z_1^2 z_2 z_3 + 16z_2^2 - 144z_1 z_2^2 + 270z_1^2 z_2^2 \\ & - 1512z_1^2 z_2^2 z_3 + 216z_1^3 z_2^2 - 864z_1^3 z_2^2 z_3 + 216z_1^2 z_2^3 + 864z_1^2 z_2^3 z_3 - 972z_1^3 z_2^3 + 3888z_1^3 z_2^3 z_3 \\ & + 729z_1^4 z_2^4 - 5832z_1^4 z_2^4 z_3 + 11664z_1^4 z_2^4 z_3^2, \\ \Delta_2 =& 4z_3 - 1. \end{aligned} \tag{5.68}$$

Here $\Delta_2$ corresponds to a locus with enhanced $SU(2)$ gauge symmetry where the theory stays regular, compare also the discussion in [KM96; KMP96].

The three Picard Fuchs operators that correspond to the Mori vectors of the toric diagram

$$\begin{aligned} \mathcal{L}_1 =& \Theta_1(\Theta_1 - 2\Theta_2 - 2\Theta_3) + 4\Theta_2\Theta_3 - z_1(2\Theta_1 - \Theta_2 + 1)(2\Theta_1 - \Theta_2), \\ \mathcal{L}_2 =& \Theta_2(\Theta_2 - 2\Theta_1) + z_2(2\Theta_2 - \Theta_1 + 1)(2\Theta_2 - \Theta_1), \\ \mathcal{L}_3 =& \Theta_3^2 + z_3(2\Theta_3 - \Theta_1 + 1)(2\Theta_3 - \Theta_1), \end{aligned} \tag{5.69}$$

need also in this case to be supplemented by a fourth operator corresponding to the sum of the three mori cone vectors

$$\mathcal{L}_4 = \Theta_2\Theta_3^2 + z_1 z_2 z_3(\Theta_1 - 2\Theta_2)(\Theta_1 - 2\Theta_3)(2\Theta_1 - \Theta_2). \tag{5.70}$$

The latter discards one double-logarithmic and a triple-logarithmic solution of the three operators (5.69). In addition, we observe again that the system of solutions of (5.69) takes the form $(1, t^i, \partial_i F, 2F - t^i \partial_i F)$. Having remarked this we present the solution to the complete system that includes (5.70),

$$\begin{aligned}
\sigma_0 =&1,\\
\sigma_1 =&\log(z_1)+2z_1+3z_1^2-z_3-\frac{3}{2}z_3^2-z_2-\frac{3}{2}z_2^2+\mathcal{O}(z^3),\\
\sigma_2 =&\log(z_2)-z_1-\frac{3}{2}z_1^2+2z_2+3z_2^2+\mathcal{O}(z^3),\\
\sigma_3 =&\log(z_3)+2z_3+3z_3^2+\frac{20}{3}z_3^3+\frac{35}{2}z_3^4+\frac{252}{5}z_3^5+\mathcal{O}(z^6),\\
\sigma_4 =&-2z_1-2z_1\log(z_1)-z_1\log(z_3)+\frac{2}{3}z_3\log(z_3)-\frac{2}{3}\log(z_1)\log(z_3)-2z_2\log(z_2)\\
&-\frac{2}{3}\log(z_1)\log(z_1)-\frac{1}{3}\log(z_3)\log(z_2)-\frac{2}{3}\log(z_2)^2+\mathcal{O}(z^2),\\
\sigma_5 =&-2z_2-2z_2\log(z_1)-\frac{1}{3}\log(u)^2+\frac{1}{3}z_3\log(z_3)-z_2\log(z_3)-\frac{1}{3}\log(z_1)\log(z_3)-4z_2\log(z_2)\\
&-\frac{4}{3}\log(z_1)\log(z_2)-\frac{2}{3}\log(z_3)\log(z_2)-\frac{4}{3}\log(z_2)^2+\mathcal{O}(z^2),
\end{aligned}\tag{5.71}$$

with

$$\left(\partial_1 F^{(0,0)},\partial_2 F^{(0,0)},t_1,t_2,t_3\right)=\left(\frac{1}{(2\pi i)^2}\sigma_4,\frac{1}{(2\pi i)^2}\sigma_5,\frac{1}{2\pi i}\sigma_1,\frac{1}{2\pi i}\sigma_2,\frac{1}{2\pi i}\sigma_3\right).\tag{5.72}$$

Using the monodromy invariant variables $Q_1=e^{\sigma_1}$, $Q_2=e^{\sigma_2}$ and $Q_3=e^{\sigma_3}$ the mirror map is given by

$$\begin{aligned}
z_1(Q_1,Q_2,Q_3)=&Q_1-2Q_1^2+3Q_1^3+Q_1Q_3-4Q_1^2Q_3+Q_1Q_2-3Q_1^2Q_2+Q_1Q_2Q_3+\mathcal{O}(Q^4),\\
z_2(Q_1,Q_2,Q_3)=&Q_2+Q_1Q_2+Q_1Q_2Q_3-2Q_2^2-3Q_1Q_2^2+3Q_2^3+\mathcal{O}(Q^4),\\
z_3(Q_1,Q_2,Q_3)=&Q_3-2Q_3^2+3Q_3^3-4Q_3^4+5Q_3^5-6Q_3^6+7Q_3^7+\mathcal{O}(Q^8),
\end{aligned}\tag{5.73}$$

where $z_3(Q_1,Q_2,Q_3)$ can be expressed as

$$z_3(Q_3)=\frac{Q_3}{(1+Q_3)^2}.\tag{5.74}$$

Integrating $\partial_1 F^{(0,0)}$ and $\partial_2 F^{(0,0)}$ we obtain the prepotential

$$F^{(0,0)}=-2Q_1-2Q_2-\frac{2}{9}T_1^3-\frac{1}{3}T_1^2T_3-\frac{1}{3}T_1^2T_2-\frac{1}{3}T_1T_2T_3-\frac{2}{3}T_1T_2^2-\frac{1}{3}T_2^2T_3-\frac{4}{9}T_2^3+\mathcal{O}(Q^2).\tag{5.75}$$

We list the corresponding instanton numbers in appendix E. Having computed the prepotential, we can determine the Yukawa couplings

$$\begin{aligned}
C_{111}=&-\frac{1}{3z_1^3\Delta_1}(4-8z_1+16z_1^2-64z_1^2z_3-32z_2+80z_1z_2-168z_1^2z_2+672z_1^2z_2z_3+64z_2^2-192z_1z_2^2+540z_1^2z_2^2-2160z_1^2z_2^2z_3-432z_1^2z_2^3+1728z_1^2z_2^3z_3),\\
C_{112}=&-\frac{1}{3z_1^2z_2\Delta_1}(2-16z_1+32z_1^2-128z_1^2z_3-16z_2+124z_1z_2-264z_1^2z_2+1056z_1^2z_2z_3+32z_2^2-240z_1z_2^2+594z_1^2z_2^2-2376z_1^2z_2^2z_3-216z_1^2z_2^3+864z_1^2z_2^3z_3),\\
C_{122}=&-\frac{1}{3z_1z_2^2\Delta_1}(4-32z_1+64z_1^2-256z_1^2z_3-20z_2+176z_1z_2-384z_1^2z_2+1536z_1^2z_2z_3+16z_2^2-192z_1z_2^2+540z_1^2z_2^2-2160z_1^2z_2^2z_3-108z_1^2z_2^3+432z_1^2z_2^3z_3),\\
C_{222}=&-\frac{1}{3z_2^3\Delta_1}(8-64z_1+128z_1^2-512z_1^2z_3-22z_2+208z_1z_2-480z_1^2z_2+1920z_1^2z_2z_3+8z_2^2-132z_1z_2^2+432z_1^2z_2^2-1728z_1^2z_2^2z_3-54z_1^2z_2^3+216z_1^2z_2^3z_3).
\end{aligned}\tag{5.76}$$

We also use the knowledge of the prepotential in order to successfully check the relation (2.46). Here, the two derivatives have to be taken with respect to the two true moduli $t_1$ and $t_2$. The toric data (5.66) provides us with the following intersection matrix

$$C=\begin{pmatrix}-2 & 1\\ 1 & -2\end{pmatrix}.\tag{5.77}$$

We obtain

$$\begin{aligned}
q_1(z_1,z_2,z_3) &= z_1^4 z_2^6 z_3^2 + 4 z_1^5 z_2^6 z_3^2 + 16 z_1^6 z_2^6 z_3^2 + 16 z_1^4 z_2^7 z_3^2 + 56 z_1^5 z_2^7 z_3^2 + 160 z_1^4 z_2^8 z_3^2 + \mathcal{O}(z^{15}),\\
q_2(z_1,z_2,z_3) &= z_1^4 z_3^2 + 16 z_1^5 z_3^2 + 160 z_1^6 z_3^2 - 2 z_1^4 z_2 z_3^2 - 40 z_1^5 z_2 z_3^2 + z_1^4 z_2^2 z_3^2 + \mathcal{O}(z^9),\\
r(z_1,z_2,z_3) &= z_1^2 z_3 + 8 z_1^3 z_3 + 48 z_1^4 z_3 + 2 z_1^2 z_2 z_3 + 6 z_1^3 z_2 z_3 + 9 z_1^2 z_2^2 z_3 + \mathcal{O}(z^6),
\end{aligned} \tag{5.78}$$

and plugging these expressions into the respective Fourier expansions of the absolute Igusa invariants (3.16) we find perfect agreement with the rational expressions obtained from the mirror curve (5.67). In addition, we also verify our conjecture for the expression of the determinant of the topological metric (4.11)

$$|G_{ij}| = \sqrt{\frac{E_6(\tau)B(z_i)}{16 z_1^2 z_2^2 E_4(\tau) C'(z_i)}}, \tag{5.79}$$

where the indices $i$, $j$ are again only running over true moduli $t_1$, $t_2$.

As a next step, we use the known GV invariants for local $\mathbb{F}^2$ [Ali09; HKP13] to fix the ambiguities for the free energies at genus one

$$F^{(1,0)} = \frac{1}{24}\log\left(\Delta_1 z_1^{-4} z_2^{-4} z_3^{-2}\right), \qquad \mathcal{F}^{(0,1)} = -\frac{1}{12}\log\left(\Delta_1 z_1^8 z_2^8 z_3\right) - \frac{1}{2}\log\left(|G_{ij}|\right).$$

The corresponding refined GV invariants can be found in Appendix E. Finally using the logarithmic derivatives of $\chi_{10}$ we obtain the propagator from (4.13) as

$$\begin{aligned}
\mathcal{S}_{11} &= \frac{3z_1^2}{10} - \frac{6z_1^3}{5} - 8 z_1^4 z_3 - 32 z_1^5 z_3 - \frac{3}{10} z_1^2 z_2 + \frac{9}{5} z_1^3 z_2 + 16 z_1^4 z_2 z_3 + \mathcal{O}(z^7),\\
\mathcal{S}_{12} &= -\frac{3}{20} z_1 z_2 + \frac{3}{5} z_1^2 z_2 + 4 z_1^3 z_2 z_3 + 16 z_1^4 z_2 z_3 + \frac{3}{5} z_1 z_2^2 - \frac{9}{4} z_1^2 z_2^2 - 14 z_1^3 z_2^2 z_3 + \mathcal{O}(z^7),\\
\mathcal{S}_{22} &= \frac{3z_2^2}{10} - \frac{3}{10} z_1 z_2^2 - 2 z_1^2 z_2^2 z_3 - 8 z_1^3 z_2^2 z_3 - \frac{6z_2^3}{5} + \frac{9}{5} z_1 z_2^3 + 10 z_1^2 z_2^3 z_3 + \mathcal{O}(z^7).
\end{aligned} \tag{5.80}$$

We have successfully checked that this propagator obeys the over-determined system of equations (2.66) and (2.67).

# 6 | Almost Holomorphic Siegel Modular Forms

The propagator that was described in (4.13) is an almost meromorphic Siegel modular form. To subsume it under a satisfying structure theory, we first develop a theory of almost holomorphic Siegel modular forms. Shimura studied them very early on [Shi86] and called them nearly holomorphic. Subsequent work by Zagier [Zag94] and Kaneko-Zagier [KZ98] employed "almost holomorphic" as terminology, and defined quasi modular form as the "constant part" of almost holomorphic ones.

From a geometric and representation theoretic point of view, it is clear how to define almost holomorphic Siegel modular forms. We discuss this in Remark 6.14. A classical, explicit description, however, is not available in the literature, and we start the mathematical part of this work by filling that gap. In fact, not even all covariant differential operators are explicitly known. They are the theme of Section 6.2. In Section 6.3, we provide a definition of almost holomorphic functions. It serves as a foundation for Section 6.4, which contains the definition and basic properties of almost holomorphic Siegel modular forms.

**§6.1 Preliminaries.** We revisit some basic aspects of the theory of Siegel modular forms. The reader can find a good introduction, which goes more into detail, in [Fre83] or [BGHZ08]. Siegel's foundational work [Sie51] provides one of the most general contexts in which Siegel modular forms can be understood.

**The upper half spaces.** We write $\mathbb{H}^{(n)} = \{\tau \in \mathrm{Mat}^{\mathrm{t}}_n(\mathbb{C}) : \mathfrak{Im}(\tau) \text{ is positive definite}\}$ for the degree $n$ Siegel upper half space. The real and imaginary parts of $\tau$, which are matrices of size $n \times n$, are written $x = \mathfrak{Re}(\tau)$, $y = \mathfrak{Im}(\tau)$, throughout. Coordinates of $\tau$, $x$, and $y$ are denoted respectively by $\tau_{ij}$, $x_{ij}$, and $y_{ij}$. We use the ambiguous notation $y^{-1}_{ij} = \big(y^{-1}\big)_{ij}$ for the $ij$-th entry of $y^{-1}$ (as opposed to the inverse of $y_{ij}$).

**The symplectic group.** The Siegel upper half space carries an action of the symplectic group. Let $J_n = \left(\begin{smallmatrix} 0 & \mathbb{1}_n \\ -\mathbb{1}_n & 0 \end{smallmatrix}\right)$ be the standard symplectic form, where $\mathbb{1}_n$ is the $n \times n$ identity matrix. The symplectic group is its stabilizer

$$\mathrm{G}^{(n)} = \mathrm{Sp}_n(\mathbb{R}) = \big\{g \in \mathrm{Mat}_{2n}(\mathbb{R}) : {}^{\mathrm{t}}g J_n g = J_n\big\}. \tag{6.1}$$

Elements of $\mathrm{G}^{(n)}$ are written as $\left(\begin{smallmatrix} a & b \\ c & d \end{smallmatrix}\right)$ with $a, b, c, d \in \mathrm{Mat}_n(\mathbb{R})$. The action of $\mathrm{G}^{(n)}$ on $\mathbb{H}^{(n)}$ is then defined by

$$g\tau = (a\tau + b)\,(c\tau + d)^{-1}. \tag{6.2}$$

We fix special elements $\mathrm{inv} = \left(\begin{smallmatrix} 0 & -\mathbb{1}_n \\ \mathbb{1}_n & 0 \end{smallmatrix}\right) \in \mathrm{G}^{(n)}$ and $\mathrm{trans}(b) = \left(\begin{smallmatrix} \mathbb{1}_n & b \\ 0 & \mathbb{1}_n \end{smallmatrix}\right) \in \mathrm{G}^{(n)}$ for $b \in \mathrm{Mat}^{\mathrm{t}}_n(\mathbb{R})$, where $\mathrm{Mat}^{\mathrm{t}}_n(\mathbb{R})$ denotes the space of $n \times n$ symmetric matrices with real entries. Note that inv and trans$(b)$, for $b$ running trough $\mathrm{Mat}^{\mathrm{t}}_n(\mathbb{R})$, generate $\mathrm{G}^{(n)}$.

The integral points of $\mathrm{G}^{(n)}$ are denoted by $\Gamma^{(n)} = \mathrm{Sp}_n(\mathbb{Z})$. In analogy to the previous observation, $\Gamma^{(n)}$ is generated by inv and all trans$(b)$ with $b \in \mathrm{Mat}^{\mathrm{t}}_n(\mathbb{Z})$.

**Slash actions.** Given a representation $\sigma$ of $\mathrm{GL}_n(\mathbb{C})$, we set

$$\big(f|_\sigma\, g\big)(\tau) = \sigma(c\tau + d)^{-1}\, f(g\tau)$$

for $f \in C^\infty\big(\mathbb{H}^{(n)} \to V(\sigma)\big)$ and $g \in \mathrm{G}^{(n)}$. We call $|_\sigma$ the slash action associated to the weight $\sigma$, or less strictly, the weight $\sigma$ slash action.

If we do not state otherwise, $\sigma$ is assumed to be finite dimensional and complex. In the special case of $\sigma = \det^k$, we write $f|_k\, g$ for $f|_{\det^k}\, g$.

**Weights.** Write std for the standard representation of $\mathrm{GL}_n(\mathbb{C})$, defined by $(g,v) \mapsto gv$. Let $\mathrm{std}^\vee$ be its contragredient, given by $(g,v) \mapsto {}^\mathrm{t}g^{-1}v$. Bases for them will be denoted by $\mathfrak{e}_i$ and $\mathfrak{e}_i^\vee$ with $1 \le i \le n$. We identify the symmetric square $\mathrm{sym}^2$ of std and its contragredient $\mathrm{sym}^{\vee 2}$ with the representations $(g,m) \mapsto gm{}^\mathrm{t}g$ and $(g,m) \mapsto {}^\mathrm{t}g^{-1}mg^{-1}$ on $\mathrm{Mat}_n^\mathrm{t}(\mathbb{C})$. Bases in this case, are $\mathfrak{e}_{ij}$ and $\mathfrak{e}_{ij}^\vee$ with $1 \le i,j \le n$.

Since every representation of $\mathrm{SL}_n(\mathbb{C})$ is a subrepresentation of $\mathrm{std}^l$ for some $0 \le l \in \mathbb{Z}$ (cf. Theorem 5.5.11 of [GW09]), we find that every representation of $\mathrm{GL}_n(\mathbb{C})$ can be embedded into $\mathrm{std}^{l_1} \otimes \mathrm{std}^{\vee\, l_2}$ for some $0 \le l_1, l_2 \in \mathbb{Z}$. We will often write tensor products as $\sigma\sigma' = \sigma \otimes \sigma'$, if $\sigma$ and $\sigma'$ are representations of the same group. The outer tensor product is denoted by $\boxtimes$.

The space of homomorphisms between two representations is written $\mathrm{Hom}(\sigma,\sigma')$. If $\sigma$ and $\sigma'$ are irreducible, then by Schur's Lemma either $\mathrm{Hom}(\sigma,\sigma') \cong \mathbb{C}$ or $\mathrm{Hom}(\sigma,\sigma') = 0$. Do mind the difference between $\mathrm{Hom}(\sigma,\sigma')$ and the generally large space of vector space homomorphisms $\mathrm{Hom}\big(V(\sigma), V(\sigma')\big)$. To simplify notation, if we are given a $\mathrm{Hom}\big(V(\sigma), V(\sigma')\big)$-valued function $m$ on $\mathbb{H}^{(n)}$ and a function $f$ taking values in $V(\sigma)$ then we write $mf$ for $\tau \mapsto m(\tau)\, f(\tau)$.

Since we often swap components of tensor product, we fix notation once and for all. We let

$$\mathrm{t}_{ij} : \sigma_1 \otimes \cdots \otimes \sigma_a \longrightarrow \sigma_1 \otimes \cdots \otimes \sigma_{i-1} \otimes \sigma_j \otimes \sigma_{i+1} \otimes \cdots \otimes \sigma_{j-1} \otimes \sigma_i \otimes \sigma_{j+1} \otimes \cdots \otimes \sigma_a \tag{6.3}$$

be the canonical isomorphism, if, say, $1 \le i < j \le a$. Now fix distinct integers $i_1, \ldots, i_b$ and $j_1, \ldots, j_c$ between 1 and $a$. The average of components is denoted by

$$\mathrm{t}_{(i_1,\ldots,i_b)\,(j_1,\ldots,j_c)} = \frac{1}{bc} \sum_{k=1}^{b} \sum_{l=1}^{c} \mathrm{t}_{i_k j_l}\,. \tag{6.4}$$

In what follows, we will few $\mathrm{sym}^2$ and $\mathrm{sym}^{\vee 2}$ as subrepresentations of $\mathrm{std}^2$ and $\mathrm{std}^{\vee 2}$, if they originate in the application of lowering or raising operators.

**§6.2 Differential operators for Siegel modular forms.** A differential operator D from $C^\infty\big(\mathbb{H}^{(n)} \to V(\sigma)\big)$ to $C^\infty\big(\mathbb{H}^{(n)} \to V(\sigma')\big)$ is called covariant if it intertwines the associated slash actions:

$$D\big(f|_\sigma\, g\big) = \big(Df\big)|_{\sigma'}\, g$$

for all $f \in C^\infty\big(\mathbb{H}^{(n)} \to V(\sigma)\big)$ and all $g \in \mathrm{G}^{(n)}$. The space of such differential operators is denoted by $\mathbb{D}(\sigma,\sigma')$.

Differential operators for Siegel modular forms in the case of $\sigma = \det^k$ were found by Maass [Maa79] and by Shimura [Shi73]. For our purpose we need a basic understanding of the analytic properties of covariant differential operators for any (holomorphic) weight.

We proceed in three steps to compute them. We first give the lowering operators (in $\mathbb{D}\big(\sigma, \mathrm{sym}^{\vee 2} \otimes \sigma\big)$) for arbitrary holomorphic weight. This amounts to a direct verification. Second, we deduce from this the raising operators (in $\mathbb{D}\big(\sigma, \mathrm{sym}^2 \otimes \sigma\big)$) in case that $\sigma$ is the standard representation or its contragredient. As a third step, we deduce all properties of covariant differential operators that we need from the previous computations and the fact that the standard representation generates the representation ring of $\mathrm{SL}_n(\mathbb{C})$.

**Notation for differentials.** Given $1 \le i,j \le n$, we set

$$\partial_{ij} = \partial_{\tau,ij} = \partial_{\tau_{ij}} = \frac{d}{d\tau_{ij}}, \quad \partial_{\overline{ij}} = \partial_{\overline{\tau},ij} = \partial_{\overline{\tau_{ij}}} = \frac{d}{d\overline{\tau}_{ij}}. \tag{6.5}$$

Differentials with respect to $y$ are analogously denoted by $\partial_{y,ij}$.

Following both Maass and Shimura, we define the following matrices of differential operators.

$$\partial_\tau = \Big(\frac{1+\delta_{ij}}{2}\,\partial_{\tau,ij}\Big)_{i,j} \quad \text{and} \quad \partial_{\overline{\tau}} = \Big(\frac{1+\delta_{ij}}{2}\,\partial_{\overline{\tau},ij}\Big)_{i,j},$$

where $\delta_{ij}$ is the Dirac symbol.

**Lowering operators.** For holomorphic weight $\sigma$, we define the lowering operator

$$\mathrm{L} = \mathrm{L}^{(n)} = \mathrm{L}^{(n)}_{\sigma} : f \mapsto y\,{}^{\mathrm{t}}\big(y\partial_{\overline{\tau}}\big) \otimes f. \tag{6.6}$$

**Proposition 6.1.** *Let $\sigma$ be a complex representation of $\mathrm{GL}_n(\mathbb{C})$. Then $\mathrm{L}_{\sigma} \in \mathbb{D}\big(\sigma, \mathrm{sym}^{\vee 2} \otimes \sigma\big)$.*

*Remark 6.2.* Note that in the genus 1 case, $\mathrm{sym}^{\vee 2}$ is the same as $\det^{-2}$. Thus (6.6) reproduces the classical Maass raising operator (see p. 177 of [Maa71]), which is $\mathrm{L}_k = y^2\,\partial_{\overline{\tau}}$ up to normalizing, multiplicative scalars.

**Lemma 6.3.** *We have*

$$y \circ \mathrm{inv} = \overline{\tau}^{-1} y \tau^{-1} = \tau^{-1} y \overline{\tau}^{-1}.$$

*Proof.* Since $y = \frac{1}{2i}(\tau - \overline{\tau})$, we have $y \circ \mathrm{inv} = \frac{1}{2i}\big(-\tau^{-1} + \overline{\tau}^{-1}\big) = \tau^{-1} y \overline{\tau}^{-1}$. ∎

*Proof of Proposition 6.1.* Since $\sigma$ is, by assumption, holomorphic, application of the cocycle associated with $\sigma$ commutes with application of $\partial_{\overline{\tau}}$. We can therefore reduce our considerations to $f \in C^{\infty}\big(\mathbb{H}^{(n)}\big)$ and the trivial weight $\sigma = \mathbb{1}$. This, in particular, simplifies notation: Instead of $y\,{}^{\mathrm{t}}\big(y\partial_{\overline{\tau}}\big) \otimes f$ we can write $y\big(\partial_{\overline{\tau}} f\big) y$.

It is clear that L intertwines $\mathrm{trans}(b) \in \mathrm{G}^{(n)}$ for all $b \in \mathrm{Mat}^{\mathrm{t}}_n(\mathbb{R})$. Compatibility of L with the action of inv thus implies the statement. Using $y \circ \mathrm{inv} = \overline{\tau}^{-1} y \tau^{-1} = \tau^{-1} y \overline{\tau}^{-1}$ (see Lemma 6.3) and $\big(\partial_{\overline{\tau}} f\big) \circ \mathrm{inv} = \overline{\tau}\big(\partial_{\overline{\tau}}(f \circ \mathrm{inv})\big)\overline{\tau}$, verified in Equation (6.7) below, we find that

$$y\big(\partial_{\overline{\tau}} f\big) y\big|_{\mathrm{sym}^{\vee 2}}\,\mathrm{inv} = \tau\Big(\tau^{-1} y \overline{\tau}^{-1}\,\overline{\tau}\big(\partial_{\overline{\tau}}(f \circ \mathrm{inv})\big)\overline{\tau}\,\overline{\tau}^{-1} y \tau^{-1}\Big)\tau = y\big(\partial_{\overline{\tau}}(f \circ \mathrm{inv})\big) y.$$

This proves the proposition after we have checked

$$\big(\partial_{\tau} f\big) \circ \mathrm{inv} = \tau\big(\partial_{\tau}(f \circ \mathrm{inv})\big)\tau. \tag{6.7}$$

Consider

$$\big(\partial_{\tau}(f \circ \mathrm{inv})\big)(\tau) = \partial_{\overline{\tau}} f\big(-\tau^{-1}\big) = \sum_{i,j} \mathfrak{e}_{ij} \tfrac{1+\delta_{ij}}{2} \overline{\partial}_{ij} f\big(-\tau^{-1}\big) = \sum_{i,j,k,l} \mathfrak{e}_{ij} \big(\tfrac{1+\delta_{kl}}{2} \overline{\partial}_{kl} f\big)\big(-\tau^{-1}\big)\, \tfrac{1+\delta_{ij}}{2} \overline{\partial}_{ij}\big(-\overline{\tau}^{-1}\big).$$

The inner differential $\partial_{\overline{\tau}}\big(-\overline{\tau}^{-1}\big)$ can be computed using the product rule, giving

$$0 = \partial_{\tau} \otimes \big(\tau\tau^{-1}\big) = \big(\partial_{\tau} \otimes \tau\big)\big(\mathbb{1}_n \otimes \tau^{-1}\big) + \big(\mathbb{1}_n \otimes \tau\big)\big(\partial_{\tau} \otimes \tau^{-1}\big),$$

which yields

$$\partial_{\tau} \otimes \tau^{-1} = -\big(\mathbb{1}_n \otimes \tau^{-1}\big)\Big(\sum_{i,j} \mathfrak{e}_{ij} \otimes \tfrac{1}{2}(\mathfrak{e}_{ij} + \mathfrak{e}_{ji})\Big)\big(\mathbb{1}_n \otimes \tau^{-1}\big).$$

Symmetry of $\tau$ allows us to shift the occurrence of $\tau^{-1}$ from the second to the first tensor component:

$$\partial_{\tau} \otimes \tau^{-1} = -\big(\tau^{-1} \otimes \mathbb{1}_n\big)\Big(\sum_{i,j} \mathfrak{e}_{ij} \otimes \tfrac{1}{2}(\mathfrak{e}_{ij} + \mathfrak{e}_{ji})\Big)\big(\tau^{-1} \otimes \mathbb{1}_n\big).$$

This establishes the transformation behavior in (6.7). ∎

**The raising operator for the standard representation.** To deduce raising operators from lowering operators, we will use the intertwining operators

$$\mathrm{C}_{\mathrm{std}}\, f = y\overline{f} \quad \text{and} \quad \mathrm{C}_{\mathrm{sym}^2}\, f = y\overline{f}y. \tag{6.8}$$

**Lemma 6.4.** *The operator* $\mathrm{C}_{\mathrm{std}}$ *is covariant from* std *to* $\mathrm{std}^\vee$*, and* $\mathrm{C}_{\mathrm{sym}^2}$ *is covariant from* $\mathrm{sym}^2$ *to* $\mathrm{sym}^{\vee 2}$.

*Proof.* Covariance of $\mathrm{C}_{\mathrm{std}}$ follows from

$$y\overline{f}|_{\mathrm{std}^\vee}\, g = {}^{\mathrm{t}}(c\tau + d)\ {}^{\mathrm{t}}(c\tau + d)^{-1} y(c\overline{\tau} + d)^{-1}\,\overline{f(g\tau)} = y\,\overline{(c\tau + d)^{-1} f(g\tau)}.$$

Covariance of $\mathrm{C}_{\mathrm{sym}^2}$ can be established by a similar computation. ∎

*Remark 6.5.* The inverses of the above operators are given explicitly by

$$\mathrm{C}_{\mathrm{std}^\vee}\, f = \mathrm{C}_{\mathrm{std}}^{-1} f = y^{-1}\overline{f} \quad \text{and} \quad \mathrm{C}_{\mathrm{sym}^{\vee 2}}\, f = \mathrm{C}_{\mathrm{sym}^2}^{-1} f = y^{-1}\overline{f}y^{-1}. \tag{6.9}$$

We set

$$R_{\mathrm{std}} = \mathrm{C}_{\mathrm{sym}^2}^{-1} \otimes \mathrm{C}_{\mathrm{std}}^{-1} \circ \mathrm{L}_{\mathrm{std}^\vee} \circ \mathrm{C}_{\mathrm{std}} \quad \text{and} \quad R_{\mathrm{std}^\vee} = \mathrm{C}_{\mathrm{sym}^2}^{-1} \otimes \mathrm{C}_{\mathrm{std}} \circ \mathrm{L}_{\mathrm{std}} \circ \mathrm{C}_{\mathrm{std}}^{-1}. \tag{6.10}$$

From the covariance properties of the defining operators, we deduce that

$$R_{\mathrm{std}} \in \mathbb{D}\big(\mathrm{std}, \mathrm{sym}^{\vee 2} \otimes \mathrm{std}\big) \quad \text{and} \quad R_{\mathrm{std}^\vee} \in \mathbb{D}\big(\mathrm{std}^\vee, \mathrm{sym}^{\vee 2} \otimes \mathrm{std}^\vee\big). \tag{6.11}$$

Recall that we denote the canonical basis of $V(\mathrm{std})$ and $V(\mathrm{sym}^2)$ by $\mathfrak{e}_i$ $(1 \le i \le n)$ and $\mathfrak{e}_{ij}$ $(1 \le i, j \le n)$, respectively. In a formal way, we write $\mathrm{t}_{(12)3} : V\big(\mathrm{sym}^2 \otimes \mathrm{std}\big) \to V\big(\mathrm{sym}^2 \otimes \mathrm{std}\big)$ for the map that descends from

$$\begin{aligned} \tfrac{1}{2}(\mathrm{t}_{13} + \mathrm{t}_{23}) :\ V(\mathrm{std} \otimes \mathrm{std} \otimes \mathrm{std}) &\longrightarrow V(\mathrm{std} \otimes \mathrm{std} \otimes \mathrm{std}) \\ \mathfrak{e}_l \otimes \mathfrak{e}_m \otimes \mathfrak{e}_n &\longmapsto \tfrac{1}{2}\big(\mathfrak{e}_n \otimes \mathfrak{e}_m \otimes \mathfrak{e}_l + \mathfrak{e}_l \otimes \mathfrak{e}_n \otimes \mathfrak{e}_m\big). \end{aligned} \tag{6.12}$$

**Proposition 6.6.** *For* $f \in C^\infty\big(\mathbb{H}^{(n)} \to V(\mathrm{std})\big)$*, we have*

$$\mathrm{R}_{\mathrm{std}}\, f = \partial_\tau \otimes f - \tfrac{i}{2}\mathrm{t}_{(12)3}\big(y^{-1} \otimes f\big), \tag{6.13}$$

*and for* $f \in C^\infty\big(\mathbb{H}^{(n)} \to V(\mathrm{std}^\vee)\big)$*, we have*

$$\mathrm{R}_{\mathrm{std}^\vee}\, f = \partial_\tau \otimes f + \tfrac{i}{2}\mathrm{t}_{(12)3}\big(\mathbb{1}_n \otimes y^{-1} f\big). \tag{6.14}$$

*Proof.* Denote coordinates of $f$ corresponding to the $\mathfrak{e}_i$ by $f_i$. To compute $\mathrm{R}_{\mathrm{std}}$, we start by computing $L_{\mathrm{std}^\vee} \circ \mathrm{C}_{\mathrm{std}}\, f$, which equals

$$\begin{aligned} &\sum_{i,j,k,l} \mathfrak{e}_{ij}\, y_{ik} y_{jl} \tfrac{1+\delta_{kl}}{2} \overline{\partial}_{kl} \sum_{m,n} \mathfrak{e}_m y_{mn} \overline{f_n} \\ &\qquad = \sum_{i,j,m} \mathfrak{e}_{ij}\mathfrak{e}_m \Big( \sum_{k,l,n} y_{ik} y_{jl} y_{mn}\, \tfrac{1+\delta_{kl}}{2} \overline{\partial}_{kl} \overline{f_n} + \tfrac{i}{4} \sum_n \big(y_{im} y_{jn} + y_{in} y_{jm}\big)\overline{f_n} \Big). \end{aligned}$$

Recall that we write $y_{ij}^{-1}$ for the $ij$-th entry of $y^{-1}$. Now, $\mathrm{C}_{\mathrm{sym}^{\vee 2}} \otimes \mathrm{C}_{\mathrm{std}^\vee}$ maps the previous expression to

$$\begin{aligned} &\sum_{i',j',m'} \mathfrak{e}_{i'j'}\mathfrak{e}_{m'}\, y_{ii'}^{-1} y_{jj'}^{-1} y_{mm'}^{-1} \sum_{i,j,m} \Big( \sum_{k,l,n} y_{ik} y_{jl} y_{mn}\, \tfrac{1+\delta_{kl}}{2} \partial_{kl} f_n - \tfrac{i}{2} \sum_n \tfrac{1}{2}\big(y_{im} y_{jn} + y_{in} y_{jm}\big) f_n \Big) \\ &= \sum_{k,l,n} \mathfrak{e}_{kl}\mathfrak{e}_n\, \tfrac{1+\delta_{kl}}{2} \partial_{kl} f_n - \tfrac{i}{4} \sum_{m,n,m'} \mathfrak{e}_{mn}\mathfrak{e}_{m'}\, \big(y_{mm'}^{-1} f_n + y_{nm'}^{-1} f_m\big). \end{aligned}$$

The expression for $\mathrm{R}_{\mathrm{std}^\vee}$ can be derived in an analogous way. ∎

**Differential operators for tensor products of weights.** To extend our considerations to arbitrary representations, we have to study how raising operators combine when taking the tensor product of two weights. For a representation $\sigma$, the order 1 raising operator $\mathrm{R}_\sigma$ is of the form

$$\mathrm{R}_\sigma f = \partial_\tau f + \mathrm{K}_\sigma f\,, \quad \text{for} \quad \mathrm{K}_\sigma \in \mathrm{C}^\infty\big(\mathbb{H}^{(n)} \to \mathrm{Hom}\big(V(\sigma), V(\mathrm{sym}^2 \otimes \sigma)\big)\big).$$

Given two representations $\sigma$ and $\sigma'$, we write $\iota_\sigma$ for the inclusion

$$\begin{aligned}\iota_\sigma : \mathrm{Hom}\big(V(\sigma), V(\mathrm{sym}^2 \otimes \sigma)\big) &\longrightarrow \mathrm{Hom}\big(V(\sigma \otimes \sigma'), V(\mathrm{sym}^2 \otimes \sigma \otimes \sigma')\big)\\ \phi &\longmapsto \phi \otimes \mathrm{id}.\end{aligned}$$

Similarly, $\iota_{\sigma'}$ is defined as an inclusion of $\mathrm{Hom}\big(V(\sigma'), V(\mathrm{sym}^2 \otimes \sigma')\big)$ into the same codomain.

**Lemma 6.7.** *Using above notation, we have*

$$\mathrm{K}_{\sigma\otimes\sigma'} = \iota_\sigma\big(\mathrm{K}_\sigma\big) + \iota_{\sigma'}\big(\mathrm{K}_{\sigma'}\big).$$

*Proof.* A computation gives for $\mathrm{K}_{\sigma\otimes\sigma'}\big(f|_{\sigma\sigma'}\, g\big)$ the expression

$$\begin{aligned}&\big(\partial_\tau \otimes \sigma(c\tau+d)^{-1}\big) \otimes \sigma'(c\tau+d)^{-1} f(g\tau) + \mathrm{t}_{12}\Big(\sigma(c\tau+d)^{-1} \otimes \big(\partial_\tau \otimes \sigma'(c\tau+d)^{-1}\big)\Big) f(g\tau)\\ &+ \mathrm{K}_{\sigma\otimes\sigma'}\big(\sigma(c\tau+d)^{-1} \otimes \sigma'(c\tau+d)^{-1} f(g\tau)\big).\end{aligned}$$

The transposition $\mathrm{t}_{12}$, defined in (6.3), switches the first and second component of $\sigma \otimes \mathrm{sym}^2 \otimes \sigma'$.

By definition, we have

$$\big(\partial_\tau \otimes \sigma(c\tau+d)^{-1}\big) f(g\tau) + \mathrm{K}_\sigma \big(\sigma(c\tau+d)^{-1} f(g\tau)\big) = \big(\mathrm{K}_\sigma f\big)|_{\mathrm{sym}^2\otimes\sigma}\, g,$$

implying that

$$\big(\partial_\tau \otimes \sigma(c\tau+d)^{-1}\big) \otimes \sigma'(c\tau+d)^{-1} f(g\tau) + \iota_\sigma(\mathrm{K}_\sigma)\big(\sigma(c\tau+d)^{-1} \otimes \sigma'(c\tau+d)^{-1} f(g\tau)\big)$$

equals $\big(\iota_\sigma(\mathrm{K}_\sigma) f\big)|_{\mathrm{sym}^2\otimes\sigma\otimes\sigma'}\, g$. The analogue equation holds for $\mathrm{K}_{\sigma'}$ and its image under $\iota_{\sigma'}$. After plugging this into the first equality and comparing terms, this implies the lemma. ■

Let $a, b, c$, $i_1, \ldots, i_b$, and $j_1, \ldots, j_c$ be as in (6.4). The raising operators for powers of $\mathrm{std}^\vee$ involve the following modification of $\mathrm{t}_{(i_1,\ldots,i_b)\,(j_1,\ldots,j_c)}$:

$$\mathrm{t}^y_{(i_1,\ldots,i_b)\,(j_1,\ldots,j_c)} = \frac{1}{bc} \sum_{k=1}^{b} \sum_{l=1}^{c} \mathrm{t}_{i_k j_l} \circ \mu^y_{j_l}\,, \tag{6.15}$$

where $\mu^y_j$ is multiplication with $y^{-1}$ in the $j$-th component.

**Corollary 6.8.** *Explicit expressions for the corvariant differential operators*

$$\begin{aligned}&\mathrm{R}_{\mathrm{std}^l} \in \mathbb{D}\big(\mathrm{std}^l, \mathrm{sym}^{\vee 2} \otimes \mathrm{std}^l\big), \quad \mathrm{R}_{\mathrm{std}^{\vee l}} \in \mathbb{D}\big(\mathrm{std}^{\vee l}, \mathrm{sym}^{\vee 2} \otimes \mathrm{std}^{\vee l}\big), \quad \textit{and}\\ &\mathrm{R}_{\mathrm{std}^{l_1}\mathrm{std}^{\vee l_2}} \in \mathbb{D}\big(\mathrm{std}^{l_1}\mathrm{std}^{\vee l_2}, \mathrm{sym}^{\vee 2} \otimes \mathrm{std}^{l_1}\mathrm{std}^{\vee l_2}\big)\end{aligned} \tag{6.16}$$

*are*

$$\begin{aligned}&\mathrm{R}_{\mathrm{std}^l} f = \partial_\tau \otimes f - \tfrac{il}{2} \mathrm{t}_{(12)(3\cdots2+l)}(y^{-1} \otimes f), \qquad \mathrm{R}_{\mathrm{std}^{\vee l}} f = \partial_\tau \otimes f + \tfrac{il}{2} \mathrm{t}^y_{(12)(3\cdots2+l)}(\mathbb{1}_n \otimes f), \quad \textit{and}\\ &\mathrm{R}_{\mathrm{std}^{l_1}\mathrm{std}^{\vee l_2}} = \partial_\tau \otimes f - \tfrac{il_1}{2} \mathrm{t}_{(12)(3\cdots2+l_1)}(y^{-1} \otimes f) + \tfrac{il_2}{2} \mathrm{t}^y_{(12)(3+l_1\cdots2+l_1+l_2)}(\mathbb{1}_n \otimes f).\end{aligned} \tag{6.17}$$

**The Lie algebra of $\mathrm{Sp}_n$.** To describe $\mathbb{D}(\sigma, \mathrm{sym}^2\sigma)$ and $\mathbb{D}(\sigma, \mathrm{sym}^{\vee 2}\sigma)$, we employ the Lie theoretic description of differential operators given by Helgason [Hel59; Hel92] and, in context of discrete series representations, by Schmid [Sch76].

Let $\mathfrak{sp}_n$ denote the complexified Lie algebra of $\mathrm{G}^{(n)} = \mathrm{Sp}_n(\mathbb{R})$. Fix

$$\mathfrak{k}_n = \left\{ \begin{pmatrix} a & b \\ -b & a \end{pmatrix} : a = -{}^{\mathrm{t}}a \in \mathrm{Mat}_n(\mathbb{C}),\, b = {}^{\mathrm{t}}b \in \mathrm{Mat}_n(\mathbb{C}) \right\}, \tag{6.18}$$

which is the complexified Lie algebra of the maximal compact subgroup

$$\mathrm{K}_n = \left\{ \begin{pmatrix} a & b \\ -b & a \end{pmatrix} : a, b \in \mathrm{Mat}_n(\mathbb{R}),\, ai + b \in \mathrm{U}_n(\mathbb{R}) \right\} \subseteq \mathrm{Sp}_n(\mathbb{R}).$$

Denote the center of $\mathfrak{k}_n$ by $\mathfrak{k}_{n,\mathbb{C}}$ or, suppressing the genus, $\mathfrak{k}_\mathbb{C}$. To a representation $\sigma$ of $\mathrm{GL}_n(\mathbb{C})$ we can attach a representation of $\mathfrak{k}_n$ by taking the differential of its restriction to $\mathrm{K}_n$. Since $\mathrm{K}_n$ is connected, we can and will refer to representations of $\mathfrak{k}_n$ by corresponding representations of $\mathrm{GL}_n(\mathbb{C})$.

As a $\mathfrak{k}_n$ module, we have the decomposition

$$\mathfrak{sp}_n \cong \mathfrak{k}_n \oplus \mathfrak{p}_n^+ \oplus \mathfrak{p}_n^-, \tag{6.19}$$

where $\mathfrak{p}_n^\pm$ is the differential of the $K_n$ representations $\mathrm{sym}^2$ and $\mathrm{sym}^{\vee 2}$. A suitable generator of $\mathfrak{k}_\mathbb{C}$ acts on them by $\pm 2$. We will throughout refer to eigenvalues of this generator as $\mathfrak{k}_\mathbb{C}$-eigenvalues.

We denote invariants of a Lie algebra representation $V$ of $\mathfrak{k}_n$ by $\mathrm{H}^0(\mathfrak{k}_n, V)$. Fix two representation $\sigma$ and $\sigma'$ of $\mathrm{GL}_n(\mathbb{C})$. Write $\mathbb{D}_o(\sigma,\sigma')$ for the space of order $o$ differential operators that are covariant from $|_\sigma$ to $|_{\sigma'}$. In Theorem 10 of [Hel59], Helgason described them by the following vector space isomorphism:

$$\mathbb{D}_o(\sigma,\sigma') \cong \mathrm{H}^0\big(\mathfrak{k}_n, \mathrm{sym}^o\,(\mathfrak{p}_n^+ \oplus \mathfrak{p}_n^-) \otimes \sigma^\vee \otimes \sigma'\big). \tag{6.20}$$

We reiterate Helgson's remark that this isomorphism is not compatible with composition of differential operators.

Helgason's result allows us to restrict the order of covariant differential operators.

**Proposition 6.9.** *Fix irreducible representations $\sigma$ and $\sigma'$ of $\mathrm{GL}_n(\mathbb{C})$.*

*(i) If the $\mathfrak{k}_\mathbb{C}$-eigenvalues of $\sigma$ and $\sigma'$ agree, then $\mathbb{D}_1(\sigma,\sigma') = \mathbb{D}_0(\sigma,\sigma')$.*

*(ii) If the $\mathfrak{k}_\mathbb{C}$-eigenvalues of $\sigma$ and $\sigma'$ differ by $\pm 2$, then $\mathbb{D}_2(\sigma,\sigma') = \mathbb{D}_1(\sigma,\sigma')$.*

*(iii) We have $\mathbb{D}_0(\sigma,\sigma') \cong \mathrm{Hom}(\sigma,\sigma')$.*

*(iv) The space $\mathbb{D}_1(\sigma,\sigma')$ is spanned by operators $\pi_\mathrm{L} \circ \mathrm{L}_\sigma$ and $\pi_\mathrm{R} \circ \mathrm{R}_{\sigma'}$, where $\pi_\mathrm{L} \in \mathrm{Hom}\big(\mathrm{sym}^2\sigma, \sigma'\big)$ and $\pi_\mathrm{R} \in \mathrm{Hom}\big(\mathrm{sym}^{\vee 2}\sigma, \sigma'\big)$, respectively.*

*Proof.* Case (iii) is a direct consequence of (6.20). To prove case (i) and (ii), it suffices to compute $\mathfrak{k}_\mathbb{C}$-eigenvalues in $\mathrm{sym}^o(\mathfrak{p}_n^+ \oplus \mathfrak{p}_n^-) \otimes V(\sigma)^\vee \otimes \sigma^\vee \otimes \sigma'$ for $o \in \{1,2\}$.

The $\mathfrak{k}_\mathbb{C}$-eigenvalues of $\mathbb{1} \cong \mathrm{sym}^0(\mathfrak{p}_n^+ \oplus \mathfrak{p}_n^-)$ and $\mathfrak{p}_n^+ \otimes \mathfrak{p}_n^- \subset \mathrm{sym}^2(\mathfrak{p}_n^+ \oplus \mathfrak{p}_n^-)$ are 0. On $\mathfrak{p}_n^+$ and $\mathfrak{p}_n^-$ they are $\pm 2$, by definition, and on $\mathrm{sym}^2 \mathfrak{p}_n^+$ and $\mathrm{sym}^2 \mathfrak{p}_n^-$ they equal $\pm 4$. In case (i) the $\mathfrak{k}_\mathbb{C}$-eigenvalue of $\sigma^\vee \otimes \sigma'$ equals $-a + a \pm 2 = \pm 2$ for some $a \in \mathbb{C}$, and in case (ii) it equals 0. Combining these values we deduce all cases but (iv).

To establish the remaining statement (iv), note that the case of lowering operators is implied by the isomorphism

$$\mathrm{H}^0\big(\mathfrak{k}_n, \mathrm{sym}^{\vee 2} \otimes \sigma^\vee \otimes \sigma'\big) \cong \operatorname*{Hom}_{\mathfrak{k}_n}\big(\mathrm{sym}^2 \otimes \sigma, \sigma'\big).$$

The case of raising operators can be dealt with analogously. ■

**Commutation relations.** By abuse of notation, we denote two projections by the same symbol:

$$\pi_{\mathrm{sym}^2} : \mathrm{sym}^2 \otimes \mathrm{sym}^{\vee 2} \otimes \sigma,\, \mathrm{sym}^{\vee 2} \otimes \mathrm{sym}^2 \otimes \sigma \longrightarrow \sigma. \tag{6.21}$$

To concisely state commutation relations of differential operators, we suppress the composition symbol: $\mathrm{LR}_\sigma = \mathrm{L} \circ \mathrm{R}_\sigma$. Powers of operators will be understood as repeated composition: $\mathrm{L}^d = \mathrm{L} \circ \cdots \circ \mathrm{L}$ for $0 \le d \in \mathbb{Z}$.

**Lemma 6.10.** *We have*

$$\pi_{\mathrm{sym}^2}\, \mathrm{L} \mathrm{R}_{\mathrm{sym}^{\vee 2} \otimes \sigma} \mathrm{L}_\sigma - \mathrm{L} \pi_{\mathrm{sym}^2}\, \mathrm{R}_{\mathrm{sym}^{\vee 2} \otimes \sigma} \mathrm{L}_\sigma \in \mathbb{D}_1\big(\sigma, \mathrm{sym}^2 \otimes \sigma\big).$$

*In particular, it can be expressed as a linear combination of constituents of* $\mathrm{L}_\sigma$.

*Proof.* It is straightforward to see that the given expression is a differential operator of order at most 2, since symbols of differential operators commute. We may assume that $\sigma$ is irreducible. Denote its $\mathfrak{k}_\mathbb{C}$ eigenvalue by $a$. Further, observe that the $\mathfrak{k}_\mathbb{C}$ eigenvalues of $\mathrm{sym}^{\vee 2} \otimes \mathrm{sym}^2 \otimes \mathrm{sym}^{\vee 2} \otimes \sigma$ equals $a-1$. By Helgason's theory as stated in Proposition 6.9, the left hand side is an order 1 differential operator, which is a linear combination of the constituents of $\mathrm{L}_\sigma$ by the same proposition. ∎

**Lemma 6.11.** *We have*

$$\mathrm{L}_{\mathrm{sym}^2 \sigma} \mathrm{R}_\sigma - \mathrm{t}_{12} \mathrm{R}_{\mathrm{sym}^{\vee 2} \sigma} \mathrm{L}_\sigma \in \mathrm{Hom}\big(\sigma, \mathrm{sym}^2 \mathrm{sym}^{\vee 2} \sigma\big).$$

*Proof.* The left hand side is a covariant differential operator of order at most 1 that preserves $\mathfrak{k}_\mathbb{C}$ eigenvalues. Therefore, we can deduce the statement from Proposition 6.9. ∎

**Lemma 6.12.** *Let $\sigma$ be a representation with $\mathfrak{k}_\mathbb{C}$ eigenvalue $a$. Then*

$$\pi_{\mathrm{sym}^2}\big(\mathrm{L}_{\mathrm{sym}^2 \sigma} \mathrm{R}_\sigma - \mathrm{t}_{12} \mathrm{R}_{\mathrm{sym}^{\vee 2} \sigma} \mathrm{L}_\sigma\big) = \frac{-(1+n)\, a}{8}.$$

*Proof.* By Lemma 6.11, the left hand side lies in $\mathrm{Hom}(\sigma,\sigma)$. We may assume that $\sigma$ is irreducible, so that $\mathrm{Hom}(\sigma,\sigma) \cong \mathbb{C}$. To compute the precise value, note that in the following expression for $\mathrm{LR} - \mathrm{t}_{12}\mathrm{RL}$

$$\big(y\,{}^{\mathrm{t}}(y\partial_{\overline{\tau}})\big)\big(\partial_\tau + \mathrm{K}_\sigma\big) - \mathrm{t}_{12}\big(\big(\partial_\tau + \mathrm{K}_{\mathrm{sym}^{\vee 2} \sigma}\big)\big(y\,{}^{\mathrm{t}}(y\partial_{\overline{\tau}})\big)\big)$$

order 0 terms can only arise from $y\,{}^{\mathrm{t}}(y\partial_{\overline{\tau}})\mathrm{K}_\sigma$. Using Lemma 6.7, we reduce ourselves to the cases $\sigma = \mathrm{std}$ and $\sigma = \mathrm{std}^\vee$. We focus on the former one, and leave the latter one to the reader. For any $1 \le o \le n$, we have

$$y\,{}^{\mathrm{t}}(y\partial_{\overline{\tau}}) \otimes \frac{-i}{2} \mathrm{t}_{(12)3}(y^{-1} \otimes \mathfrak{e}_o) = \frac{-i}{2} \sum_{i,j,k,l} \mathfrak{e}_{ij}\, y_{ik} y_{jl}\, \frac{1+\delta_{kl}}{2} \overline{\partial}_{kl} \sum_{m,n,o} \frac{1}{2}\big(\mathfrak{e}_{mo}\mathfrak{e}_n + \mathfrak{e}_{on}\mathfrak{e}_m\big) y^{-1}_{mn}$$
$$= \frac{1}{4} \sum_{i,j,k,l,m,n} \mathfrak{e}_{ij}\, y_{ik} y_{jl}\, \frac{1}{2}\big(\mathfrak{e}_{mo}\mathfrak{e}_n + \mathfrak{e}_{on}\mathfrak{e}_m\big) \frac{-1}{2}\big(y^{-1}_{km} y^{-1}_{ln} + y^{-1}_{kn} y^{-1}_{lm}\big).$$

We execute multiplication of $y$ and $y^{-1}$ twice

$$\frac{-1}{16} \sum_{i,j,m,n} \mathfrak{e}_{ij}\big(\mathfrak{e}_{mo}\mathfrak{e}_n + \mathfrak{e}_{on}\mathfrak{e}_m\big)\big(\delta_{im}\delta_{jn} + \delta_{in}\delta_{jm}\big) = \frac{-1}{16} \sum_{i,j} \mathfrak{e}_{ij}\big(\mathfrak{e}_{io}\mathfrak{e}_j + \mathfrak{e}_{oj}\mathfrak{e}_i + \mathfrak{e}_{jo}\mathfrak{e}_i + \mathfrak{e}_{oi}\mathfrak{e}_j\big).$$

Applying $\pi_{\mathrm{sym}^2}$, we obtain

$$\frac{-1}{16} \sum_{i,j} \big(\delta_{jo} + \delta_{io} + \delta_{ij}\delta_{jo}\mathfrak{e}_o + \delta_{io}\delta_{ij}\big)\, \mathfrak{e}_o = \frac{-(1+n)}{8} \mathfrak{e}_o.$$ ∎

**§6.3 Almost holomorphic functions.** We throughout write $\mathscr{O}\big(\mathbb{H}^{(n)}\big)$ for the space of holomorphic functions on $\mathbb{H}^{(n)}$.

**Definition 6.13.** A smooth function $f : \mathbb{H}^{(n)} \to V(\sigma)$ is almost holomorphic of depth $d$ if $\mathrm{L}^{d+1} f = 0$.

*Remark 6.14.* This definition is motivated by the theory of holomorphic discrete series, whose behavior under differential operators is well-known [Sch76]. Very recently, a work that focuses on this perspective has appear as a preprint: [PSS15].

The following are basic properties of almost holomorphic functions on $\mathbb{H}^{(n)}$ which are straight forward to prove.

**Proposition 6.15.** *Suppose that $f : \mathbb{H}^{(n)} \to V(\sigma)$ and $g : \mathbb{H}^{(n)} \to V(\sigma')$ are almost holomorphic functions of depth $d$ and $d'$. Then the tensor product $f \otimes g$ is almost holomorphic of depth $d + d'$.*

*Proof.* We abbreviate $d'' = d + d'$. It suffices to inspect the equation $\mathrm{L}^{d''+1} f \otimes g = \sum_{t=0}^{d''+1} \mathrm{t}_t\big((\mathrm{L}^t f) \otimes (\mathrm{L}^{d''+1-t} g)\big)$ with suitable permutations of tensor components $\mathrm{t}_t : (\mathrm{sym}^2)^t \sigma (\mathrm{sym}^2)^{d''+1-t} \sigma' \to (\mathrm{sym}^2)^{d''+1} \sigma\sigma'$. ■

**Lemma 6.16.** *If $f$ is almost holomorphic of depth $d$, then $\pi_{\mathrm{sym}^2} \mathrm{LRL} f - \mathrm{L} \pi_{\mathrm{sym}^2} \mathrm{RL} f$ has depth $d - 1$.*

*Proof.* This is a consequence of Lemma 6.10, seen that $f$, which has depth $d$, vanishes after applying $d + 1$ times arbitrary constituents of L. ■

**Analytic properties.** Almost holomorphic functions in the setting of elliptic modular forms can be easily recognized as a polynomial in $y^{-1}$ whose coefficients are holomorphic functions. Their total degree coincides with the depth.

**Theorem 6.17.** *A function $f \in C^\infty\big(\mathbb{H} \to V(\sigma)\big)$ is almost holomorphic of depth $d$ if and only if it is a polynomial in the entries of $y^{-1}$ of degree at most $d$ and with holomorphic coefficients.*

*Proof.* Note that $\mathrm{L}\, y^{-1}$ is constant. After embedding $\sigma$ into $\big(\mathrm{sym}^2 \mathrm{sym}^{\vee 2}\big)^d \otimes \sigma$, it follows that every polynomial in the entries of $y^{-1}$ of at most degree $d$ is indeed almost holomorphic of depth $d$.

We proceed as in the case of elliptic modular forms to establish the converse. Note that it suffices to treat irreducible $\sigma$. Since the lowering operator is the same for representations $\sigma$ and $\det^k \sigma$ for any $k \in \mathbb{Z}$, we may always modify $\sigma$ correspondingly. In particular, we can assume that the $\mathfrak{k}_{\mathbb{C}}$-eigenvalue of $\sigma$ is arbitrarily large. Then Lemma 6.12, shows that the space of almost holomorphic functions of depth $d$ is contained in the span of depth $d - 1$ functions and their image under $\pi_{\mathrm{sym}^2} \mathrm{R}_{\mathrm{sym}^2 \sigma}$. Since the case $d = 0$ is clear, the result follows by induction. ■

Theorem 6.17 tells us that an almost holomorphic function $f$ is a polynomial in the entries of $y^{-1}$. For $t \in \mathrm{Mat}^{\mathrm{t}}_n(\frac{1}{2}\mathbb{Z})$ with integral diagonal entries, set $y^{-t} = \exp\big(\mathrm{trace}\big(\mathrm{t} \log(\mathrm{y}^{-1})\big)\big)$. Then $f$ can be written as

$$f(\tau) = \sum_t f_t(\tau)\, y^{-t} \tag{6.22}$$

with holomorphic $f_t$, finitely many of which are non-zero. We call $f_t$ the $t$-th part of $f$, and $f_0$ is called the constant part of $f$.

**§6.4 Definition of almost holomorphic Siegel modular forms.** Seen that one goal of this work is to settle basic tools for the treatment of almost holomorphic Siegel modular forms, we give a slightly more general definition, including vector valued modular Siegel modular forms with respect to representations of $\Gamma^{(n)} = \mathrm{Sp}_n(\mathbb{Z})$, than the one that is necessary to prove Theorem 8.5. Note also that statements in this paper can be extended to representations of the metaplectic double cover $\mathrm{Mp}_{2n}(\mathbb{Z})$ of $\mathrm{Sp}_n(\mathbb{Z})$ without difficulties, since they depend only on analytic properties of corresponding modular forms.

**Definition 6.18.** Let $\rho$ be a finite dimensional, complex representation of $\Gamma^{(n)}$ and $\sigma$ a finite dimensional, complex representation of $\mathrm{GL}_n(\mathbb{C})$. An almost holomorphic function $f : \mathbb{H}^{(n)} \to V(\sigma) \otimes V(\rho)$ of depth $d$ that satisfies

$$f|_{\sigma,\rho}\,\gamma = f \quad \text{for all } \gamma \in \Gamma^{(n)},$$

and, if $n = 1$, is bounded by some power of $y^{-1}$ as $y \to \infty$, is called an almost holomorphic Siegel modular form of genus $n$, depth $d$, weight $\sigma$, and type $\rho$.

We write $\mathrm{M}^{(n)}\big(\sigma^{[d]} \boxtimes \rho\big)$ for the space of such functions; notation which is inspired by the fact that almost holomorphic Siegel modular forms correspond to automorphic forms whose "$K$-type at the infinite place", related to $\sigma$, is not the highest possible, but $d$ steps away from it. Proposition 6.15 implies that

$$\mathrm{M}^{(n)}\big(\sigma^{[d]} \boxtimes \rho\big) \otimes \mathrm{M}^{(n)}\big(\sigma'^{[d']} \boxtimes \rho'\big) \subseteq \mathrm{M}^{(n)}\big((\sigma\sigma')^{[d+d']} \boxtimes \rho\rho'\big). \tag{6.23}$$

An almost holomorphic Siegel modular form $f$ has Fourier expansion

$$f(\tau) = \sum_{t \in \mathrm{Mat}^{\mathrm{t}}_n(\mathbb{Q})} c(f;\, t, y)\, \exp\big(2\pi i\, \mathrm{trace}(t\tau)\big) \tag{6.24}$$

where every Fourier coefficient $c(f;\, t, y)$ is a polynomial in the entries of $y^{-1}$ with coefficients in $\mathbb{C}$.

# 7 | Fourier-Jacobi expansions

Fourier-Jacobi expansions of Siegel modular forms have developed to a major tool to study Siegel modular forms. They have abstractly been considered in [EZ85]. Roughly, a Siegel modular form $f$ can be written as

$$f(\tau) = \sum_{m \in \mathbb{Z}} \phi_m(\tau_1, z)\, \exp(2\pi i\, m\tau_2)$$

where we use, here and throughout, the decomposition $\tau = \left(\begin{smallmatrix} \tau_1 & {}^{\mathrm{t}}z \\ z & \tau_2 \end{smallmatrix}\right)$ with $\tau_1 \in \mathbb{H}^{(n-1)}$, $\tau_2 \in \mathbb{H}^{(1)}$, and $z \in \mathbb{C}^{n-1}$. Each $\phi_m$ is a Jacobi form, which can be related to Siegel modular forms of genus $n-1$. The theory of Fourier-Jacobi expansions has recently culminate in [BWR14], where the last named author and his coauthor prove that such formal expansions converge, if they exhibit the symmetry relations of Fourier coefficients that hold for Fourier expansions of Siegel modular forms.

In this section, we present the foundations of Fourier Jacobi expansions of almost holomorphic Siegel modular forms. In particular, we attach almost holomorphic Fourier-Jacobi coefficients to almost holomorphic Siegel modular forms and show that they determine them uniquely. In the case of genus $n = 2$, we characterize the spaces of almost holomorphic Jacobi forms using the approach in [IK11]. Note that it is possible to extend this to arbitrary genera, but the focus of this paper lies on $n = 2$.

**§7.1 Preliminaries.** Jacobi forms were defined in [EZ85] and generalized in [Zie89]. Here, we revisit basic notation.

**The Jacobi upper half space.** Write $\mathbb{H}^{(n)\mathrm{J}} = \mathbb{H}^{(n)} \times \mathbb{C}^n = \{(\tau, z) : \tau \in \mathbb{H}^{(n)}, z \in \mathbb{C}^n\}$ for the Jacobi upper half space of genus $n$ (and cogenus 1). The real and imaginary part of $\tau$ and $z$ are denoted by $\tau = x + iy$ and $z = u + iv$. If $n = 1$, then we suppress the genus, writing $\mathbb{H}^{\mathrm{J}}$ instead of $\mathbb{H}^{(n)\mathrm{J}}$.

**The Heisenberg group.** Fix the Heisenberg group

$$\mathrm{H}(\mathbb{R}^n) = \big\{h = (\lambda, \mu, \kappa) : \lambda, \mu \in \mathbb{R}^n, \kappa \in \mathrm{Mat}_n(\mathbb{R})\big\}$$

with addition

$$(\lambda_1, \mu_1, \kappa_1) + (\lambda_2, \mu_2, \kappa_2) = \big(\lambda_1 + \lambda_2, \mu_1 + \mu_2, \kappa_1 + \kappa_2 + \lambda_1\, {}^{\mathrm{t}}\mu_2 - \mu_1\, {}^{\mathrm{t}}\lambda_2\big).$$

It carries a right action of $\mathrm{G}^{(n)}$ defined by the natural action on row vectors of length $2n$:

$$(\lambda, \mu, \kappa) \left(\begin{smallmatrix} a & b \\ c & d \end{smallmatrix}\right) = \big(\lambda a + \mu c, \lambda b + \mu d, \kappa\big).$$

**The Jacobi group.** The real Jacobi group $\mathrm{G}^{(n)\mathrm{J}} = \mathrm{G}^{(n)\mathrm{J}}(\mathbb{R})$ is the non-trivial extension $0 \to \mathrm{H}(\mathbb{R}^n) \to \mathrm{G}^{(n)\mathrm{J}} \to \mathrm{G}^{(n)} \to 1$ of $\mathrm{G}^{(n)}$ by $\mathrm{H}(\mathbb{R}^n)$. Multiplication in $\mathrm{G}^{(n)\mathrm{J}}$ is defined by

$$\big(g_1, h_1\big)\big(g_2, h_2\big) = \big(g_1 g_2, h_1 g_2 + h_1\big).$$

We often consider elements of $\mathrm{H}(\mathbb{R}^n)$ as elements of $\mathrm{G}^{(n)\mathrm{J}}$ via the inclusion $\mathrm{H}(\mathbb{R}^n) \subset \mathrm{G}^{(n)\mathrm{J}}$. Shorthand notation for elements of $\mathrm{G}^{(n)}$ considered as elements of $\mathrm{G}^{(n)\mathrm{J}}$ is provided by the following section to $\mathrm{G}^{(n)\mathrm{J}} \twoheadrightarrow \mathrm{G}^{(n)}$:

$$g \in \mathrm{G}^{(n)} \longmapsto \big(g, (0,0,0)\big) \in \mathrm{G}^{(n)\mathrm{J}}.$$

If $n = 1$, we drop the superscript $(n)$, writing $\mathrm{G}^{\mathrm{J}}$ instead of $\mathrm{G}^{(1)\mathrm{J}}$. Recall the shorthand notation $a, b, c, d$ for the entries of $g \in \mathrm{G}^{(n)}$. Similarly, the entries of $h \in \mathrm{H}(\mathbb{R}^n)$ will be denoted by $\lambda$, $\mu$, and $\kappa$ without further mentioning it.

We have an action of $\mathrm{G}^{(n)\mathrm{J}}$ on $\mathbb{H}^{(n)\mathrm{J}}$:

$$g^{\mathrm{J}}\,(\tau, z) = \big(g, (\lambda, \mu, \kappa)\big)\,(\tau, z) = \Big((a\tau + b)(c\tau + d)^{-1},\, (z + \lambda\tau + \mu)(c\tau + d)^{-1}\Big). \tag{7.1}$$

*Remark 7.1.* The connection between Jacobi forms and Siegel modular forms, that we are going to exhibit, is based on the embedding

$$\mathrm{G}^{(n)\mathrm{J}}_{\mathrm{sym}} \hookrightarrow \mathrm{G}^{(n)}, \quad \Big(\big(\begin{smallmatrix} a & b \\ c & d \end{smallmatrix}\big), (\lambda, \mu, \kappa)\Big) \longmapsto \begin{pmatrix} a & 0 & b & a\,{}^{\mathrm{t}}\mu - b\,{}^{\mathrm{t}}\lambda \\ \lambda & 1 & \mu & \kappa \\ c & 0 & d & c\,{}^{\mathrm{t}}\mu - d\,{}^{\mathrm{t}}\lambda \\ 0 & 0 & 0 & 1 \end{pmatrix}, \tag{7.2}$$

where $\mathrm{G}^{(n)\mathrm{J}}_{\mathrm{sym}}$ is the subgroup of $\mathrm{G}^{(n)\mathrm{J}}$ such that $\kappa + \mu\,{}^{\mathrm{t}}\lambda$ is symmetric. Details can be found in [Zie89].

The (discrete) Jacobi group $\Gamma^{(n)\mathrm{J}} \subset \mathrm{G}^{(n)\mathrm{J}}$ is defined as

$$\Gamma^{(n)\mathrm{J}} = \big\{\gamma^{\mathrm{J}} = \big(\gamma, (\lambda, \mu, 0)\big) \,:\, g \in \Gamma^{(n)},\, \lambda, \mu \in \mathbb{Z}^n\big\}.$$

**Slash actions.** Let $\sigma$ be a finite dimensional, complex representation of the affine group

$$\mathrm{Aff}_n(\mathbb{C}) = \mathrm{GL}_n(\mathbb{C}) \ltimes \mathbb{C}^n. \tag{7.3}$$

We attach to it a slash action on $C^\infty\big(\mathbb{H}^{(n)\mathrm{J}} \to V(\sigma)\big)$ by

$$\begin{aligned}\big(\phi|^{\mathrm{J}}_{\sigma,m}\, g^{\mathrm{J}}\big)(\tau, z) = {} & \sigma\big(c\tau + d, c\,{}^{\mathrm{t}}z + c\,{}^{\mathrm{t}}\mu - d\,{}^{\mathrm{t}}\lambda\big)^{-1} \\ & \cdot \exp\Big(2\pi i m\,\big(-(z + \lambda\tau + \mu)c(c\tau + d)^{-1}\,{}^{\mathrm{t}}(z + \lambda\tau + \mu) + 2\lambda\,{}^{\mathrm{t}}z + \lambda\tau\,{}^{\mathrm{t}}\lambda + \mathrm{trace}(\kappa)\big)\Big)\,\phi\big(\mathrm{g}^{\mathrm{J}}(\tau, \mathrm{z})\big).\end{aligned} \tag{7.4}$$

This is called the Jacobi slash action of weight $\sigma$ and index $m$. If $\sigma = \det^k$ for some $k \in \mathbb{Z}$, then we write $|^{\mathrm{J}}_{k,m}$ for the corresponding slash action.

**Weights.** Via the embeddings

$$\begin{aligned} \mathrm{GL}_n(\mathbb{C}) &\hookrightarrow \mathrm{Aff}_n(\mathbb{C}), & g &\longmapsto (g, 0), \quad \text{and} \\ \mathrm{Aff}_n(\mathbb{C}) &\hookrightarrow \mathrm{GL}_{n+1}(\mathbb{C}), & (g, h) &\longmapsto \big(\begin{smallmatrix} g & h \\ 0 & 1 \end{smallmatrix}\big) \end{aligned}$$

representations of $\mathrm{GL}_n(\mathbb{C})$ and $\mathrm{GL}_{n+1}(\mathbb{C})$ give rise to $\mathrm{Aff}_n(\mathbb{C})$-representations. We ambiguously use $\mathrm{std}^j$ and $\mathrm{sym}^j$ for (symmetric) powers of the standard representation of $\mathrm{GL}_n(\mathbb{C})$ for any $n$. When considering them as representations of $\mathrm{Aff}_n(\mathbb{C})$, we will add $n$ or $n+1$ as a subscript to distinguish them: $\mathrm{std}^j_n$ and $\mathrm{std}^j_{n+1}$ are then the representations of $\mathrm{Aff}_n(\mathbb{C})$ that arise from the first and second inclusion, respectively.

**§7.2 Differential operators for genus 1 Jacobi forms.** The first, physical, part of this paper only made use of genus 2 Siegel modular form. We have so far formulated the theory of Jacobi forms in the case of arbitrary genus $n$. For the rest of this section, we will focus on the case of genus 1 Jacobi forms (which arise from genus 2 Siegel modular forms). The main reason is that the explicit theory of differential operators has not yet been fully developed for all genera. It will be the theme of a separate paper.

We revisit differential operators, following closely the exposition in [CR10], but use different notation for the raising and lowering operators. To define differential operators, we use the following notation:

$$\partial_\tau := \tfrac{\partial}{\partial\tau} = \tfrac{1}{2}\big(\tfrac{\partial}{\partial x} - i\tfrac{\partial}{\partial y}\big), \qquad \partial_{\overline{\tau}} := \tfrac{\partial}{\partial\overline{\tau}} = \tfrac{1}{2}\big(\tfrac{\partial}{\partial x} + i\tfrac{\partial}{\partial y}\big),$$

$$\partial_z := \tfrac{\partial}{\partial z} = \tfrac{1}{2}\big(\tfrac{\partial}{\partial u} - i\tfrac{\partial}{\partial v}\big), \qquad \partial_{\overline{z}} := \tfrac{\partial}{\partial\overline{z}} = \tfrac{1}{2}\big(\tfrac{\partial}{\partial u} + i\tfrac{\partial}{\partial v}\big).$$

The heat operator, normalized as in [IK11], is

$$\mathbb{L}_m = \frac{4m}{2\pi i}\partial_\tau - \frac{1}{4\pi^2}\partial_z^2. \tag{7.5}$$

The raising operators and lowering operators with respect to Jacobi slash actions $|^{\mathrm{J}}_{k,m}$ are given by

$$\mathrm{R}^{\mathrm{J}}_{k,m} := 2i\big(\partial_\tau + \tfrac{v}{y}\partial_z + 2\pi i m\tfrac{v^2}{y^2}\big) + \tfrac{k}{y}, \qquad \mathrm{L}^{\mathrm{J}}_{k,m} := -2iy\big(y\partial_{\overline{\tau}} + v\partial_{\overline{z}}\big),$$

$$\mathrm{R}^{\mathrm{JH}}_{k,m} := i\partial_z - 4\pi m\tfrac{v}{y}, \qquad \mathrm{L}^{\mathrm{JH}}_{k,m} : -iy\partial_{\overline{z}}.$$

Bernd and Schmidt described their covariance properties in [BS98]. For $g \in \mathrm{G}^{\mathrm{J}}$ and $\phi \in C^\infty(\mathbb{H}^{\mathrm{J}})$, we have

$$\mathrm{L}^{\mathrm{J}}_{k,m}\big(\phi|_{k,m}\,g\big) = \big(\mathrm{L}^{\mathrm{J}}_{k,m}\phi\big)|_{k-2,m}\,g, \qquad \mathrm{R}^{\mathrm{J}}_{k,m}\big(\phi|_{k,m}\,g\big) = \big(\mathrm{R}^{\mathrm{J}}_{k,m}\phi\big)|_{k+2,m}\,g$$
$$\mathrm{L}^{\mathrm{JH}}_{k,m}\big(\phi|_{k,m}\,g\big) = \big(\mathrm{L}^{\mathrm{JH}}_{k,m}\phi\big)|_{k-1,m}\,g, \qquad \mathrm{R}^{\mathrm{JH}}_{k,m}\big(\phi|_{k,m}\,g\big) = \big(\mathrm{R}^{\mathrm{JH}}_{k,m}\phi\big)|_{k+1,m}\,g.$$

As usual, we suppress the subscripts $k$ and $m$. Note, however, that $(\mathrm{R}^{\mathrm{J}})^d = \mathrm{R}^{\mathrm{J}}_{k+2d-2,m}\cdots\mathrm{R}^{\mathrm{J}}_{k,m}$ and $(\mathrm{R}^{\mathrm{JH}})^{d^{\mathrm{H}}} = \mathrm{R}^{\mathrm{JH}}_{k+d^{\mathrm{H}}-1,m}\cdots\mathrm{R}^{\mathrm{JH}}_{k,m}$. The commutation rules for the raising and lowering operator are:

$$\big[\mathrm{L}^{\mathrm{J}},\mathrm{L}^{\mathrm{JH}}\big] = \big[\mathrm{R}^{\mathrm{J}},\mathrm{R}^{\mathrm{JH}}\big] = 0, \quad \big[\mathrm{L}^{\mathrm{J}},\mathrm{R}^{\mathrm{J}}\big] = -k, \quad \big[\mathrm{L}^{\mathrm{JH}},\mathrm{R}^{\mathrm{JH}}\big] = im, \quad \big[\mathrm{L}^{\mathrm{J}},\mathrm{R}^{\mathrm{JH}}\big] = -\mathrm{L}^{\mathrm{JH}}, \quad \big[\mathrm{L}^{\mathrm{JH}},\mathrm{R}^{\mathrm{J}}\big] = \mathrm{R}^{\mathrm{JH}}. \tag{7.6}$$

**§7.3 Almost holomorphic functions (genus 1).** Let $\mathscr{O}\big(\mathbb{H}^{\mathrm{J}}\big)$ be the space of holomorphic functions on $\mathbb{H}^{\mathrm{J}}$.

**Definition 7.2.** A smooth function $\phi : \mathbb{H}^{\mathrm{J}} \to \mathbb{C}$ is almost holmorphic of depth $(d, d^{\mathrm{H}})$ with $0 \le d, d^{\mathrm{H}} \in \mathbb{Z}$ if $\big(\mathrm{L}^{\mathrm{J}}\big)^d\,\phi = \big(\mathrm{L}^{\mathrm{JH}}\big)^{d^{\mathrm{H}}}\,\phi = 0$.

The analogues of Proposition 6.15 and Theorem 6.17 are as follows.

**Proposition 7.3.** *Suppose that $\phi : \mathbb{H}^{\mathrm{J}} \to \mathbb{C}$ and $\psi : \mathbb{H}^{\mathrm{J}} \to \mathbb{C}$ are almost holomorphic functions of depth $(d, d^{\mathrm{H}})$ and $(d', d'^{\mathrm{H}}$. Then the product $fg$ is almost holomorphic of depth $\big(d + d', d^{\mathrm{H}} + d'^{\mathrm{H}}\big)$.*

**Theorem 7.4.** *A function $\phi \in C^\infty\big(\mathbb{H}^{\mathrm{J}}\big)$ is almost holomorphic of depth $(d, d^{\mathrm{H}})$ if and only if it is a polynomial in $y^{-1}$ and $vy^{-1}$ of degree at most $d$ and $d^{\mathrm{H}}$, respectively, with holomorphic coefficients.*

*Proof.* This follows from the expressions for $\mathrm{L}^{\mathrm{J}}$ and $\mathrm{L}^{\mathrm{JH}}$, using induction. Indeed, up to non-zero multiplicative constants, $\mathrm{L}^{\mathrm{J}}\,y^{-a}(vy^{-1})^b$ equals $ay^{-(a-1)}(vy^{-1})^b$, and $\mathrm{L}^{\mathrm{JH}}\,y^{-a}(vy^{-1})^b$ is $by^{-a}(vy^{-1})^{b-1}$. ■

**§7.4 Almost holomorphic Jacobi forms (genus 1).** We define almost holomorphic Jacobi forms only for weight $\sigma = \det^k$, since in the general case, we have not studied the corresponding lowering operators.

**Definition 7.5.** An almost holomorphic function $\phi \in C^\infty(\mathbb{H}^{\mathrm{J}})$ is called an almost holomorphic Jacobi form of depth $(d, d^{\mathrm{H}})$, weight $k$, and index $m$ if it satisfies

(i) $\phi|^{\mathrm{J}}_{k,m}\gamma = \phi$ for all $\gamma \in \Gamma^{\mathrm{J}}$.

(ii) For all $\alpha, \beta \in \mathbb{Q}$, $\phi(\tau, \alpha\tau + \beta)$ is bounded by some power of $y$ as $\tau \to i\infty$.

We write $\mathrm{J}^{[d,d^{\mathrm{H}}]}_{k,m}$ for the space of such function.

The following proposition generalizes Lemma 6.3 in [BRR12]. It says that there is no analogue of the exceptional almost holomorphic elliptic modular form $E_2$.

**Proposition 7.6.** *For every $k, m \in \mathbb{Z}$ and every pair of positive integers $d$ and $d^{\mathrm{H}}$, we have*

$$\mathrm{J}^{[d,d^{\mathrm{H}}]}_{k,m} = \bigoplus_{t=0}^{d} \bigoplus_{t^{\mathrm{H}}=0}^{d^{\mathrm{H}}} (\mathrm{R}^{\mathrm{J}})^t (\mathrm{R}^{\mathrm{JH}})^{t^{\mathrm{H}}} \mathrm{J}_{k,m}.$$

*Proof.* It suffices to know that there is no Jacobi form of integral weight and index that is either constant in $\tau$ or $z$. Then the statement follows along the lines of the proof of Lemma 6.3 in [BRR12].∎

**§7.5 Covariant operators from vector valued weights (genus 1).** Recall that we focus on the case of genus $n = 1$. In this section, we reinterpret a result by Ibukiyama and Kyomura [IK11] in terms of covariant operators

$$\mathrm{C}^{\mathrm{J}}_{\det^k \mathrm{sym}^l_2} : C^\infty\big(\mathbb{H}^{\mathrm{J}} \to \det^k \mathrm{sym}^l_2\big) \longrightarrow C^\infty\Big(\mathbb{H}^{\mathrm{J}} \to \bigoplus_{j=0}^{l} V(\det^{k+j})\Big), \quad \phi \longmapsto \big(\iota_0(\phi), \ldots, \iota_l(\phi)\big), \tag{7.7}$$

where $\iota_j$ is defined below.

Fix a basis $\mathfrak{f}_j$ $(0 \le j \le l)$ for $V(\det^k \mathrm{sym}^l_2)$, where $\left(\begin{smallmatrix} a & 0 \\ 0 & 1 \end{smallmatrix}\right) \in \mathrm{G}^{\mathrm{J}}$ acts on $\mathfrak{f}_j$ by multiplication with $a^j$. Coordinates of a function $\phi : \mathbb{H}^{\mathrm{J}} \to V(\mathrm{sym}^l_2)$ with respect to this basis are denoted by $\langle \phi, \mathfrak{f}_j \rangle$. Ibukiyama and Kyomura defined the following maps $\iota_j$ on $C^\infty\big(\mathbb{H}^{\mathrm{J}} \to V(\mathrm{sym}^l_2)\big)$ on page 795 of [IK11]:

$$\iota_j(\phi) = \sum_{t=0}^{\lfloor j/2 \rfloor} \sum_{j'=0}^{j-2t} \binom{l-j'}{j-2t-j'} \binom{l-j+2t}{2t} \frac{(2k+2j-2t-5)!!\,(2t-1)!!}{(-2m)^{j-j'}\,(2k+2j-5)!!}\, \partial_z^{j-2t-j'}\, \mathbb{L}^t_m \langle \phi, \mathfrak{f}_{j'} \rangle, \tag{7.8}$$

where $n!! = (2n)!/(2^n n!)$. They prove in Theorem 2.1 that

$$\iota_v\big(\phi|_{\det^k \mathrm{sym}^l_2, m}\, g^{\mathrm{J}}\big) = \iota_v(\phi)|_{k+v,m}\, g^{\mathrm{J}}$$

for all $g^{\mathrm{J}} \in \mathrm{G}^{\mathrm{J}}$. This, in particular, implies covariance of $\mathrm{C}_{\det^k \mathrm{sym}^l}$.

**§7.6 Fourier Jacobi coefficients.** In this section, we briefly return to the case of general genus $n$. We attach Fourier Jacobi coefficients to every almost holomorphic Siegel modular form.

Decompose $\tau$ as

$$\begin{pmatrix} \tau_1 & {}^{\mathrm{t}}z \\ z & \tau_2 \end{pmatrix}, \quad \tau_1 \in \mathbb{H}^{(n-1)}, \tau_2 \in \mathbb{H}, \quad \text{and} \quad z \in \mathbb{C}^{n-1}$$

Write $y_1$ and $y_2$ for the imaginary part of $\tau_1$ and $\tau_2$. The ambiguous notation $y_1^{-1}$ will be used to denote $(y^{-1})_1$. Every almost holomorphic Siegel modular form has a weak Fourier Jacobi expansion

$$f(\tau) = \sum_{m \in \mathbb{Q}} \widetilde{\phi}_m(\tau_1, z, y_2)\, \exp(2\pi i\, m \tau_2). \tag{7.9}$$

We define Fourier Jacobi coefficients of $f$ by

$$\phi_m(\tau_1, z) = \lim_{y_2 \to \infty} \widetilde{\phi}_m(\tau_1, z, y_2), \tag{7.10}$$

which is well-defined, since $y^{-1} \to \begin{pmatrix} y_1^{-1} & 0 \\ 0 & 0 \end{pmatrix}$ as $y_2 \to \infty$. If $f$ has trivial type and weight $\sigma$, then $m \in \mathbb{Z}$ if $\widetilde{\phi}_m \neq 0$, and for every $y_2$, $\widetilde{\phi}_m(\tau_1, z, y_2)$ is invariant under the $|^{\mathrm{J}}_{\sigma,m}$ action of $\Gamma^{(\mathrm{n}-1)\mathrm{J}}$.

*Remark 7.7.* Observe that $\det y^{-1} \to 0$ as $y_2 \to \infty$. In particular, it is possible to find subexpressions of, for example, $\mathrm{R}^n f$ for a Siegel modular form $f$, whose term of highest depth vanishes when passing to the Fourier Jacobi expansion.

# 8 | Classification of Almost Holomorphic Siegel Modular Forms

The graded ring of elliptic modular forms is defined as

$$\mathrm{M}(\bullet) = \mathrm{M}^{(1)}(\bullet) = \bigoplus_{0 \le k \in \mathbb{Z}} \mathrm{M}^{(1)}\big(\det{}^k \boxtimes \mathbb{1}\big).$$

A minimal set of (algebraically independent) generators is given by the Eisenstein series $E_4$ and $E_6$, where

$$E_k(\tau) = 1 - \frac{2k}{B_k} \sum_{m=1}^{\infty} \sigma_{k-1}(m) \exp(2\pi i\, m\tau) \qquad k \ge 4.$$

When extending this to a differential graded algebra with derivative R acting as $\mathrm{R}_k = \partial_\tau - \frac{ki}{2} y^{-2}$ on the $k$-th graded piece, we obtain

$$\mathrm{R}^\infty \mathrm{M}(\bullet) = \bigoplus_{d=0}^{\infty} \mathrm{R}^d \mathrm{M}(\bullet) \subset \bigcup_{0 \le d \in \mathbb{Z}} \bigoplus_{0 \le k \in \mathbb{Z}} \mathrm{M}^{(1)}\big(\det{}^{k[d]} \boxtimes \mathbb{1}\big).$$

This carries an action of the lowering operator L, which is not surjective. Instead, we have

$$\operatorname*{coker}_{\mathrm{R}^\infty \mathrm{M}(\bullet)} \mathrm{L} = \mathrm{R}^\infty \mathrm{M}(\bullet) \,/\, \mathrm{L}\big(\mathrm{R}^\infty \mathrm{M}(\bullet)\big) = \operatorname{span} 1. \tag{8.1}$$

The (modular) weight 2 Eisenstein series is

$$E_2(\tau) = \frac{-3}{\pi} y^{-1} + 1 - 24 \sum_{n=1}^{\infty} \sigma_1(n) \exp(2\pi i\, n\tau) \in \mathrm{M}^{(1)}\big(\det{}^{2[1]}\big). \tag{8.2}$$

While $E_2$ is not in the range of R, the Ramanujan relation

$$\tfrac{1}{2\pi i} \mathrm{R}\, E_2 = \tfrac{1}{2\pi i} \mathrm{R}_{\det^2} E_2 = \tfrac{1}{12}\big(E_2^2 - E_4\big) \tag{8.3}$$

implies that

$$\widehat{\mathrm{M}}^{(1)}(\bullet) = \bigcup_{0 \le d} \bigoplus_{0 \le k \in \mathbb{Z}} \mathrm{M}^{(1)}\big(\det{}^{k[d]}\big) = \bigoplus_{d=0}^{\infty} \mathrm{R}^d \big(\mathbb{C}[E_4, E_6] + E_2 \mathbb{C}[E_4, E_6]\big) = \mathbb{C}\big[E_2, E_4, E_6\big]$$

is a differential algebra. Since $\mathrm{L} E_2^d = \frac{3d}{2\pi i} E_2^{d-1}$ we have an exact sequence

$$0 \longrightarrow \mathrm{M}^{(1)}(\bullet) \longrightarrow \widehat{\mathrm{M}}^{(1)}(\bullet) \xrightarrow{\ \mathrm{L}\ } \widehat{\mathrm{M}}^{(1)}(\bullet) \longrightarrow 0. \tag{8.4}$$

The goal of this section is to show that the situation is more complicated for higher genera. We focus on the case $n = 2$, and find that

$$\operatorname*{coker}_{\widehat{\mathrm{M}}^{(2)}} \mathrm{L} = \widehat{\mathrm{M}}^{(2)} \,/\, \mathrm{L}\widehat{\mathrm{M}}^{(2)} = \operatorname{span} 1.$$

In particular, we prove that there is no analogue of $E_2$ in genus 2. In Section 9, we will then provide a replacement for it.

**§8.1 The non-commutative algebra of Siegel modular forms.** It is common to consider the graded ring of Siegel modular forms, summing over all spaces of scalar valued Siegel modular forms like so

$$\bigoplus_{k \in \mathbb{Z}} \mathrm{M}^{(n)}\big(\det{}^k\big). \tag{8.5}$$

Note that the sum runs over all isomorphism classes of complex 1-dimensional representations of $\mathrm{GL}_n(\mathbb{C})$. The above is, indeed, a ring: The tensor product of two Siegel modular forms of weight $k$ and $k'$ is a modular form of weight $\det^k \otimes \det^{k'}$, which is isomorphic to $\det^{k+k'}$. Further, it is possible to consistently choose bases of $V(\det^k)$ for all $k$, which turn all the isomorphisms $\det^k \otimes \det^{k'} \cong \det^{k+k'}$ into isomorphisms of framed representations. One simply chooses the unit vector $\mathfrak{e}^k \in V(\det^k)$ as a basis, and uses isomorphisms $\mathfrak{e}^k \otimes \mathfrak{e}^{k'} \mapsto \mathfrak{e}^{k+k'}$.

What seems like excess of precision in the case of scalar valued Siegel modular forms, becomes important in the vector valued setting. Given two representations $\sigma$ and $\sigma'$ of $\mathrm{GL}_n(\mathbb{C})$ the tensor product $\sigma \otimes \sigma'$ in general decomposes into many irreducibles with multiplicities possibly larger than 1. Also, it is not trivial to consistently choose bases of all irreducibles that behave well with respect to tensor products. A similar problem arises from types of Siegel modular forms.

There are two approaches to accommodate this problem, neither of which seems to have appeared in the literature. It is possible to sum over tensor powers of the standard representation. One would consider

$$\bigoplus_{0 \le l \in \mathbb{Z}} \mathrm{M}^{(n)}\big(\mathrm{std}^l\big) \tag{8.6}$$

By a result of Weissauer [Wei83], every irreducible representation $\sigma$ for which $\mathrm{M}^{(n)}(\sigma \boxtimes \rho)$ is not trivial for some $\rho$ is contained in some tensor power of $\mathrm{std}^l$, so that (8.6) contains all Siegel modular forms of level 1. More specifically, $\det^k \subseteq \mathrm{std}^{nk}$. In particular, the non-commutative graded ring (8.6) of Siegel modular forms contains the classical one (8.5).

The type of Siegel modular forms, if $n \ge 2$, can be expressed in terms of Weyl representations $\rho_{\mathscr{L}}$ attached to (positive definite and integral) lattices $\mathscr{L}$. ONe can thus appy a similar construction, to obtain a non-commutative ring that contains all Siegel modular forms of arbitrary weights and types. We arbitrary will not go into details, though.

The second approach, which seems most suitable to study the analogue of (8.1), is to sacrifice the algebra structure in favor of a grading with respect to irreducible representations of $\mathrm{GL}_n(\mathbb{C})$ and $\Gamma^{(n)}$. The direct sum

$$\mathrm{M}^{(n)}\big(\bullet \boxtimes \bullet\big) = \bigoplus \mathrm{M}^{(n)}\big(\sigma \boxtimes \rho\big), \tag{8.7}$$

where $\sigma$ and $\rho$ run through fixed representatives of isomorphism classes of representations, is a commutative hyper-algebra.

**§8.2 Hyper-algebras.** A hyper-group is group whose multiplication is multi-valued [Wal37]. We take the idea of multi-valued operators to define hyper-algebras, which we could not find in the literature.

Let $\mathrm{SubVec}(A)$ be the set of all finite rank submodules of a vector space $A$.

**Definition 8.1.** A triple $(A, +, \cdot)$ with $(A, +)$ a $K$-vector space and a binary operator

$$\cdot : A \times A \longrightarrow \mathrm{SubVec}(A)$$

is called a hyper-algebra (with strong identity) if

(i) there is a unique element $e \in A$ such that $e \cdot r = r \cdot e = \mathrm{span}\, r$ for all $r \in A$, and for every $a \in A$ there is $a'$ such that $e \in aa', a'a$.

(ii) for $k \in K \setminus \{0\}$ and $a, a' \in A$, we have $aa' = (ka)a' = a(ka')$.

(iii) for $a, a', a'' \in A$, we have $a(a' + a'') \subseteq aa' + aa''$.

(iv) for $a, a', a'' \in A$, we have $a(a'a'') = (aa')a''$, where the product is linearly extended to $\mathrm{SubVec}(A) \times \mathrm{SubVec}(A)$.

Further, we say that

(1) $A$ is commutative if for $a, a' \in A$, we have $aa' = a'a$.

(2) $A$ is $G$-graded for a hyper-group $G$ if $A = \bigoplus_g A_g$ and $A_g A_{g'} \subseteq \sum_{h \in gg'} A_h$.

(3) $A$ is a differential hyper-algebra, if there is a $K$-linear map

$$\partial :\, A \longrightarrow \mathrm{SubVec}(A)$$

satisfying the Leibniz rule $\partial(aa') \subseteq (\partial a)a' + a(\partial a')$ for all $a, a' \in A$. If $A$ is graded by $G$, then we say that $\partial$ has grading $g \in G$, if $\partial A_{g'} \subseteq \sum_{h \in gg'} A_h$ for all $g' \in G$.

*Remark 8.2.* The notion of hyper-groups makes it obvious to call the structure that we have defined a hyper-algebra. But the reader should be warned that this is not standard. In fact, it has nothing in common with hyper-algebras associated with groups in, for example, [Sul78].

**§8.3 The differential graded hyper-algebra of almost holomorphic Siegel modular forms.** Recall that isomorphism classes of irreducible, finite-dimensional, complex representations of $\mathrm{GL}_n(\mathbb{C})$ form a hyper-group $\mathrm{Rep}(\mathrm{GL}_n(\mathbb{C}))$ with multiplication given by the decomposition of tensor products. Typically, this is formulated in terms of fusion algebras (see [Eho95] for one instance) or Grothendieck rings, but Definition 8.1 suggests that it is more convenient to work with hyper-groups. Similarly, irreducible, finite-dimensional representations of $\Gamma^{(n)}$ with finite index kernel, yield a hyper-group $\mathrm{Rep}(\Gamma^{(n)})$. By fixing representatives once and for all, we consider elements of $\mathrm{Rep}(\mathrm{GL}_n(\mathbb{C}))$ and $\mathrm{Rep}(\Gamma^{(n)})$ as actual representations.

Set

$$\widehat{\mathrm{M}}^{(n)} = \bigcup_{0 \le d \in \mathbb{Z}} \bigoplus_{\substack{\sigma \in \mathrm{Rep}(\mathrm{GL}_n(\mathbb{C})) \\ \rho \in \mathrm{Rep}(\Gamma^{(n)})}} \mathrm{M}^{(n)}\big(\sigma^{[d]} \boxtimes \rho\big). \tag{8.8}$$

For $f \in \mathrm{M}^{(n)}\big(\sigma^{[d]} \boxtimes \rho\big)$ and $g \in \mathrm{M}^{(n)}\big(\sigma'^{[d']} \boxtimes \rho'\big)$, set

$$f \otimes g = \operatorname*{span}_{\substack{\kappa_\sigma : \sigma \otimes \sigma' \to \sigma_{\mathrm{irr}} \\ \kappa_\rho : \rho \otimes \rho' \to \rho_{\mathrm{irr}}}} \big(\kappa_\sigma \boxtimes \kappa_\rho\big)\big(f \otimes g\big), \tag{8.9}$$

where the span runs through all homomorphism representations with irreducible $\sigma_{\mathrm{irr}}$ and $\rho_{\mathrm{irr}}$. The $\mathbb{C}$-module $\widehat{\mathrm{M}}^{(n)}$ carries a graded commutative hyper-$\mathbb{C}$-algebra structure when equipped with multiplication as in (8.9).

The raising operators $\mathrm{R}_{\mathrm{std}^l}$ can be extended to $\widehat{\mathrm{M}}^{(n)}$. Given $\sigma$ choose an embedding $\iota : \sigma \hookrightarrow \mathrm{std}^l$ for some $0 \le l \in \mathbb{Z}$. Then

$$\mathrm{R} f = \operatorname*{span}_{\kappa : \mathrm{sym}^2 \otimes \mathrm{std}^l \to \sigma_{\mathrm{irr}}} \kappa\big(\mathrm{R}_{\mathrm{std}^l} f\big). \tag{8.10}$$

When equipped with the differential R, then $\widehat{\mathrm{M}}^{(n)}$ becomes a differential graded hyper-algebra, and R has grading $\mathrm{sym}^2$. Similarly, we obtain an action of L on $\widehat{\mathrm{M}}^{(n)}$.

**§8.4 Exception Siegel modular forms.** We finish the classification of almost holomorphic Siegel modular forms of genus 2 that was initiated by Pitale, Saha, and Schmidt in [PSS15]. The goal of this section is to prove the following theorem.

**Theorem 8.3.** *Given $\sigma$, $\rho$, and $0 \le d$, we have*

$$\mathrm{M}^{(2)}\big(\sigma^{[d]} \boxtimes \rho\big) = \sum_{t=0}^{d} \big(\pi_{\mathrm{sym}^2}\mathrm{R}\big)^t\, \mathrm{M}^{(n)}\big((\mathrm{sym}^{\vee 2})^t \sigma \boxtimes \rho\big).$$

*That is, Siegel modular forms generate $\widehat{\mathrm{M}}^{(2)}$ as a differential hyper-algebra.*

*Proof.* This was almost proved in [PSS15], except for the cases $\sigma = \det^3 \mathrm{sym}^l$ for some $0 \le l \in \mathbb{Z}$. As noted there, it suffices to show that $\dim \mathrm{M}^{(2)}\big(\mathrm{detsym}^l \boxtimes \rho\big) = 0$ for all $l$ and all $\rho$ to rule out the remaining cases.

Suppose that $f \in \mathrm{M}^{(2)}\big(\mathrm{detsym}^l \boxtimes \rho\big)$ is a non-zero Siegel modular form. We consider its Fourier Jacobi expansion

$$f(\tau) = \sum_{0 \le m \in \mathbb{Z}} \phi_m(\tau_1, z)\, e\big(m\tau_2\big).$$

Applying the operator $\mathrm{C}^{\mathrm{J}}_{\mathrm{detsym}^l_2}$ and the map $\langle \cdot, \mathfrak{f}_l \rangle$ studied in Section 7.7, we find that

$$\langle \phi_m\big(\tau_1, z\big), \mathfrak{f}_l \rangle$$

is a classical Jacobi form of weight 1 and index $m$. By results of Ibukiyama and Skoruppa it vanishes [IS07]. Applying the next Lemma 8.4 yields the result. ∎

**Lemma 8.4.** *Let $f \in \mathrm{M}^{(n)}\big(\sigma \boxtimes \rho\big)$ and $v \in \sigma^\vee$. If $v \circ f = 0$ then $(\mathrm{GL}_n(\mathbb{C})\, v) \circ f = 0$.*

*Proof.* This follows from invariance of $f$ under the action of the Levi factor. For $u \in \mathrm{GL}_n(\mathbb{Z})$, we have

$$(uv)\, f(\tau) = (uv)\, u f\big({}^{\mathrm{t}}u \tau u\big) = v\, f\big({}^{\mathrm{t}}u^{-1} \tau u^{-1}\big) = 0\,.$$

The $uv$ for $u \in \mathrm{GL}_n(\mathbb{Z})$ span $\mathrm{GL}_n(\mathbb{C})\, v \subseteq V(\sigma^\vee)$, and this implies the lemma. ∎

The classification by Pitale, Saha, and Schmidt, which appeared shortly before the present manuscript was finished, shows that there is no modular level 1 Siegel modular form of weight $\mathrm{sym}^2$ if $n = 2$. We extend this to all genera by applying a variant of the argument in the proof of Theorem 8.3.

**Proposition 8.5.** *If $n \ge 2$, then for any $\rho$, we have*

$$\mathrm{M}^{(n)}\big(\mathrm{sym}^{2\,[1]} \boxtimes \mathbb{1}\big) = \mathrm{M}^{(n)}\big(\mathrm{sym}^2 \boxtimes \mathbb{1}\big).$$

*Proof.* We can and will assume that $\rho$ is irreducible. We claim that every $f \in \mathrm{M}^{(n)}\big(\mathrm{sym}^{2[1]}\big)$ is of the form $f(\tau) = g(\tau) + c y^{-1}$ for some constant $c \in \mathbb{C}$ and holomorphic $g$. Indeed, assume that $f(\tau) = g(\tau) + h(y^{-1})$ for a holomorphic function $g$ and a linear function $h \in \mathrm{sym}^2 \otimes \mathrm{sym}^{\vee 2}$ without constant term. A direct computation shows that the image of $f$ under L equals $-\mathrm{t}_{12}\, h$. It is a modular form of weight $\mathrm{sym}^{\vee 2}\mathrm{sym}^2$, which by Lemma 8.7 is supported on the span of $\sum_{ij} \mathfrak{e}^\vee_{ij} \frac{1}{2}(\mathfrak{e}_{ij} + \mathfrak{e}_{ji})$. This proves the claim.

It suffices to show that there is no $f^{(n)} \in \mathrm{M}^{(n)}\big(\mathrm{sym}^{2[1]}\big)$ of the form $f^{(n)}(\tau) = g^{(n)}(\tau) + y^{-1}$. The 0-th Fourier Jacobi coefficient of $f^{(n)}$, defined in (7.10), is a Siegel modular form of genus $n-1$. Since $\dim \mathrm{M}^{(n-1)}(\mathrm{sym}^2) = 0$, we find that it equals $f^{(n-1)}$. We are therefore reduced to the case $n = 2$.

Consider the case $n = 2$. We employ the Fourier Jacobi expansion. As a next step we show that vanishing of the 0-th Fourier Jacobi coefficient of $f^{(2)}$ contradicts its existence. Assuming that $f^{(2)}$ exists, let $m$ be minimal subject to the condition that the $m$-th Fourier Jacobi coefficient, say $\phi_m$ of $f^{(2)}$ is non-zero. Combining (7.10) and (7.7), we see that $C^{\mathrm{J}}_{\mathrm{sym}^2}\,\phi_m$ has Fourier expansion starting at $m$:

$$\sum_{m\le n;\,r} c(n,r)\exp\big(2\pi i(n\tau_1 + rz)\big).$$

Its components are elements of $\mathrm{J}^{[0,0]}_{0,m}$, $\mathrm{J}^{[0,0]}_{1,m}$, and $\mathrm{J}^{[1,0]}_{2,m}$. Using the same argument as in [BWR14], these correspond via the maps $\mathscr{D}_{2\nu}$ in (9) of [EZ85] to almost modular forms modular forms of weight at most $2+2m$, whose Fourier expansions starts at $m$. If $m \ge 1$, they are zero by the theory of almost holomorphic elliptic modular forms. This implies that $\phi_0 \ne 0$, if $f^{(2)}$ exists.

It suffices to show that the 0-th Fourier Jacobi coefficient of $f^{(2)}$ would have to vanish, contradicting its existence. The Fourier Jacobi expansions of $f^{(2)}$ starts like

$$f^{(2)}(\tau) = \begin{pmatrix}\phi_{0,3}(\tau_1,z) & \phi_{0,2}(\tau_1,z)\\ \phi_{0,2}(\tau_1,z) & \phi_{0,1}(\tau_1,z)\end{pmatrix} + y^{-1} + \begin{pmatrix}\phi_{1,3}(\tau_1,z) & \phi_{1,2}(\tau_1,z)\\ \phi_{1,2}(\tau_1,z) & \phi_{1,1}(\tau_1,z)\end{pmatrix} e(\tau_2) + O\big(e(2\tau_2)\big),$$

where $\tau = \left(\begin{smallmatrix}\tau_1 & z\\ z & \tau_2\end{smallmatrix}\right)$ as in Section 7. The covariant operator $\mathrm{C}^{\mathrm{J}}_{\mathrm{sym}^2}$ in (7.7) yields Jacobi forms in

$$\mathrm{J}_{0,0} = \{0\},\quad \mathrm{J}_{0,1} = \{0\},\quad \mathrm{J}^{[1,0]}_{0,2} = \mathrm{span}(E_2),\quad \mathrm{J}_{1,0} = \{0\},\quad \mathrm{J}_{1,1} = \{0\},\quad \mathrm{J}^{[1,0]}_{1,2} = \{0\}.$$

This implies that

$$\phi_{0,1} = c,\quad \phi_{0,2} = 0,\quad \phi_{0,3}(\tau_1,z) + y_1^{-1} = \tfrac{-\pi}{3}E_2(\tau_1),\quad \phi_{1,1} = 0,\quad \phi_{1,2} = 0,\quad \phi_{1,3} = 0$$

for some constant $c \in \mathbb{C}$. The symmetry relations $c(f;\,t,y) = {}^{\mathrm{t}}u c(f;\,{}^{\mathrm{t}}utu, y)u$ for $u \in \mathrm{GL}_2(\mathbb{Z})$, when applied to $t = \left(\begin{smallmatrix}1&0\\0&0\end{smallmatrix}\right)$ and $u = \left(\begin{smallmatrix}0&1\\1&0\end{smallmatrix}\right)$ contradicts the given equality for $\phi_{0,3}$. This establishes the statement. ■

*Remark 8.6.* In the case $n \ge 4$, one could prove Proposition 8.5 by combining [Wei87] and [Wei86].

**Lemma 8.7.** *Suppose that $\sigma$ and $\rho$ are irreducible. If the $\mathfrak{k}_{\mathbb{C}}$-eigenvalue of $\sigma$ is negative, or if it is zero and $\sigma \ne \mathbb{1}$ or $\rho \ne \mathbb{1}$, then* $\dim \mathrm{M}^{(n)}\big(\sigma \boxtimes \rho\big) = 0$.

*Proof.* This is a consequence of Theorem 2 in [Wei83]. Indeed, for negative $\mathfrak{k}_{\mathbb{C}}$-eigenvalue, we obtain the statement directly. If it vanishes, than the corank of $\sigma$ in the sense defined in [Wei83] equals $n$, so that $\sigma$ is a power of det, implying $\sigma = \mathbb{1}$. Then $\rho = \mathbb{1}$ is a consequence of the fact $\rho$ is irreducibe, and hence its invariant subspaces are all trivial. ■

# 9 | A meromorphic replacement for $E_2$

Throughout this section, we focus on the case $n = 2$. Physical considerations naturally gave rise to the logarithmic derivative $\partial_\tau \log(\phi^{(2)}_{10})$ of $\phi^{(2)}_{10} = 4\chi_{10}$ for Igusa's $\chi_{10} \in \mathrm{M}^{(2)}\big(\det^{10}\big)$, defined on page 195 of [Igu62]. The goal of this section is to describe $\partial_\tau \log(\chi_{10})$ as a meromorphic replacement of the constant part of the weight 2 Eisenstein series $E_2$ in the genus 1 case. In particular, we will find a ring-like structure equipped with the (vector valued) differential $\partial_\tau$ acting on it.

**§9.1 An almost meromorphic Siegel modular form.** In (4.13), we find the following definition:

$$S(\tau) = S^{(2)}(\tau) = \tfrac{1}{10}\partial_\tau \log(\phi^{(2)}_{10}) - \tfrac{i}{2}y^{-1}, \tag{9.1}$$

where $\partial_\tau$ is the vector valued derivative defined in Section 6.2. We can conveniently express $S$ as

$$S(\tau) = \tfrac{1}{10}\phi^{(2)\,-1}_{10}\,\mathrm{R}_{\det^{10}}\,\phi^{(2)}_{10} = \tfrac{1}{10}\phi^{(2)\,-1}_{10}\big(\partial_\tau - 5iy^{-1}\big)\phi^{(2)}_{10}, \tag{9.2}$$

from which modularity of $S$ becomes immediately clear.

In analogy with Definition 6.13, we define almost meromorphic functions for complex representations $\sigma$ of $\mathrm{GL}_n(\mathbb{C})$, subsuming $S$.

**Definition 9.1.** Let $D \subset \mathbb{H}^{(n)}$ be a smooth, locally closed submanifold of codimension 1. A smooth function $f : \mathbb{H}^{(n)} \setminus D \to V(\sigma)$ is almost meromorphic of depth $d$ if $\mathrm{L}^{d+1} f = 0$.

It is immediately clear that Theorem 6.17 generalizes. That is, almost meromorphic functions are polynomials in the entries of $y^{-1}$ whose coefficients are meromorphic functions. The constant term of this polynomial is called its constant part.

Let us summarize immediate consequences of Equation (9.2).

**Proposition 9.2.** *The function $S$ defined in (9.1) is an almost meromorphic function, whose singularities lie on the submanifolds $z = 0$ and its $\Gamma^{(2)}$ translates. For any $\gamma \in \Gamma^{(2)}$ we have $S|_{\mathrm{sym}^2}\,\gamma = S$.*

The Ramanujan relation (8.3) is generalized by the next proposition. Before stating it, recall the averaged transposition $\mathrm{t} = \mathrm{t}_{(12)(34)}$ defined in (6.4). When applying it to $\mathrm{sym}^2\mathrm{sym}^2$, as below, it has the form

$$\mathrm{t}(\mathfrak{e}_{11}\mathfrak{e}_{11}) = \mathfrak{e}_{11}\mathfrak{e}_{11}\,,\, \mathrm{t}(\mathfrak{e}_{11}\mathfrak{e}_{12}) = \tfrac{1}{2}\mathfrak{e}_{11}\widetilde{\mathfrak{e}}_{12} + \tfrac{1}{2}\widetilde{\mathfrak{e}}_{12}\mathfrak{e}_{11}\,,\, \mathrm{t}(\mathfrak{e}_{11}\mathfrak{e}_{22}) = \tfrac{1}{4}\widetilde{\mathfrak{e}}_{12}\widetilde{\mathfrak{e}}_{12}\,,$$
$$\mathrm{t}(\widetilde{\mathfrak{e}}_{12}\widetilde{\mathfrak{e}}_{12}) = \mathfrak{e}_{11}\mathfrak{e}_{22} + \mathfrak{e}_{22}\mathfrak{e}_{11} + \tfrac{1}{2}\widetilde{\mathfrak{e}}_{12}\widetilde{\mathfrak{e}}_{12}\,, \qquad \text{where} \quad \widetilde{\mathfrak{e}}_{ij} = \frac{\mathfrak{e}_{ij} + \mathfrak{e}_{ji}}{1 + \delta_{ij}}.$$

**Proposition 9.3.** *We have*

$$\mathrm{R}_{\mathrm{sym}^2}\, S = \mathrm{t}(S \otimes S) \,+\, f_{\mathrm{RS}}(\tau), \tag{9.3}$$

*where $f_{\mathrm{RS}}$ is a meromorphic Siegel modular forms of weight $\mathrm{sym}^2 \otimes \mathrm{sym}^2$, which has a pole of order 2 at $z = 0$; that is, $\phi^{(2)}_{10} f_{\mathrm{RS}}$ is holomorphic.*

*Proof.* We prove this by first comparing non-constant parts and then checking Fourier coefficients. To summarize, we have

$$100\, S \otimes S = \phi^{(2)\,-2}_{10}\, \partial_\tau \phi^{(2)}_{10} \otimes \partial_\tau \phi^{(2)}_{10} \,-\, 5i \phi^{(2)\,-1}_{10}\, \partial_\tau \phi^{(2)}_{10} \otimes y^{-1} \,-\, 5i \phi^{(2)\,-1}_{10}\, y^{-1} \otimes \partial_\tau \phi^{(2)}_{10} \,-\, 25\, y^{-1} \otimes y^{-1}\,,$$
$$\begin{aligned} 10\,\mathrm{R}_{\mathrm{sym}^2}\, S = {} & -\phi^{(2)\,-2}_{10} \big(\partial_\tau \phi^{(2)}_{10} \otimes \partial_\tau \phi^{(2)}_{10}\big) \,+\, \phi^{(2)\,-1}_{10} \big(\partial_\tau \otimes \partial_\tau\big) \phi^{(2)}_{10} \,-\, 5i \tfrac{-i}{2}\, \partial_y y^{-1} \\ & -\, i \phi^{(2)\,-1}_{10}\, \mathrm{t}_{(12)(34)} \big(y^{-1} \otimes \partial_\tau \phi^{(2)}_{10}\big) \,-\, 5\, \mathrm{t}_{(12)(34)} \big(y^{-1} \otimes y^{-1}\big). \end{aligned}$$

Set $\Delta = \sum_{ij} \mathfrak{e}_{ij} \frac{1}{2} (\mathfrak{e}_{ij} + \mathfrak{e}_{ji})$. We have $\partial_y y^{-1} = -(y^{-1} \otimes \mathbb{1}_2) \Delta (y^{-1} \otimes \mathbb{1}_2)$, and a straightforward computation reveals that $\mathrm{t}_{(12)(34)} (y^{-1} \otimes y^{-1}) = (y^{-1} \otimes \mathbb{1}_2) \Delta (y^{-1} \otimes \mathbb{1}_2)$. It is obvious from the definition of $\mathrm{t}$ that $\mathrm{t}_{(12)(34)} (y^{-1} \otimes \partial_\tau \phi^{(2)}_{10}) = \mathrm{t}_{(12)(34)} (\partial_\tau \phi^{(2)}_{10} \otimes y^{-1})$. Taking these equalities, we find that $\mathrm{R}\, S - \mathrm{t}(S \otimes S)$ is meromorphic.

To finish the proof consider the holomorphic modular forms

$$100 \phi^{(2)\,2}_{10} \big(\mathrm{R}_{\mathrm{sym}^2}\, S - \mathrm{t}(S \otimes S)\big) = -\,10 \big(\partial_\tau \phi^{(2)}_{10} \otimes \partial_\tau \phi^{(2)}_{10}\big) \,+\, 10 \phi^{(2)}_{10} \big(\partial_\tau \otimes \partial_\tau\big) \phi^{(2)}_{10} \,-\, \mathrm{t}_{(12)(34)} \big(\partial_\tau \phi^{(2)}_{10} \otimes \partial_\tau \phi^{(2)}_{10}\big). \tag{9.4}$$

We compute its expansion around $z = 0$. Let $\Delta_{12} = \Delta(\tau_1) \Delta(\tau_2)$ be the product of the modular discriminant depending on $\tau_1$ and $\tau_2$. Then $\frac{100}{(2\pi i)^4} \phi^{(2)\,2}_{10} \big(\mathrm{R}_{\mathrm{sym}^2}\, S - \mathrm{t}(S \otimes S)\big)$ has expansion

$$\begin{aligned} & -\,10 \big(\widetilde{\mathfrak{e}}_{12} z \Delta_{12} + \mathscr{O}(z^3)\big) \big(\widetilde{\mathfrak{e}}_{12} z \Delta_{12} + \mathscr{O}(z^3)\big) - \mathrm{t}\Big(\big(\widetilde{\mathfrak{e}}_{12} z \Delta_{12} + \mathscr{O}(z^3)\big) \big(\widetilde{\mathfrak{e}}_{12} z \Delta_{12} + \mathscr{O}(z^3)\big)\Big) \\ & +\, 10 \big(z^2 \Delta_{12} + \mathscr{O}(z^4)\big) \Big(\mathfrak{e}_{11} \widetilde{\mathfrak{e}}_{12} z \partial_{\tau_1} \Delta_{12} + \mathfrak{e}_{22} \widetilde{\mathfrak{e}}_{12} z \partial_{\tau_2} \Delta_{12} + \widetilde{\mathfrak{e}}_{12} \mathfrak{e}_{11} z \partial_{\tau_1} \Delta_{12} \\ & \qquad\qquad + \widetilde{\mathfrak{e}}_{12} \mathfrak{e}_{22} z \partial_{\tau_2} \Delta_{12} + \widetilde{\mathfrak{e}}_{12} \widetilde{\mathfrak{e}}_{12} \tfrac{1}{2} \Delta_{12} + \mathscr{O}(z^3)\Big). \end{aligned}$$

The leading term is therefore

$$z^2 \Delta^2_{12} \big(\tfrac{-11}{2} \widetilde{\mathfrak{e}}_{12} \widetilde{\mathfrak{e}}_{12} - \tfrac{1}{4} \mathfrak{e}_{11} \mathfrak{e}_{22} - \tfrac{1}{4} \mathfrak{e}_{22} \mathfrak{e}_{11}\big) + 10 z^3 \Delta_{12} \big(\partial_{\tau_1} \Delta_{12} (\mathfrak{e}_{11} \widetilde{\mathfrak{e}}_{12} + \widetilde{\mathfrak{e}}_{12} \mathfrak{e}_{11}) + \partial_{\tau_2} \Delta_{12} (\mathfrak{e}_{22} \widetilde{\mathfrak{e}}_{12} + \widetilde{\mathfrak{e}}_{12} \mathfrak{e}_{22}\big),$$

which, in particular, implies that $\phi^{(2)}_{10} f$ is holomorphic. ∎

**§9.2 Fourier coefficients of $f_{\mathrm{RS}}$.** Our goal is to compute Fourier coefficients of $f_{\mathrm{RS}}$ defined in (9.3). We use an Ansatz to find an expression for $\phi^{(2)}_{10}\, f_{\mathrm{RS}}$ in terms of usual generators. This expression is displayed in Equation (9.5). We obtain it by first decomposing $\det^{10}\mathrm{sym}^2\mathrm{sym}^2$ explicitly, finding a basis for the corresponding spaces of Siegel modular forms, and then comparing coefficients with (9.4). Multiplication with $\phi^{(2)\,-1}_{10}$ then yields Fourier coefficients of $f_{\mathrm{RS}}$ (in a prescribed Weyl chamber).

In this Section we will frequently refer to Sage [Ste+14] scripts. They can all be found on the last named author's homepage. Note that several of them make us of the implementation of Jacobi forms [Rau12] that is currently only available at Sage-Trac, i.e., as a branch u/mraum/ticket/16448 at Sage-Git.

We will use polynomials $p$ in two variables $v_1$ and $v_2$ with the action $(gp)(v_1\ v_2) = p\big((v_1\ v_2)g\big)$ as a basis for symmetric power representations of $\mathrm{SL}_2(\mathbb{C})$. An isomorphism to our previous basis $\mathfrak{e}_{11}$, $\mathfrak{e}_{12}+\mathfrak{e}_{21}$, $\mathfrak{e}_{22}$ of $\mathrm{sym}^2$ is given by

$$\mathfrak{e}_{11} \mapsto v_1^2, \quad \mathfrak{e}_{12}+\mathfrak{e}_{21} \mapsto 2v_1v_2, \quad \mathfrak{e}_{22} \mapsto v_2^2.$$

**A decomposition of $\mathrm{sym}^2 \otimes \mathrm{sym}^2$.** By the Clebsch-Gordan rules we know that $\mathrm{sym}^2\otimes\mathrm{sym}^2 \cong \mathrm{sym}^0 \oplus \mathrm{sym}^2 \oplus \mathrm{sym}^4$ as a representation of $\mathrm{SL}_2(\mathbb{C})$. Let us find an explicit basis in terms of $v_1$ and $v_2$. To simplify notation, we will denote by $u_1$, $u_2$ and $v_1$, $v_2$ the variables of the polynomials in the first and second tensor component. Those in the image will be denoted by $w_1$ and $w_2$.

The trivial representation, one checks, is spanned by

$$u_1^2v_2^2 - 2u_1u_2v_1v_2 + u_2^2v_1^2\,.$$

Bases for $\mathrm{sym}^2$ and $\mathrm{sym}^4$, respectively, are given by

$$w_1^2 = u_1^2v_1v_2 - u_1u_2v_1^2, \quad w_1w_2 = \tfrac{1}{2}\big(u_1^2v_2^2 - u_2^2v_1^2\big), \quad w_2^2 = u_1u_2v_2^2 - u_2^2v_1v_2\,,$$

and

$$w_1^4 = u_1^2v_1^2, \quad w_1^3w_2 = \tfrac{1}{2}\big(u_1^2v_1v_2 + u_1u_2v_1^2\big), \quad w_1^2w_2^2 = \tfrac{1}{6}\big(u_2^2v_1^2 + 4u_1u_2v_1v_2 + u_1^2v_2^2\big),$$
$$w_1w_2^3 = \tfrac{1}{2}\big(u_2^2v_1v_2 + u_1u_2v_2^2\big), \quad w_2^4 = u_2^2v_2^2.$$

**Table of basic Fourier coefficients.** In Table 13, we give the first few Fourier coefficients of classical scalar valued Siegel modular forms that have been taken from [LMF15]; Or equivalently computed by the Sage script basic_fourier_expansions.sage. On the left of Table 13, we give names of Siegel modular forms, which are all explained in below. On the right, we give their Fourier coefficients according to the scheme dictated by the first 9 lines.

The space $\mathrm{M}^{(2)}(\det^{12})$ has dimension 3. A natural basis consists of the Siegel Eisenstein series $E^{(2)}_{12}$, the Klingen Eisenstein series $E^{(2)}_{\Delta}$ attached to the unique weight 12 cusp form $\Delta$ in genus 1, and the Maass lift $\phi^{(2)}_{12}$ of the unique Jacobi cusp form $\phi_{12}$ of weight 12 and index 1.

The space $\mathrm{M}^{(2)}(\det^{10}\mathrm{sym}^4)$ is one dimensional. A generator was found by Ibukiyama in [Ibu12] using $E^{(2)}_4$ and $E^{(2)}_6$ as input data. The Ibukiyama-Rankin-Cohen bracket is

$$\{f,g\}_{\mathrm{sym}^4} \in \mathrm{M}^{(2)}\big(\det^{k+k'}\mathrm{sym}^4\big) \quad \text{for} \quad f \in \mathrm{M}^{(2)}\big(\det^{k}\big) \text{ and } g \in \mathrm{M}^{(2)}\big(\det^{k'}\big).$$

With respect to our choice of basis $w_1^4, \ldots, w_2^4$, it equals

$$\begin{aligned}
&\tfrac{1}{2}k'(k'+1)\,g\,\Big(\partial_{11}^2 f\, w_1^4 + 2\partial_{11}\partial_{12} f\, w_1^3w_2 + \big(\partial_{12}^2 f + 2\partial_{11}\partial_{22} f\big)\, w_1^2w_2^2 + 2\partial_{12}\partial_{22} f\, w_1w_2^3 + \partial_{22}^2 f\, w_2^4\Big)\\
&-(k+1)(k'+1)\,\Big((\partial_{11}f)(\partial_{11}g)\, w_1^4 + \big((\partial_{11}f)(\partial_{12}g) + (\partial_{12}f)(\partial_{11}g)\big)\, w_1^3w_2\\
&\qquad + \big((\partial_{11}f)(\partial_{22}g) + (\partial_{12}f)(\partial_{12}g) + (\partial_{22}f)(\partial_{11}g)\big)\, w_1^2w_2^2\\
&\qquad + \big((\partial_{22}f)(\partial_{12}g) + (\partial_{12}f)(\partial_{22}g)\big)\, w_1w_2^3 + (\partial_{22}f)(\partial_{22}g)\, w_1^4\Big)\\
&+\tfrac{1}{2}k(k+1)\,f\,\Big(\partial_{11}^2 2\, w_1^4 + 2\partial_{11}\partial_{12} g\, w_1^3w_2 + \big(\partial_{12}^2 g + 2\partial_{11}\partial_{22} g\big)\, w_1^2w_2^2 + 2\partial_{12}\partial_{22} g\, w_1w_2^3 + \partial_{22}^2 g\, w_2^4\Big).
\end{aligned}$$

We do not reproduce the Fourier coefficients of $\{E_4^{(2)}, E_6^{(2)}\}_{\mathrm{sym}^4}$, but refer the reader to the Sage script mentioned above.

**An explicit expression for $\phi_{10}^{(2)} f_{\mathrm{RS}}$.** Using the Sage script fRS_phi2_10_fourier_expansion.sage, we compute an explicit representation of $\phi_{10}^{(2)} f_{\mathrm{RS}}$. We obtain

$$\phi_{10}^{(2)} f_{\mathrm{RS}} = \frac{-1}{72576000}\{E_4^{(2)}, E_6^{(2)}\}_{\mathrm{sym}^4} + \frac{1}{400}\phi_{12}^{(2)}, \tag{9.5}$$

where the embedding of $\det^{10}\mathrm{sym}^4$ and $\det^{12}$ into $\det^{10}\mathrm{sym}^2\mathrm{sym}^2$ is as above. From this expression, we obtain the Fourier coefficients in Table 14.

**Fourier expansion of $\phi_{10}^{(2)\,-1}$.** The Fourier expansion of $\phi_{10}^{(2)\,-1}$ is not unique, but depends on a Weyl chamber as explained in [Bor98]. The zero locus of $\phi_{10}^{(2)}$ is $\Gamma^{(2)}\{\tau : z = 0\}$, and a Weyl chamber is a connected component of its complement in $\mathbb{H}^{(2)}$. We consider the one that contains $i\begin{pmatrix} 1 & t^2 \\ t^2 & t \end{pmatrix}$ for all sufficiently small, positive $t$. In this Weyl chamber, $\phi_{10}^{(2)\,-1}$ has Fourier expansion

$$q_1 \zeta q_2 \prod_{(n,r,m)>0} \big(1 - q_1^n \zeta^r q_2^m\big)^{c(nm-r^2)},$$

where $q_1 = \exp(2\pi i \tau_1)$, $\zeta = \exp(2\pi i z)$, $q_2 = \exp(2\pi i \tau_2)$, and

$$(n,r,m) > 0 \Leftrightarrow (n > 0) \vee (n = 0 \wedge m > 0) \vee (n = m = 0 \wedge r > 0).$$

The $c(n)$ are the Fourier coefficients of the weak Jacobi form of weight 0 and index 1 that is uniquely determined by $c(-1) = 2$. Its Fourier expansion is computed to the necessary extend in the Sage script phi2_10_inv_fourier_expansion.sage.

To compute the Fourier expansion of $\phi_{10}^{(2)\,-1}$ correctly the bound on $(n,r,m)$ that were found in [GKR13] have to be taken into account. This, eventually yields the Fourier coefficients, some of which are displayed in Table 15. The reader is also referred to the file phi2_10_inv_fe.sobj. This file is computed by phi2_10_inv_fourier_expansion.sage and contains a Sage-readable dictionary of Fourier coefficients

$$\text{chi10inv_fe.sobj} : (n,r,m) \mapsto c\big(\phi_{10}^{(2)\,-1}, \big(\begin{smallmatrix} n & r/2 \\ r/2 & m \end{smallmatrix}\big)\big)$$

for many more $(n,r,m)$.

**Fourier coefficients of $f_{\mathrm{RS}}$.** Computing Fourier coefficients of $f_{\mathrm{RS}}$ in the Weyl chamber that was previously defined, we simply have to multiply $\phi_{10}^{(2)} f_{\mathrm{RS}}$ by $\phi_{10}^{(2)\,-1}$. This is performed in the Sage script fRS_fourier_expansion.sage. The resulting Fourier coefficients are displayed in Table 16. The previously mentioned Sage script will produce a file fRS_fe.sobj, which contains a dictionary of Fourier coefficients

$$\text{fRS_fe.sobj} : (n,r,m) \mapsto c\big(f_{\mathrm{RS}}, \big(\begin{smallmatrix} n & r/2 \\ r/2 & m \end{smallmatrix}\big)\big).$$

# A | The Ambiguities for $\mathbb{C}^3/\mathbb{Z}_5$

$$A_1 = \frac{1}{30z_2\Delta}\left(500z_1z_2^2+16z_2^2-450z_1z_2-16z_2+81z_1+3\right)$$

$$A_2 = \frac{1}{30z_1\Delta}\left(500z_1z_2^2+32z_2^2-225z_1z_2-16z_2+27z_1+2\right) \tag{A.1}$$

$$f_1^{11} = -\frac{1}{400\Delta}\left(-100000z_1^2z_2^5+150000z_1^2z_2^4-16800z_1z_2^4-576z_2^4-1000z_1^2z_2^3+30160z_1z_2^3+1008z_2^3-3600z_1^2z_2^2-11190z_1z_2^2-396z_2^2+1215z_1z_2+45z_2\right) \tag{A.2}$$

$$f_1^{12} = -\frac{1}{400\Delta}\left(50000z_1^3z_2^4+25000z_1^3z_2^3+6900z_1^2z_2^3+288z_1z_2^3-12000z_1^3z_2^2+3695z_1^2z_2^2+96z_1z_2^2+10800z_1^3z_2-2555z_1^2z_2-102z_1z_2+405z_1^2+15z_1\right) \tag{A.3}$$

$$f_1^{22} = -\frac{1}{400z_2\Delta}\left(-25000z_2^4z_1^4+81000z_2^2z_1^4-32400z_2z_1^4-2700z_2^3z_1^3+6265z_2^2z_1^3-2460z_2z_1^3+135z_1^3-144z_2^3z_1^2+152z_2^2z_1^2-49z_2z_1^2+5z_1^2\right) \tag{A.4}$$

$$f_2^{11} = -\frac{1}{400z_1\Delta}\left(-200000z_1^2z_2^6+50000z_1^2z_2^5-31600z_1z_2^5-1152z_2^5+3000z_1^2z_2^4+21320z_1z_2^4+864z_2^4-1200z_1^2z_2^3-4815z_1z_2^3-216z_2^3+365z_1z_2^2+18z_2^2\right) \tag{A.5}$$

$$f_2^{12} = -\frac{1}{400\Delta}\left(100000z_1^2z_2^5-12500z_1^2z_2^4+12800z_1z_2^4+576z_2^4-14000z_1^2z_2^3-7360z_1z_2^3-432z_2^3+3600z_1^2z_2^2+1220z_1z_2^2+108z_2^2-45z_1z_2-9z_2\right) \tag{A.6}$$

$$f_2^{22} = -\frac{1}{400\Delta}\left(-50000z_1^3z_2^4+125000z_1^3z_2^3-4900z_1^2z_2^3-288z_1z_2^3+57000z_1^3z_2^2+31405z_1^2z_2^2+1216z_1z_2^2-10800z_1^3z_2-12985z_1^2z_2-554z_1z_2+1485z_1^2+67z_1\right) \tag{A.7}$$

$$\tilde{f}_{11}^1 = \frac{1}{20z_2\Delta}\left(-105000z_1^2z_2^3-100z_1z_2^3-4500z_1^2z_2^2-16535z_1z_2^2-528z_2^2+6957z_1z_2+248z_2-783z_1-29\right) \tag{A.8}$$

$$\tilde{f}_{11}^2 = \frac{1}{20z_2^2\Delta}\left(15000z_2^3z_1^3+13500z_2^2z_1^3+50z_2^3z_1^2+2855z_2^2z_1^2-216z_2z_1^2-81z_1^2+84z_2^2z_1-9z_2z_1-3z_1\right) \tag{A.9}$$

$$\tilde{f}_{12}^1 = \frac{1}{20z_1\Delta}\left(-10000z_1^2z_2^3-200z_1z_2^3-1500z_1^2z_2^2-1470z_1z_2^2-48z_2^2+664z_1z_2+24z_2-81z_1-3\right) \tag{A.10}$$

$$\tilde{f}_{12}^2 = \frac{1}{20z_2\Delta}\left(-7500z_1^2z_2^3+100z_1z_2^3+4500z_1^2z_2^2-465z_1z_2^2-16z_2^2+243z_1z_2+8z_2-27z_1-1\right) \tag{A.11}$$

$$\tilde{f}_{22}^1 = \frac{1}{20z_1\Delta}\left(5000z_1z_2^4-400z_2^4-500z_1z_2^3+760z_2^3-297z_2^2+33z_2\right) \tag{A.12}$$

$$\tilde{f}_{22}^2 = \frac{1}{20z_1\Delta}\left(-90000z_1^2z_2^3+200z_1z_2^3+1500z_1^2z_2^2-13530z_1z_2^2-400z_2^2+6086z_1z_2+200z_2-729z_1-25\right) \tag{A.13}$$

Any other combination of indices follows by symmetry. We find the following holomorphic ambiguities:

$$\begin{aligned} f^{(2,0)} = -\frac{1}{\Delta^2}\Big(&\frac{31}{9216}-\frac{151}{2880}z_1+\frac{49}{160}z_1^2-\frac{143}{180}z_1^3+\frac{139}{180}z_1^4+\frac{51}{512}z_2-\frac{2431}{1536}z_1z_2+\frac{10943}{1152}z_1^2z_2-\frac{2557}{96}z_1^3z_2+\frac{2185}{72}z_1^4z_2+\frac{13851}{5120}z_2^2-\frac{23553}{512}z_1z_2^2+\frac{309015}{1024}z_1^2z_2^2 \\ &-\frac{4593875}{4608}z_1^3z_2^2+\frac{453125}{288}z_1^4z_2^2-\frac{179375}{288}z_1^5z_2^2-\frac{264375}{512}z_1^3z_2^3+\frac{1571875}{1536}z_1^4z_2^3+\frac{671875}{128}z_1^5z_2^3+\frac{212890625}{3072}z_1^6z_2^4\Big) \end{aligned} \tag{A.14}$$

$$\begin{aligned} f^{(1,1)} = -\frac{1}{\Delta^2}\Big(&-\frac{331}{57600}+\frac{257}{3600}z_1-\frac{61}{200}z_1^2+\frac{109}{225}z_1^3-\frac{7}{45}z_1^4-\frac{9}{640}z_2-\frac{549}{640}z_1z_2+\frac{3493}{288}z_1^2z_2-\frac{1205}{24}z_1^3z_2+\frac{1175}{18}z_1^4z_2-\frac{2511}{1280}z_2^2+\frac{2673}{128}z_1z_2^2-\frac{16035}{256}z_1^2z_2^2+\frac{310625}{1152}z_1^3z_2^2 \\ &-\frac{113375}{72}z_1^4z_2^2+\frac{190625}{72}z_1^5z_2^2+\frac{406875}{128}z_1^3z_2^3-\frac{2615625}{128}z_1^4z_2^3+\frac{3015625}{96}z_1^5z_2^3+\frac{1953125}{256}z_1^6z_2^4\Big) \end{aligned} \tag{A.15}$$

$$\begin{aligned} f^{(0,2)} = -\frac{1}{\Delta^2}\Big(&\frac{167}{14400}-\frac{121}{720}z_1+\frac{271}{300}z_1^2-\frac{479}{225}z_1^3+\frac{416}{225}z_1^4+\frac{1419}{3200}z_2-\frac{60691}{9600}z_1z_2+\frac{13353}{400}z_1^2z_2-\frac{27973}{360}z_1^3z_2+\frac{610}{9}z_1^4z_2+\frac{3807}{640}z_2^2-\frac{28197}{320}z_1z_2^2+\frac{63135}{128}z_1^2z_2^2 \\ &-\frac{1513675}{1152}z_1^3z_2^2+\frac{119125}{72}z_1^4z_2^2-\frac{16375}{24}z_1^5z_2^2-\frac{38625}{64}z_1^3z_2^3+\frac{610625}{192}z_1^4z_2^3-\frac{828125}{288}z_1^5z_2^3+\frac{5078125}{384}z_1^6z_2^4\Big) \end{aligned} \tag{A.16}$$

$$
\begin{aligned}
f^{(3,0)} = \frac{1}{\Delta^4}\Big(& -\frac{45893}{2322432000} + \frac{76957}{145152000} z_1 - \frac{5771}{907200} z_1^2 + \frac{82289}{1814400} z_1^3 - \frac{193997}{907200} z_1^4 + \frac{391327}{567000} z_1^5 - \frac{105361}{70875} z_1^6 + \frac{27334}{14175} z_1^7 - \frac{2294}{2025} z_1^8 + \frac{4703}{4300800} z_2 - \frac{2092763}{46448640} z_1 z_2 \\
& + \frac{14325557}{19353600} z_1^2 z_2 - \frac{30369853}{4838400} z_1^3 z_2 + \frac{110692871}{3628800} z_1^4 z_2 - \frac{27287759}{302400} z_1^5 z_2 + \frac{13058167}{75600} z_1^6 z_2 - \frac{2544593}{11340} z_1^7 z_2 + \frac{3407}{21} z_1^8 z_2 - \frac{1088733}{1638400} z_2^2 + \frac{74510981}{3440640} z_1 z_2^2 \\
& - \frac{32289113111}{103219200} z_1^2 z_2^2 + \frac{51377486471}{19353600} z_1^3 z_2^2 - \frac{281433165299}{19353600} z_1^4 z_2^2 + \frac{25441364153}{483840} z_1^5 z_2^2 - \frac{5963971939}{48384} z_1^6 z_2^2 + \frac{270034265}{1512} z_1^7 z_2^2 - \frac{148713575}{1008} z_1^8 z_2^2 + \frac{3117875}{54} z_1^9 z_2^2 \\
& + \frac{20151747}{1638400} z_2^3 - \frac{703924371}{1638400} z_1 z_2^3 + \frac{75851055759}{11468800} z_1^2 z_2^3 - \frac{171022252321}{2949120} z_1^3 z_2^3 + \frac{2992829314891}{9289728} z_1^4 z_2^3 - \frac{200845661185}{165888} z_1^5 z_2^3 + \frac{306851519075}{96768} z_1^6 z_2^3 - \frac{269782983125}{48384} z_1^7 z_2^3 \\
& + \frac{23012309375}{4032} z_1^8 z_2^3 - \frac{1362475000}{567} z_1^9 z_2^3 - \frac{68914557}{2867200} z_2^4 + \frac{2996919}{3584} z_1 z_2^4 - \frac{1514923803}{114688} z_1^2 z_2^4 + \frac{69556453935}{458752} z_1^3 z_2^4 - \frac{879922720925}{688128} z_1^4 z_2^4 + \frac{29843009687875}{4128768} z_1^5 z_2^4 \\
& - \frac{24771987600625}{884736} z_1^6 z_2^4 + \frac{10351512453125}{129024} z_1^7 z_2^4 - \frac{32090023234375}{193536} z_1^8 z_2^4 + \frac{156671796875}{768} z_1^9 z_2^4 - \frac{7648528515625}{72576} z_1^{10} z_2^4 + \frac{949714875}{7168} z_1^3 z_2^5 - \frac{15414913125}{7168} z_1^4 z_2^5 \\
& + \frac{1470856303125}{114688} z_1^5 z_2^5 - \frac{2262255265625}{43008} z_1^6 z_2^5 + \frac{94437331796875}{387072} z_1^7 z_2^5 - \frac{307776826953125}{387072} z_1^8 z_2^5 + \frac{58100248046875}{48384} z_1^9 z_2^5 - \frac{16720126953125}{24192} z_1^{10} z_2^5 \\
& - \frac{9189205078125}{57344} z_1^6 z_2^6 + \frac{11576087890625}{6144} z_1^7 z_2^6 - \frac{1381279521484375}{172032} z_1^8 z_2^6 + \frac{1885945947265625}{110592} z_1^9 z_2^6 - \frac{3183334228515625}{145152} z_1^{10} z_2^6 + \frac{261849365234375}{20736} z_1^{11} z_2^6 \\
& + \frac{3382801513671875}{258048} z_1^9 z_2^7 - \frac{16896783447265625}{258048} z_1^{10} z_2^7 + \frac{4187774658203125}{64512} z_1^{11} z_2^7 - \frac{42278289794921875}{1032192} z_1^{12} z_2^8\Big)
\end{aligned}
\tag{A.17}
$$

$$
\begin{aligned}
f^{(2,1)} = \frac{1}{\Delta^4}\Big(& \frac{625649}{3870720000} - \frac{961951}{241920000} z_1 + \frac{255053}{6048000} z_1^2 - \frac{772259}{3024000} z_1^3 + \frac{1507181}{1512000} z_1^4 - \frac{2560261}{945000} z_1^5 + \frac{624748}{118125} z_1^6 - \frac{161746}{23625} z_1^7 + \frac{2902}{675} z_1^8 - \frac{590831}{43008000} z_2 + \frac{217453951}{387072000} z_1 z_2 \\
& - \frac{300910403}{32256000} z_1^2 z_2 + \frac{73848433}{896000} z_1^3 z_2 - \frac{2595602309}{6048000} z_1^4 z_2 + \frac{692727839}{504000} z_1^5 z_2 - \frac{22889351}{8400} z_1^6 z_2 + \frac{60546149}{18900} z_1^7 z_2 - \frac{189563}{105} z_1^8 z_2 + \frac{14282469}{4096000} z_2^2 - \frac{1599168133}{14336000} z_1 z_2^2 \\
& + \frac{409397602771}{258048000} z_1^2 z_2^2 - \frac{191931715301}{14336000} z_1^3 z_2^2 + \frac{13258897251}{179200} z_1^4 z_2^2 - \frac{442311668783}{1612800} z_1^5 z_2^2 + \frac{8998561657}{13440} z_1^6 z_2^2 - \frac{4141160927}{4032} z_1^7 z_2^2 + \frac{115010095}{126} z_1^8 z_2^2 - \frac{13846175}{36} z_1^9 z_2^2 \\
& - \frac{1027647}{40960} z_2^3 + \frac{9840393}{10240} z_1 z_2^3 - \frac{22718585481}{1433600} z_1^2 z_2^3 + \frac{173273342717}{1228800} z_1^3 z_2^3 - \frac{11697769909093}{15482880} z_1^4 z_2^3 + \frac{296490468547}{110592} z_1^5 z_2^3 - \frac{221032737295}{32256} z_1^6 z_2^3 + \frac{100661606225}{8064} z_1^7 z_2^3 \\
& - \frac{53858794375}{4032} z_1^8 z_2^3 + \frac{15902575625}{3024} z_1^9 z_2^3 + \frac{97437411}{1433600} z_2^4 - \frac{139889997}{71680} z_1 z_2^4 + \frac{3753997083}{143360} z_1^2 z_2^4 - \frac{5726780109}{14336} z_1^3 z_2^4 + \frac{264980590145}{57344} z_1^4 z_2^4 - \frac{7967478867275}{258048} z_1^5 z_2^4 \\
& + \frac{12468528248875}{98304} z_1^6 z_2^4 - \frac{23258817959375}{64512} z_1^7 z_2^4 + \frac{47254204909375}{64512} z_1^8 z_2^4 - \frac{261136421875}{288} z_1^9 z_2^4 + \frac{11303918515625}{24192} z_1^{10} z_2^4 - \frac{15532942725}{14336} z_1^3 z_2^5 + \frac{280236067875}{14336} z_1^4 z_2^5 \\
& - \frac{2025211175625}{14336} z_1^5 z_2^5 + \frac{107575149209375}{172032} z_1^6 z_2^5 - \frac{289179227984375}{129024} z_1^7 z_2^5 + \frac{42752775390625}{7168} z_1^8 z_2^5 - \frac{8761371484375}{1008} z_1^9 z_2^5 + \frac{19897498046875}{4032} z_1^{10} z_2^5 \\
& + \frac{21033648984375}{28672} z_1^6 z_2^6 - \frac{18082766015625}{2048} z_1^7 z_2^6 + \frac{9956001431640625}{258048} z_1^8 z_2^6 - \frac{489531435546875}{6144} z_1^9 z_2^6 + \frac{265013720703125}{3024} z_1^{10} z_2^6 - \frac{146752197265625}{3456} z_1^{11} z_2^6 \\
& - \frac{1767578857421875}{43008} z_1^9 z_2^7 + \frac{3002840576171875}{14336} z_1^{10} z_2^7 - \frac{7132171630859375}{32256} z_1^{11} z_2^7 + \frac{36258697509765625}{516096} z_1^{12} z_2^8\Big)
\end{aligned}
\tag{A.18}
$$

$$
\begin{aligned}
f^{(1,2)} = \frac{1}{\Delta^4}\Big(& -\frac{127}{193536} + \frac{107687}{6720000} z_1 - \frac{2554393}{15120000} z_1^2 + \frac{772703}{756000} z_1^3 - \frac{254039}{63000} z_1^4 + \frac{532729}{47250} z_1^5 - \frac{2709046}{118125} z_1^6 + \frac{1213208}{39375} z_1^7 - \frac{66784}{3375} z_1^8 + \frac{1117523}{80640000} z_2 - \frac{92939087}{96768000} z_1 z_2 \\
& + \frac{19622311}{1008000} z_1^2 z_2 - \frac{1150991209}{6048000} z_1^3 z_2 + \frac{792346361}{756000} z_1^4 z_2 - \frac{6475451077}{1890000} z_1^5 z_2 + \frac{31226009}{4725} z_1^6 z_2 - \frac{33107323}{4725} z_1^7 z_2 + \frac{3047144}{945} z_1^8 z_2 - \frac{1053621}{204800} z_2^2 + \frac{542784461}{3584000} z_1 z_2^2 \\
& - \frac{43056741157}{21504000} z_1^2 z_2^2 + \frac{260795636281}{16128000} z_1^3 z_2^2 - \frac{135282624397}{1512000} z_1^4 z_2^2 + \frac{105139716709}{302400} z_1^5 z_2^2 - \frac{27852246421}{30240} z_1^6 z_2^2 + \frac{2363539459}{1512} z_1^7 z_2^2 - \frac{290206555}{189} z_1^8 z_2^2 + \frac{18496600}{27} z_1^9 z_2^2 \\
& - \frac{4003911}{204800} z_2^3 + \frac{53150553}{204800} z_1 z_2^3 + \frac{1342225737}{1433600} z_1^2 z_2^3 - \frac{14319930859}{614400} z_1^3 z_2^3 + \frac{94688637991}{1935360} z_1^4 z_2^3 + \frac{7450037635}{13824} z_1^5 z_2^3 - \frac{40072005695}{12096} z_1^6 z_2^3 + \frac{15410053175}{2016} z_1^7 z_2^3 \\
& - \frac{1485511625}{168} z_1^8 z_2^3 + \frac{664515625}{126} z_1^9 z_2^3 - \frac{84507867}{358400} z_2^4 + \frac{102906369}{17920} z_1 z_2^4 - \frac{4346713773}{71680} z_1^2 z_2^4 + \frac{40957934373}{57344} z_1^3 z_2^4 - \frac{231127656755}{28672} z_1^4 z_2^4 + \frac{3200762919325}{57344} z_1^5 z_2^4 \\
& - \frac{1591211867375}{6912} z_1^6 z_2^4 + \frac{6516814669375}{10752} z_1^7 z_2^4 - \frac{25771664196875}{24192} z_1^8 z_2^4 + \frac{55222578125}{48} z_1^9 z_2^4 - \frac{810787890625}{1512} z_1^{10} z_2^4 + \frac{7953979275}{3584} z_1^3 z_2^5 - \frac{152339916375}{3584} z_1^4 z_2^5 \\
& + \frac{4777944586875}{14336} z_1^5 z_2^5 - \frac{4026517634375}{2688} z_1^6 z_2^5 + \frac{1201057407859375}{258048} z_1^7 z_2^5 - \frac{498395713203125}{48384} z_1^8 z_2^5 + \frac{657354311328125}{48384} z_1^9 z_2^5 - \frac{44716935546875}{6048} z_1^{10} z_2^5 \\
& - \frac{6006763828125}{7168} z_1^6 z_2^6 + \frac{5262648046875}{512} z_1^7 z_2^6 - \frac{329135134765625}{7168} z_1^8 z_2^6 + \frac{2596957685546875}{27648} z_1^9 z_2^6 - \frac{8894909423828125}{96768} z_1^{10} z_2^6 + \frac{40810302734375}{1152} z_1^{11} z_2^6 \\
& + \frac{992905029296875}{32256} z_1^9 z_2^7 - \frac{1694947509765625}{10752} z_1^{10} z_2^7 + \frac{1370550537109375}{8064} z_1^{11} z_2^7 - \frac{1363372802734375}{43008} z_1^{12} z_2^8\Big)
\end{aligned}
\tag{A.19}
$$

$$
\begin{aligned}
f^{(0,3)} = \frac{1}{\Delta^4}\Big(& \frac{64675591}{58060800000} - \frac{54856777}{1814400000} z_1 + \frac{65194601}{181440000} z_1^2 - \frac{55731167}{22680000} z_1^3 + \frac{242045929}{22680000} z_1^4 - \frac{216143287}{7087500} z_1^5 + \frac{200947909}{3543750} z_1^6 - \frac{22452548}{354375} z_1^7 + \frac{1654522}{50625} z_1^8 + \frac{40894657}{645120000} z_2 \\
& - \frac{8658850111}{5806080000} z_1 z_2 + \frac{697315817}{48384000} z_1^2 z_2 - \frac{594293389}{8064000} z_1^3 z_2 + \frac{122562197}{567000} z_1^4 z_2 - \frac{975682873}{2520000} z_1^5 z_2 + \frac{54401369}{105000} z_1^6 z_2 - \frac{187652267}{283500} z_1^7 z_2 + \frac{2428696}{4725} z_1^8 z_2 + \frac{84392949}{20480000} z_2^2 \\
& - \frac{24515809627}{215040000} z_1 z_2^2 + \frac{598212759449}{430080000} z_1^2 z_2^2 - \frac{19527458020663}{1935360000} z_1^3 z_2^2 + \frac{1494655943567}{30240000} z_1^4 z_2^2 - \frac{1392743766707}{8064000} z_1^5 z_2^2 + \frac{258536031797}{604800} z_1^6 z_2^2 - \frac{944837177}{1344} z_1^7 z_2^2 \\
& + \frac{511856045}{756} z_1^8 z_2^2 - \frac{30988525}{108} z_1^9 z_2^2 + \frac{7989111}{128000} z_2^3 - \frac{858617793}{512000} z_1 z_2^3 + \frac{70359296979}{3584000} z_1^2 z_2^3 - \frac{2572634392021}{18432000} z_1^3 z_2^3 + \frac{165039233141773}{232243200} z_1^4 z_2^3 - \frac{908991091601}{331776} z_1^5 z_2^3 \\
& + \frac{737264763775}{96768} z_1^6 z_2^3 - \frac{334904956675}{24192} z_1^7 z_2^3 + \frac{58374183625}{4032} z_1^8 z_2^3 - \frac{61068818875}{9072} z_1^9 z_2^3 + \frac{15197463}{35840} z_2^4 - \frac{10461879}{896} z_1 z_2^4 + \frac{2531382111}{17920} z_1^2 z_2^4 - \frac{2039769849}{1792} z_1^3 z_2^4 \\
& + \frac{317068931875}{43008} z_1^4 z_2^4 - \frac{796055737705}{21504} z_1^5 z_2^4 + \frac{114660098778775}{884736} z_1^6 z_2^4 - \frac{38440191740875}{129024} z_1^7 z_2^4 + \frac{84196628171875}{193536} z_1^8 z_2^4 - \frac{429043328125}{1152} z_1^9 z_2^4 + \frac{9972960171875}{72576} z_1^{10} z_2^4 \\
& - \frac{1768497705}{1792} z_1^3 z_2^5 + \frac{17572939425}{896} z_1^4 z_2^5 - \frac{288041235375}{1792} z_1^5 z_2^5 + \frac{5223196121875}{7168} z_1^6 z_2^5 - \frac{1612869727140625}{774144} z_1^7 z_2^5 + \frac{1013819765140625}{258048} z_1^8 z_2^5 - \frac{143767815859375}{32256} z_1^9 z_2^5 \\
& + \frac{105542079296875}{48384} z_1^{10} z_2^5 + \frac{182184140625}{896} z_1^6 z_2^6 - \frac{967665390625}{384} z_1^7 z_2^6 + \frac{91877128515625}{8064} z_1^8 z_2^6 - \frac{4438408203125}{192} z_1^9 z_2^6 + \frac{6031614501953125}{290304} z_1^{10} z_2^6 \\
& - \frac{65818115234375}{10368} z_1^{11} z_2^6 - \frac{37890185546875}{8064} z_1^9 z_2^7 + \frac{64582275390625}{2688} z_1^{10} z_2^7 - \frac{105147705078125}{4032} z_1^{11} z_2^7 + \frac{106353759765625}{32256} z_1^{12} z_2^8\Big)
\end{aligned}
\tag{A.20}
$$

# B | The Ambiguities for $\mathbb{C}^3/\mathbb{Z}_6$

$$\begin{aligned} A_1 =& \frac{1}{15z_1\Delta_1}\Big(2-12z_1+16z_1^2-64z_3z_1^2-16z_2+102z_1z_2-144z_1^2z_2+576z_3z_1^2z_2+32z_2^2-216z_1z_2^2+270z_1^2z_2^2-1512z_3z_1^2z_2^2+108z_1^3z_2^2-432z_3z_1^3z_2^2+216z_1^2z_2^3 \\ &+864z_3z_1^2z_2^3-486z_1^3z_2^3+1944z_3z_1^3z_2^3\Big) \end{aligned} \tag{B.1}$$

$$\begin{aligned} A_2 =& \frac{1}{15z_2\Delta_1}\Big(2-16z_1+32z_1^2-128z_3z_1^2-12z_2+102z_1z_2-216z_1^2z_2+864z_3z_1^2z_2+16z_2^2-144z_1z_2^2+270z_1^2z_2^2-1512z_3z_1^2z_2^2+216z_1^3z_2^2-864z_3z_1^3z_2^2+108z_1^2z_2^3 \\ &+432z_3z_1^2z_2^3-486z_1^3z_2^3+1944z_3z_1^3z_2^3\Big) \end{aligned} \tag{B.2}$$

$$\begin{aligned} f_1^{11} =& \frac{1}{100\Delta_1}\Big(-9z_1+72z_1^2-144z_1^3+320z_3z_1^3+81z_1z_2-684z_1^2z_2+1440z_1^3z_2-3200z_3z_1^3z_2-216z_1z_2^2+1908z_1^2z_2^2-3726z_1^3z_2^2+10428z_3z_1^3z_2^2-1944z_1^4z_2^2+6720z_3z_1^4z_2^2 \\ &-4800z_3z_1^5z_2^2+19200z_3^2z_1^5z_2^2+144z_1z_2^3-1296z_1^2z_2^3+486z_1^3z_2^3-12492z_3z_1^3z_2^3+10692z_1^4z_2^3-36960z_3z_1^4z_2^3+28800z_3z_1^5z_2^3-115200z_3^2z_1^5z_2^3+1944z_1^3z_2^4+7728z_3z_1^3z_2^4 \\ &-8748z_1^4z_2^4+28656z_3z_1^4z_2^4-6561z_1^5z_2^4+6804z_3z_1^5z_2^4+77760z_3^2z_1^5z_2^4+6561z_1^5z_2^5-52164z_3z_1^5z_2^5+103680z_3^2z_1^5z_2^5\Big) \end{aligned} \tag{B.3}$$

$$\begin{aligned} f_1^{12} =& \frac{1}{200\Delta_1}\Big(9z_2-72z_1z_2+144z_1^2z_2-320z_3z_1^2z_2-108z_2^2+900z_1z_2^2-1872z_1^2z_2^2+4832z_3z_1^2z_2^2+432z_2^3-3744z_1z_2^3+7614z_1^2z_2^3-25164z_3z_1^2z_2^3+1944z_1^3z_2^3-6720z_3z_1^3z_2^3 \\ &+4800z_3z_1^4z_2^3-19200z_3^2z_1^4z_2^3-576z_2^4+5184z_1z_2^4-7776z_1^2z_2^4+51552z_3z_1^2z_2^4-16524z_1^3z_2^4+57912z_3z_1^3z_2^4-26640z_3z_1^4z_2^4+106560z_3^2z_1^4z_2^4-7776z_1^2z_2^5-30912z_3z_1^2z_2^5 \\ &+34992z_1^3z_2^5-124128z_3z_1^3z_2^5+6561z_1^4z_2^5-16524z_3z_1^4z_2^5-38880z_3^2z_1^4z_2^5-26244z_1^4z_2^6+208656z_3z_1^4z_2^6-414720z_3^2z_1^4z_2^6\Big) \end{aligned} \tag{B.4}$$

$$\begin{aligned} f_1^{22} =& \frac{1}{100z_1\Delta_1}\Big(-9z_2^2+72z_1z_2^2-144z_1^2z_2^2+512z_3z_1^2z_2^2+108z_2^3-900z_1z_2^3+1872z_1^2z_2^3-6800z_3z_1^2z_2^3-432z_2^4+3744z_1z_2^4-7614z_1^2z_2^4+31836z_3z_1^2z_2^4-1944z_1^3z_2^4+7512z_3z_1^3z_2^4 \\ &-1200z_3z_1^4z_2^4+4800z_3^2z_1^4z_2^4+576z_2^5-5184z_1z_2^5+7776z_1^2z_2^5-59040z_3z_1^2z_2^5+16524z_1^3z_2^5-63456z_3z_1^3z_2^5+6120z_3z_1^4z_2^5-24480z_3^2z_1^4z_2^5+7776z_1^2z_2^6+30912z_3z_1^2z_2^6 \\ &-34992z_1^3z_2^6+133632z_3z_1^3z_2^6-6561z_1^4z_2^6+45684z_3z_1^4z_2^6-77760z_3^2z_1^4z_2^6+26244z_1^4z_2^7-208656z_3z_1^4z_2^7+414720z_3^2z_1^4z_2^7\Big) \end{aligned} \tag{B.5}$$

$$\begin{aligned} f_2^{11} =& \frac{1}{200\Delta_1}\Big(9z_1^2-72z_1^3+144z_1^4-320z_3z_1^4-72z_1^2z_2+612z_1^3z_2-1296z_1^4z_2+1872z_3z_1^4z_2+9600z_3z_1^5z_2-19200z_3z_1^6z_2+76800z_3^2z_1^6z_2+144z_1^2z_2^2-1296z_1^3z_2^2+2430z_1^4z_2^2 \\ &-5796z_3z_1^4z_2^2+1944z_1^5z_2^2-35520z_3z_1^5z_2^2+72000z_3z_1^6z_2^2-288000z_3^2z_1^6z_2^2+1944z_1^4z_2^3+7728z_3z_1^4z_2^3-8748z_1^5z_2^3+28872z_3z_1^5z_2^3-38880z_3z_1^6z_2^3+155520z_3^2z_1^6z_2^3 \\ &+6561z_1^6z_2^4-52164z_3z_1^6z_2^4+103680z_3^2z_1^6z_2^4\Big) \end{aligned} \tag{B.6}$$

$$\begin{aligned} f_2^{12} =& \frac{1}{100\Delta_1}\Big(-9z_1z_2+72z_1^2z_2-144z_1^3z_2+512z_3z_1^3z_2+72z_1z_2^2-612z_1^2z_2^2+1296z_1^3z_2^2-4332z_3z_1^3z_2^2-2400z_3z_1^4z_2^2+4800z_3z_1^5z_2^2-19200z_3^2z_1^5z_2^2-144z_1z_2^3+1296z_1^2z_2^3 \\ &-2430z_1^3z_2^3+11628z_3z_1^3z_2^3-1944z_1^4z_2^3+13632z_3z_1^4z_2^3-15840z_3z_1^5z_2^3+63360z_3^2z_1^5z_2^3-1944z_1^3z_2^4-7728z_3z_1^3z_2^4+8748z_1^4z_2^4-31248z_3z_1^4z_2^4+4860z_3z_1^5z_2^4 \\ &-19440z_3^2z_1^5z_2^4-6561z_1^5z_2^5+52164z_3z_1^5z_2^5-103680z_3^2z_1^5z_2^5\Big) \end{aligned} \tag{B.7}$$

$$\begin{aligned} f_2^{22} =& -\frac{1}{200\Delta_1}\Big(-27z_2+216z_1z_2-432z_1^2z_2+1728z_3z_1^2z_2+252z_2^2-2124z_1z_2^2+4464z_1^2z_2^2-17792z_3z_1^2z_2^2-720z_2^3+6336z_1z_2^3-12474z_1^2z_2^3+60684z_3z_1^2z_2^3-5832z_1^3z_2^3 \\ &+25728z_3z_1^3z_2^3-4800z_3z_1^4z_2^3+19200z_3^2z_1^4z_2^3+576z_2^4-5184z_1z_2^4+3888z_1^2z_2^4-75744z_3z_1^2z_2^4+34020z_1^3z_2^4-140856z_3z_1^3z_2^4+13680z_3z_1^4z_2^4-54720z_3^2z_1^4z_2^4+7776z_1^2z_2^5 \\ &+30912z_3z_1^2z_2^5-34992z_1^3z_2^5+134496z_3z_1^3z_2^5-19683z_1^4z_2^5+156492z_3z_1^4z_2^5-311040z_3^2z_1^4z_2^5+26244z_1^4z_2^6-208656z_3z_1^4z_2^6+414720z_3^2z_1^4z_2^6\Big) \end{aligned} \tag{B.8}$$

$$\begin{aligned} \tilde{f}_{11}^1 =& \frac{1}{10z_1\Delta_1}\Big(-13+112z_1-240z_1^2+960z_3z_1^2+104z_2-949z_1z_2+2128z_1^2z_2-8672z_3z_1^2z_2+80z_1^3z_2-320z_3z_1^3z_2-208z_2^2+2008z_1z_2^2-3911z_1^2z_2^2+22932z_3z_1^2z_2^2-3900z_1^3z_2^2 \\ &+15600z_3z_1^3z_2^2-16z_1z_2^3-3252z_1^2z_2^3-13584z_3z_1^2z_2^3+15849z_1^3z_2^3-63396z_3z_1^3z_2^3+540z_1^4z_2^3-4320z_3z_1^4z_2^3+8640z_3^2z_1^4z_2^3-108z_1^3z_2^4+432z_3z_1^3z_2^4-12960z_1^4z_2^4 \\ &+103680z_3z_1^4z_2^4-207360z_3^2z_1^4z_2^4\Big) \end{aligned} \tag{B.9}$$

$$\begin{aligned} \tilde{f}_{11}^2 =& \frac{1}{5z_1\Delta_1}\Big(-2z_2+8z_1z_2-32z_3z_1z_2+17z_2^2-67z_1z_2^2+308z_3z_1z_2^2-20z_1^2z_2^2+80z_3z_1^2z_2^2-40z_2^3+128z_1z_2^3-912z_3z_1z_2^3+264z_1^2z_2^3-1056z_3z_1^2z_2^3+16z_2^4+48z_1z_2^4 \\ &+768z_3z_1z_2^4-783z_1^2z_2^4+3132z_3z_1^2z_2^4-135z_1^3z_2^4+1080z_3z_1^3z_2^4-2160z_3^2z_1^3z_2^4+108z_1^2z_2^5-432z_3z_1^2z_2^5+810z_1^3z_2^5-6480z_3z_1^3z_2^5+12960z_3^2z_1^3z_2^5\Big) \end{aligned} \tag{B.10}$$

$$\begin{aligned} \tilde{f}_{12}^1 =& \frac{1}{5\Delta_1}\Big(3z_1-32z_1^2-32z_3z_1^2+80z_1^3-320z_3z_1^3-11z_1z_2+160z_1^2z_2+240z_3z_1^2z_2-480z_1^3z_2+1920z_3z_1^3z_2-4z_1z_2^2-102z_1^2z_2^2-552z_3z_1^2z_2^2+459z_1^3z_2^2-1836z_3z_1^3z_2^2 \\ &+540z_1^4z_2^2-4320z_3z_1^4z_2^2+8640z_3^2z_1^4z_2^2-27z_1^3z_2^3+108z_3z_1^3z_2^3-810z_1^4z_2^3+6480z_3z_1^4z_2^3-12960z_3^2z_1^4z_2^3\Big) \end{aligned} \tag{B.11}$$

$$\begin{aligned} \tilde{f}_{12}^2 =& \frac{1}{10z_1\Delta_1}\Big(-3+24z_1-48z_1^2+192z_3z_1^2+24z_2-207z_1z_2+464z_1^2z_2-1696z_3z_1^2z_2-80z_1^3z_2+320z_3z_1^3z_2-48z_2^2+440z_1z_2^2-949z_1^2z_2^2+4284z_3z_1^2z_2^2-204z_1^3z_2^2+816z_3z_1^3z_2^2 \\ &+16z_1z_2^3-636z_1^2z_2^3-1968z_3z_1^2z_2^3+2619z_1^3z_2^3-10476z_3z_1^3z_2^3-540z_1^4z_2^3+4320z_3z_1^4z_2^3-8640z_3^2z_1^4z_2^3+108z_1^3z_2^4-432z_3z_1^3z_2^4-1620z_1^4z_2^4+12960z_3z_1^4z_2^4 \\ &-25920z_3^2z_1^4z_2^4\Big) \end{aligned} \tag{B.12}$$

$$\begin{aligned} \tilde{f}_{22}^1 =& \frac{1}{10z_2\Delta_1}\Big(-13z_1+124z_1^2-368z_1^3+832z_3z_1^3+320z_1^4-1280z_3z_1^4+52z_1z_2-521z_1^2z_2+1552z_1^3z_2-4128z_3z_1^3z_2-1200z_1^4z_2+4800z_3z_1^4z_2-4z_1^2z_2^2+609z_1^3z_2^2+2292z_3z_1^3z_2^2 \\ &-3024z_1^4z_2^2+12096z_3z_1^4z_2^2+2160z_1^5z_2^2-17280z_3z_1^5z_2^2+34560z_3^2z_1^5z_2^2-27z_1^4z_2^3+108z_3z_1^4z_2^3+1620z_1^5z_2^3-12960z_3z_1^5z_2^3+25920z_3^2z_1^5z_2^3\Big) \end{aligned} \tag{B.13}$$

$$\begin{aligned} \tilde{f}_{22}^2 =& \frac{1}{5z_2\Delta_1}\Big(-8+64z_1-128z_1^2+512z_3z_1^2+51840z_3z_1^4+68z_2-581z_1z_2+1256z_1^2z_2-4864z_3z_1^2z_2-80z_1^3z_2+320z_3z_1^3z_2-144z_2^2+1307z_1z_2^2-2590z_1^2z_2^2+13368z_3z_1^2z_2^2 \\ &-1464z_1^3z_2^2+5856z_3z_1^3z_2^2+4z_1z_2^3-1950z_1^2z_2^3-7656z_3z_1^2z_2^3+8775z_1^3z_2^3-35100z_3z_1^3z_2^3-540z_1^4z_2^3+4320z_3z_1^4z_2^3-8640z_3^2z_1^4z_2^3+27z_1^3z_2^4-108z_3z_1^3z_2^4-6480z_1^4z_2^4\Big) \end{aligned} \tag{B.14}$$

# C | Siegel modular forms of genus two

In this appendix we introduce the basics of Siegel modular forms that are needed. Although many statements and definitions hold for arbitrary genus, we restrict ourselves to genus two. The Siegel upper halfplane is denoted by

$$\mathbb{H}^{(2)} = \left\{\tau \in \mathbb{C}^{2\times 2} | \tau^T = \tau, \quad \operatorname{Im}\tau \geq 0\right\}, \tag{C.1}$$

on which the homogenous modular group $\Gamma_2 = \mathrm{Sp}(4,\mathbb{Z})$ operates by

$$\tau \mapsto \big(a\tau + b\big)\big(c\tau + d\big)^{-1}, \qquad \begin{pmatrix} a & b \\ c & d \end{pmatrix} \in \mathrm{Sp}(4,\mathbb{Z}). \tag{C.2}$$

The quotient $\mathbb{H}^{(2)}$ by this action is called the Siegel fundamental domain $\mathscr{F}_2 = \Gamma_2 \backslash \mathbb{H}^{(2)}$. A Siegel modular form of weight $w$ is a holomorphic function $f : \mathbb{H}^{(2)} \longrightarrow \mathbb{C}$, such that for all $\tau \in \mathbb{H}^{(2)}$ and $\gamma \in \Gamma_2$

$$f(\gamma\tau) = \det\big(c\tau + d\big)^{w} f(\tau), \quad \gamma = \begin{pmatrix} a & b \\ c & d \end{pmatrix}, \tag{C.3}$$

holds. The space of all Siegel modular forms of weight $w$ is denoted by $M_w(\mathbb{H}^{(2)})$. As $\begin{pmatrix} \mathbb{1} & \mathbb{1} \\ 0 & \mathbb{1} \end{pmatrix}$ is contained in $\Gamma_2$, any Siegel modular form $f$ admits a Fourier expansion which reads

$$f = \sum_T a(T) \exp(2\pi i \,\mathrm{Tr}(T\tau)). \tag{C.4}$$

Here the summation is over all half integer matrices $T \in \frac{1}{2}\mathbb{Z}^{2\times 2}$ which have integer diagonal elements. Next we introduce the Siegel operator $\Phi$, which is a map $M_w(\mathbb{H}^{(2)}) \longrightarrow M_w(\mathbb{H}^{(1)})$ and is defined by

$$\Phi f = \lim_{t\to\infty} f \begin{pmatrix} \tau & 0 \\ 0 & it \end{pmatrix}, \quad \tau \in \mathbb{H},\, t \in \mathbb{R}. \tag{C.5}$$

The elements of $\ker\Phi$ are called cusp forms. For $w \geq 4$ we define the Eisenstein series by

$$E_w(\tau) = \sum_{C,D} \det(c\tau + d)^{-w}. \tag{C.6}$$

The summation is over all inequivalent bottom rows $\begin{pmatrix} C & D \end{pmatrix}$ of elements of $\Gamma_2$. A classical theorem by Igusa states that the space of Siegel modular forms of genus two has a representation as

$$M = \mathbb{C}[E_4, E_6, \chi_{10}, \chi_{12}, \chi_{35}] \big/ \{\chi_{35}^2 = R\}, \tag{C.7}$$

Here $E_4$ and $E_6$ denote the Eisenstein series of degree four and six, while the cusp forms are given as follows

$$\chi_{10} = -\frac{43867}{2^{12}3^5 5^2 7\cdot 53}\big(E_4E_6 - E_{10}\big), \tag{C.8}$$
$$\chi_{12} = -\frac{131\cdot 593}{2^{13}3^7 5^3 7^2 337}\big(3^2 7^2 E_4^3 + 2\cdot 5^3 E_6^2 - 691 E_{12}\big). \tag{C.9}$$

Note that $\chi_{10}$ as defined by Igusa differs from the Maas lift $\phi_{10}^{(2)}$ considered in section 9 by a factor of four. $R$ denotes a polynomial in $E_4, E_6, \chi_{10}, \chi_{12}$ whose explicit form can be found in [Igu67].

**Fourier Expansion of Eisenstein series of genus two.** In this subsection we discuss how to compute the Fourier coefficients of Eisenstein series. We start by recalling some mathematics terminology.

- Let $d$ be a square-free integer and consider the field extension $K = \mathbb{Q}[\sqrt{d}]$ of the rational numbers. The discriminant of $K$ is given as

$$\Delta_K = \begin{cases} d & \text{if } d \equiv 1 \quad \text{modulo } 4 \\ 4d & \text{if } d \equiv 2,3 \text{ modulo } 4 \end{cases} \tag{C.10}$$

- The Möbius function $\mu : \mathbb{N} \longrightarrow \{-1, 0, 1\}$ is defined as follows

$$\mu(n) = \begin{cases} 1 \text{ if } n \text{ is a square-free, positive integer with an even number of prime factors} \\ -1 \text{ if } n \text{ is a square-free, positive integer with an odd number of prime factors} \\ 0 \text{ if } n \text{ has a squared prime factor} \end{cases} \tag{C.11}$$

- The divisor function $\sigma_k(n)$ with $k \in \mathbb{C}$ is defined as

$$\sigma_k(n) = \sum_{d|n} d^k \tag{C.12}$$

- A Dirichlet character is a function $\chi : \mathbb{Z} \longrightarrow \mathbb{C}$ with the following properties
  1. There is a $k \in \mathbb{Z}$, such that $\chi(n) = \chi(n+k)$ for all $n$
  2. $\chi(n) \begin{cases} = 0 \text{ if } \gcd(n,k) > 1 \\ \neq 0 \text{ if } \gcd(n,k) = 1 \end{cases}$
  3. $\chi(mn) = \chi(m)\chi(n)$

  In particular, a Dirichlet character $\chi$ is a group homomorphism $\left(\mathbb{Z}/(k\mathbb{Z})\right)^* \longrightarrow \mathbb{C}^*$. Vice versa, any character of the unit group of $\mathbb{Z}/(k\mathbb{Z})$ extends to a Dirichlet character by setting $\chi(n) = 0$ for $n \notin \left(\mathbb{Z}/(k\mathbb{Z})\right)^*$

- A Dirichlet $L$-function associated to a Dirichlet character is given by

$$L(s,\chi) = \sum_{n=1}^{\infty} \frac{\chi(n)}{n^s}, \quad \mathrm{Re}(s) > 1 \tag{C.13}$$

- Given a prime $p$, the Kroneckersymbol $\left(\frac{a}{p}\right)$ is for $a \in \mathbb{Z}$ defined as

$$\left(\frac{a}{p}\right) = \begin{cases} 1, \text{ if } a \text{ is a quadratic rest modulo } n \\ -1, \text{ if } a \text{ is not a quadratic rest modulo } n \\ 0, \text{ if } a \equiv 0 \text{ modulo } n \end{cases} \tag{C.14}$$

  For general $n \in \mathbb{N}$ with prime factorization $n = p_1^{\nu_1} \dots p_k^{\nu_k}$ one puts $\left(\frac{a}{n}\right) = \left(\frac{a}{p_1}\right)^{\nu_1} \dots \left(\frac{a}{p_k}\right)^{\nu_k}$. Note that $\left(\frac{\cdot}{n}\right)$ is a Dirichlet character modulo $n$.

**Theorem.** *(E. g. [EZ85]) Let $E_w$ be an Eisenstein series of weight $w$, $T = \begin{pmatrix} a & b/2 \\ b/2 & c \end{pmatrix} \in \frac{1}{2}\mathbb{Z}^{2\times 2}$ be positive semi-definite. Denote $D = b^2 - 4ac \leq 0$ and let $D_0$ be the discriminant of $\mathbb{Q}(\sqrt{D})$. Then the Fourier coefficient $a(T)$ is one, if $a = b = c = 0$ and*

$$\frac{-2w}{B_w} \sum_{d|gcd(a,b,c)} d^{w-1} \alpha(D/d^2) \tag{C.15}$$

*otherwise. Here $B_k$ denotes the kth Bernoulli number, $\alpha$ is defined by $\alpha(0) = 1$ and*

$$\alpha(D) = \frac{1}{\zeta(3-2w)} C(w-1, D). \tag{C.16}$$

*Here the Cohen function $C$ is defined by*

$$C(s-1, D) = L_{D_0}(2-s) \sum_{d|f} \mu(d) \left(\frac{D_0}{d}\right) d^{s-2} \sigma_{2s-3}(f/d), \qquad D = D_0 f^2. \tag{C.17}$$

*In this expression, $\zeta$ denotes the Dedekind zeta-function and $L_{D_0}$ is the Dirichlet L-series associated to the character $\left(\frac{\cdot}{D_0}\right)$.*

In the following we list the explicit Fourier expansions of the above generators

$$
\begin{aligned}
E_4(q_1,q_2,r) &= 1+240q_1+240q_2+2160q_1^2+30240q_1q_2+2160q_2^2+240\frac{q_1q_2}{r^2}+13440\frac{q_1q_2}{r} \\
&\quad +13440q_1q_2r+240q_1q_2r^2+6720q_1^3+181440q_1^2+181440q_1q_2^2+6720q_2^3 \\
&\quad +30240\frac{q_1^2q_2}{r^2}+30240\frac{q_1q_2^2}{r^2}+138240\frac{q_1^2q_2}{r}+138240\frac{q_1q_2^2}{r}+138240q_1^2q_2r \\
&\quad +138240q_1q_2^2r+30240q_1^2q_2r^2+30240q_1q_2^2r^2 \\
E_6(q_1,q_2,r) &= 1-504q_1-504q_2-16632q_1^2+166320q_1q_2-16632q_2^2-504\frac{q_1q_2}{r^2}+44352\frac{q_1q_2}{r} \\
&\quad +44352q_1q_2r-504q_1q_2r^2-122976q_1^3+3792096q_1^2q_2+3792096q_1q_2^2-122976q_2^3 \\
&\quad +166320\frac{q_1^2q_2}{r^2}+166320\frac{q_1q_2^2}{r^2}+2128896\frac{q_1^2q_2}{r}+2128896\frac{q_1q_2^2}{r}+2128896q_1^2q_2r \\
&\quad +2128896q_1q_2^2r+166320q_1^2q_2r^2+166320q_1q_2^2r^2 \\
\chi_{10}(q_1,q_2,r) &= \frac{1}{2}q_1q_2-9q_1^2q_2-9q_1q_2^2+\frac{q_1^2q_2}{2r^2}+\frac{q_1q_2^2}{2r^2}-\frac{q_1q_2}{4r}+\frac{4q_1^2q_2}{r}+\frac{4q_1q_2^2}{r}-\frac{1}{4}q_1q_2r \\
&\quad +4q_1^2q_2r+4q_1q_2^2r+\frac{1}{2}q_1^2q_2r^2+\frac{1}{2}q_1q_2^2r^2 \\
\chi_{12}(q_1,q_2,r) &= \frac{5}{6}q_1q_2-11q_1^2q_2-11q_1q_2^2+\frac{5}{6}\frac{q_1^2q_2}{r^2}+\frac{5}{6}\frac{q_1q_2^2}{r^2}+\frac{q_1q_2}{12r}-\frac{22}{3}\frac{q_1^2q_2}{r}-\frac{22}{3}\frac{q_1q_2^2}{r} \\
&\quad +\frac{1}{12}q_1q_2r-\frac{22}{3}q_1^2q_2r-\frac{22}{3}q_1q_2^2r+\frac{5}{6}q_1^2q_2r^2+\frac{5}{6}q_1q_2^2r^2
\end{aligned}
$$

$$
\begin{aligned}
x_1(q_1,q_2,r) &= 9+10368q_1+10368q_2+7651584q_1^2+16505856q_1q_2+7651584q_2^2+3456\frac{q_1^2}{r^2} \\
&\quad +10368\frac{q_1q_2}{r^2}+3456\frac{q_2^2}{r^2}-216\frac{q_1}{r}-238680\frac{q_1^2}{r}-216\frac{q_2}{r}-508032\frac{q_1q_2}{r}-238680\frac{q_2^2}{r} \\
&\quad -216r-238680q_1r-174078936q_1^2r-238680q_2r-382465152q_1q_2r \\
&\quad -174078936q_2^2r+3456r^2+4188672q_1r^2+3277494144q_1^2r^2+4188672q_2r^2 \\
&\quad +7253463168q_1q_2r^2+3277494144q_2^2r^2 \\
x_2(q_1,q_2,r) &= -27-34992q_1-34992q_2-25824096q_1^2-22931424q_1q_2-25824096q_2^2 \\
&\quad -21384\frac{q_1^2}{r^2}-34992\frac{q_1q_2}{r^2}-21384\frac{q_2^2}{r^2}+972\frac{q_1}{r}+910764\frac{q_1^2}{r}+972\frac{q_2}{r}+653184\frac{q_1q_2}{r} \\
&\quad +910764\frac{q_2^2}{r}+972r+910764q_1r+635595660q_1^2r+910764q_2r+838781568q_1q_2r \\
&\quad +635595660q_2^2r-21384r^2-20015424q_1r^2-14164643016q_1^2r^2-20015424q_2r^2 \\
&\quad -23598639792q_1q_2r^2-14164643016q_2^2r^2 \\
x_3(q_1,q_2,r) &= -\frac{753}{2}q_1q_2+\frac{243}{4}\frac{q_1q_2}{r}+\frac{525123}{4}q_1q_2r-3411720q_1q_2r^2
\end{aligned}
$$

# D | Refined GV invariants for $\mathbb{C}^3/\mathbb{Z}_5$

| $d_2$ \ $d_1$ | 0 | 1 | 2 | 3 | 4 | 5 | 6 | 7 | 8 | 9 | 10 |
|---|---|---|---|---|---|---|---|---|---|---|---|
| 0 | | 3 | -6 | 27 | -192 | 1695 | -17064 | 188454 | -2228160 | 27748899 | -360012150 |
| 1 | -2 | 4 | -10 | 64 | -572 | 6076 | -71740 | 909760 | -12146622 | 168604540 | -2412582616 |
| 2 | 0 | 3 | -12 | 91 | -980 | 12259 | -166720 | 2394779 | -35737460 | 548460000 | -8599208436 |
| 3 | 0 | 5 | -12 | 108 | -1332 | 18912 | -289440 | 4632120 | -76306398 | 1282295808 | -21860004816 |
| 4 | 0 | 7 | -24 | 150 | -1808 | 26983 | -443394 | 7665776 | -136440800 | 2471539911 | -45269668626 |
| 5 | 0 | 9 | -56 | 294 | -2982 | 42005 | -689520 | 12254816 | -227540162 | 4331108122 | -83626566000 |
| 6 | 0 | 11 | -140 | 675 | -5992 | 76608 | -1192644 | 20764870 | -386343036 | 7482057534 | -148451081248 |
| 7 | 0 | 13 | -324 | 1738 | -13550 | 158814 | -2322056 | 38750866 | -703362386 | 13488597425 | -268229722780 |
| 8 | 0 | 15 | -686 | 4732 | -33552 | 359898 | -4954570 | 79050699 | -1387505216 | 25992283043 | -509660731360 |
| 9 | 0 | 17 | -1328 | 12960 | -88746 | 874588 | -11327904 | 172924796 | -2932945300 | 53475853968 | -1026989105240 |
| 10 | 0 | 19 | -2394 | 34357 | -245520 | 2245125 | -27363700 | 399648535 | -6552913216 | 116272411761 | -2183695179370 |

Table 2: GV invariants at genus (0,0)

| $d_2$ \ $d_1$ | 0 | 1 | 2 | 3 | 4 | 5 | 6 | 7 | 8 | 9 | 10 |
|---|---|---|---|---|---|---|---|---|---|---|---|
| 0 | | 0 | 0 | -10 | 231 | -4452 | 80948 | -1438086 | 25301295 | -443384578 | 7760515332 |
| 1 | 0 | 0 | 0 | -18 | 576 | -13968 | 305244 | -6329628 | 127275876 | -2508961104 | 48786866820 |
| 2 | 0 | 0 | 0 | -24 | 896 | -25636 | 650852 | -15418734 | 349139480 | -7658224250 | 164057415432 |
| 3 | 0 | 0 | 0 | -28 | 1152 | -37032 | 1056780 | -27964428 | 701652588 | -16919331450 | 395589996732 |
| 4 | 0 | 0 | 0 | -30 | 1407 | -48966 | 1515448 | -43561508 | 1185905652 | -30938142568 | 779767395056 |
| 5 | 0 | 0 | 9 | -66 | 2061 | -68908 | 2174157 | -65084016 | 1863846681 | -51392001108 | 1371429906120 |
| 6 | 0 | 0 | 68 | -280 | 4500 | -119124 | 3489856 | -102704154 | 2969225052 | -83868926462 | 2312642950872 |
| 7 | 0 | 0 | 300 | -1410 | 13413 | -261576 | 6617379 | -181806634 | 5100476481 | -142921977728 | 3965155214627 |
| 8 | 0 | 0 | 988 | -6760 | 48183 | -695664 | 14702120 | -365286402 | 9681953781 | -262834540958 | 7179092104476 |
| 9 | 0 | 0 | 2698 | -29360 | 187770 | -2131298 | 37329793 | -822940764 | 20261551070 | -525109967206 | 13926533807541 |
| 10 | 0 | 0 | 6444 | -113186 | 751019 | -7150716 | 105558998 | -2048434992 | 46290047925 | -1133487473126 | 28930878789904 |

Table 3: GV invariants at genus (1,0)

| $d_2$ \ $d_1$ | 0 | 1 | 2 | 3 | 4 | 5 | 6 | 7 | 8 | 9 | 10 |
|---|---|---|---|---|---|---|---|---|---|---|---|
| 0 | | -4 | 35 | -386 | 5161 | -74368 | 1117672 | -17319898 | 274571953 | -4428464750 | 72399258543 |
| 1 | 1 | -4 | 45 | -750 | 13174 | -235148 | 4227874 | -76326692 | 1381543835 | -25051915148 | 454880537144 |
| 2 | 0 | -4 | 46 | -900 | 19554 | -420472 | 8861756 | -183661746 | 3754800426 | -75910228250 | 1520587166090 |
| 3 | 0 | -20 | 46 | -944 | 23394 | -578872 | 13923300 | -325336476 | 7413730499 | -165426093894 | 3626709524196 |
| 4 | 0 | -56 | 164 | -1370 | 29417 | -750734 | 19452681 | -494871808 | 12281325148 | -297499294404 | 7050658845633 |
| 5 | 0 | -120 | 643 | -3602 | 51118 | -1121972 | 28252291 | -735181136 | 19077844392 | -487791469432 | 12245596115880 |
| 6 | 0 | -220 | 2522 | -11456 | 121392 | -2132580 | 47798426 | -1186598986 | 30575571450 | -794192203582 | 20520959813816 |
| 7 | 0 | -364 | 8526 | -41314 | 340762 | -4920912 | 95935665 | -2184901598 | 53716745464 | -1367832764572 | 35294081485451 |
| 8 | 0 | -560 | 24835 | -154752 | 1078545 | -12985696 | 220762885 | -4561068642 | 105019097003 | -2565242151798 | 64643620048364 |
| 9 | 0 | -816 | 63278 | -566320 | 3665611 | -37876670 | 564821327 | -10560913916 | 226102977980 | -5244562425062 | 127513019275063 |
| 10 | 0 | -1140 | 145113 | -1945582 | 12930429 | -118326784 | 1568446692 | -26605876980 | 528222136235 | -11577646359018 | 269921674605741 |

Table 4: GV invariants at genus (0,1)

| $d_2$ \ $d_1$ | 0 | 1 | 2 | 3 | 4 | 5 | 6 | 7 | 8 | 9 | 10 |
|---|---|---|---|---|---|---|---|---|---|---|---|
| 0 | | 0 | 0 | 0 | -102 | 5430 | -194022 | 5784837 | -155322234 | 3894455457 | -93050366010 |
| 1 | 0 | 0 | 0 | 0 | -216 | 15012 | -659088 | 23286132 | -722619936 | 20558391372 | -549482611284 |
| 2 | 0 | 0 | 0 | 0 | -312 | 25402 | -1299948 | 52772850 | -1855049556 | 59041090140 | -1746828786552 |
| 3 | 0 | 0 | 0 | 0 | -390 | 34912 | -1994490 | 90371068 | -3526024576 | 123699660448 | -4006116067574 |
| 4 | 0 | 0 | 0 | 0 | -450 | 43803 | -2714134 | 133676408 | -5667402346 | 215531532522 | -7541030566234 |
| 5 | 0 | 0 | 0 | 0 | -564 | 56340 | -3636050 | 188364003 | -8450098622 | 341039465436 | -12674180528160 |
| 6 | 0 | 0 | -12 | 33 | -1302 | 89949 | -5397344 | 277712616 | -12685944558 | 527916004781 | -20372727166960 |
| 7 | 0 | 0 | -116 | 471 | -5616 | 205699 | -9811006 | 463444751 | -20541312540 | 851408769126 | -33210201574872 |
| 8 | 0 | 0 | -628 | 4498 | -31902 | 647792 | -22397190 | 906948972 | -37292448660 | 1490395858323 | -57261471814196 |
| 9 | 0 | 0 | -2488 | 32985 | -194232 | 2563379 | -62839000 | 2080200874 | -76550883550 | 2874492150421 | -106583214770442 |
| 10 | 0 | 0 | -8036 | 190910 | -1170468 | 11563235 | -209199272 | 5533285795 | -177007441804 | 6108659068296 | -215066543042220 |

Table 5: GV invariants at genus (2,0)

| $d_2$ \ $d_1$ | 0 | 1 | 2 | 3 | 4 | 5 | 6 | 7 | 8 | 9 | 10 |
|---|---|---|---|---|---|---|---|---|---|---|---|
| 0 | | 0 | 0 | 165 | -7448 | 239030 | -6574236 | 165498745 | -3934550232 | 89865837855 | -1992688534060 |
| 1 | 0 | 0 | 0 | 249 | -16032 | 663816 | -22351692 | 665886174 | -18286464756 | 473803348152 | -11751841988652 |
| 2 | 0 | 0 | 0 | 284 | -21788 | 1085422 | -43131672 | 1486268819 | -46417626916 | 1348910501365 | -37100685165104 |
| 3 | 0 | 0 | 0 | 294 | -24852 | 1407460 | -63647560 | 2476239454 | -86461707148 | 2782756496277 | -84053632394048 |
| 4 | 0 | 0 | 0 | 295 | -27592 | 1688645 | -83386604 | 3553493834 | -135672322024 | 4756388163712 | -155776855706696 |
| 5 | 0 | 0 | -120 | 865 | -40020 | 2221550 | -110664544 | 4919143810 | -198534576552 | 7396590575826 | -257815753553060 |
| 6 | 0 | 0 | -1484 | 5370 | -101208 | 3849720 | -169488448 | 7298405957 | -296633814860 | 11342997342789 | -409933588199632 |
| 7 | 0 | 0 | -9632 | 39827 | -381244 | 9203294 | -321797124 | 12472772327 | -484935911712 | 18313654586964 | -665820277600268 |
| 8 | 0 | 0 | -43732 | 268944 | -1812560 | 28546696 | -754793300 | 25028162865 | -894682148776 | 32327818731481 | -1151304488157956 |
| 9 | 0 | 0 | -157232 | 1579910 | -9323548 | 107088059 | -2132204912 | 58508571346 | -1866608513100 | 63045566722361 | -2156999365698592 |
| 10 | 0 | 0 | -477892 | 7945885 | -48404096 | 447651280 | -6998081720 | 157304121266 | -4377953683428 | 135534530407883 | -4387648213279700 |

Table 6: GV invariants at genus (1,1)

| $d_2$ \ $d_1$ | 0 | 1 | 2 | 3 | 4 | 5 | 6 | 7 | 8 | 9 | 10 |
|---|---|---|---|---|---|---|---|---|---|---|---|
| 0 | | 1 | -56 | 1710 | -45478 | 1108429 | -25510644 | 565560106 | -12208303534 | 258259176252 | -5377587417300 |
| 1 | 0 | 1 | -62 | 2823 | -101210 | 3123131 | -87398508 | 2285563966 | -56886814500 | 1363692509267 | -31741367003420 |
| 2 | 0 | 1 | -62 | 3083 | -135046 | 5051631 | -167324604 | 5070030051 | -143679833838 | 3866332989620 | -99855731010794 |
| 3 | 0 | 21 | -62 | 3117 | -150012 | 6401972 | -242529558 | 8329863808 | -264680950084 | 7904947970683 | -224565445050932 |
| 4 | 0 | 126 | -370 | 5015 | -187166 | 7888757 | -317686030 | 11857695191 | -411277565110 | 13383285674817 | -412600220588284 |
| 5 | 0 | 462 | -2332 | 17188 | -364870 | 12034886 | -449182560 | 16816027403 | -605858293698 | 20790005285269 | -679898745656230 |
| 6 | 0 | 1287 | -14442 | 70155 | -1022430 | 25116729 | -783594714 | 26879295676 | -941310618136 | 32492448213165 | -1089619106809056 |
| 7 | 0 | 3003 | -72430 | 338223 | -3391610 | 65165109 | -1690999402 | 51032170076 | -1653086126192 | 54861619097005 | -1817288123911204 |
| 8 | 0 | 6188 | -293502 | 1718392 | -13023266 | 195187516 | -4260040764 | 112884832140 | -3323304791332 | 103316703343033 | -3287506711857818 |
| 9 | 0 | 11628 | -991856 | 8413737 | -54852288 | 656650202 | -12066214776 | 281352005882 | -7507672419570 | 216602339119797 | -6526511388596102 |
| 10 | 0 | 20349 | -2908410 | 37685481 | -242056852 | 2402813660 | -37503651456 | 771051279956 | -18655847273504 | 498243332814881 | -14133822653295814 |

Table 7: GV invariants at genus (0,2)

| $d_2$ \ $d_1$ | 0 | 1 | 2 | 3 | 4 | 5 | 6 | 7 | 8 | 9 | 10 |
|---|---|---|---|---|---|---|---|---|---|---|---|
| 0 | | 0 | 0 | 0 | 15 | -3672 | 290853 | -15363990 | 649358826 | -23769907110 | 786400843911 |
| 1 | 0 | 0 | 0 | 0 | 28 | -9038 | 895004 | -56764044 | 2800998468 | -117271264390 | 4368276685200 |
| 2 | 0 | 0 | 0 | 0 | 39 | -14318 | 1647535 | -120373298 | 6755811733 | -317790113114 | 13156914112074 |
| 3 | 0 | 0 | 0 | 0 | 48 | -18996 | 2412872 | -196006904 | 12206895204 | -633769608952 | 28779314566504 |
| 4 | 0 | 0 | 0 | 0 | 55 | -23102 | 3156548 | -277709398 | 18773083610 | -1056971776506 | 51911426998599 |
| 5 | 0 | 0 | 0 | 0 | 60 | -27784 | 4008354 | -372807278 | 26742919550 | -1601413604094 | 83702954155532 |
| 6 | 0 | 0 | 0 | 0 | 140 | -40296 | 5522988 | -516972356 | 38066404131 | -2363089642518 | 128764263234726 |
| 7 | 0 | 0 | 15 | -56 | 1090 | -94614 | 9521920 | -811965094 | 58217079594 | -3617744837466 | 200176789418018 |
| 8 | 0 | 0 | 176 | -1538 | 11205 | -358426 | 22162654 | -1538663154 | 100876901423 | -6029485692286 | 329118801193692 |
| 9 | 0 | 0 | 1130 | -21898 | 117201 | -1881714 | 68569525 | -3571773320 | 202306292870 | -11205365856114 | 587561193934288 |
| 10 | 0 | 0 | 5232 | -201110 | 1135775 | -11788422 | 270108872 | -10102578488 | 470971719089 | -23362479231458 | 1149433429910848 |

Table 8: GV invariants at genus (3,0)

| $d_2$ \ $d_1$ | 0 | 1 | 2 | 3 | 4 | 5 | 6 | 7 | 8 | 9 | 10 |
|---|---|---|---|---|---|---|---|---|---|---|---|
| 0 | | 0 | 0 | 0 | 3646 | -332500 | 18307188 | -784000382 | 28746178815 | -947455553070 | 28883283581120 |
| 1 | 0 | 0 | 0 | 0 | 6736 | -817658 | 56229716 | -2890271136 | 123724696980 | -4664622899914 | 160128325772898 |
| 2 | 0 | 0 | 0 | 0 | 8591 | -1240914 | 100812817 | -6020174066 | 294558138571 | -12516025972770 | 478543796677390 |
| 3 | 0 | 0 | 0 | 0 | 9595 | -1541556 | 141315405 | -9507560084 | 520483728764 | -24542285301304 | 1033022927270955 |
| 4 | 0 | 0 | 0 | 0 | 10060 | -1762130 | 176472063 | -13005635874 | 779144061743 | -40074259816370 | 1832137941830936 |
| 5 | 0 | 0 | 0 | 0 | 11942 | -2095360 | 218310699 | -16998820726 | 1083695909005 | -59481598311246 | 2903077344820940 |
| 6 | 0 | 0 | 286 | -704 | 31327 | -3252496 | 304703564 | -23423999308 | 1522934680590 | -86512345468570 | 4402967053374246 |
| 7 | 0 | 0 | 4101 | -14832 | 173458 | -7932290 | 542315279 | -37210408886 | 2328821117728 | -131676086618278 | 6787311660887958 |
| 8 | 0 | 0 | 30706 | -201910 | 1335968 | -29094066 | 1282419545 | -71414914086 | 4056835940421 | -219450368091250 | 11121462357918974 |
| 9 | 0 | 0 | 160494 | -2029890 | 10923049 | -143171810 | 3973169453 | -167074649216 | 8171560102165 | -408274358839182 | 19828985332878797 |
| 10 | 0 | 0 | 660490 | -15457318 | 86650077 | -819422430 | 15457087268 | -475302472716 | 19096999811219 | -851813505320610 | 38743456874318870 |

Table 9: GV invariants at genus (2,1)

| $d_2$ \ $d_1$ | 0 | 1 | 2 | 3 | 4 | 5 | 6 | 7 | 8 | 9 | 10 |
|---|---|---|---|---|---|---|---|---|---|---|---|
| 0 | | 0 | 0 | -792 | 74640 | -4150308 | 177499683 | -6464352018 | 211097449143 | -6368926838778 | 180940547430291 |
| 1 | 0 | 0 | 0 | -1044 | 141232 | -10305512 | 547583663 | -23880243270 | 909350539113 | -31362711810712 | 1002969680533521 |
| 2 | 0 | 0 | 0 | -1100 | 173984 | -15317486 | 968113645 | -49239570774 | 2148320040446 | -83640362553570 | 2982529054781422 |
| 3 | 0 | 0 | 0 | -1106 | 185762 | -18382200 | 1322846349 | -76293814476 | 3741031288415 | -162136840377466 | 6379219588997229 |
| 4 | 0 | 0 | 0 | -1106 | 200166 | -20872804 | 1625640316 | -102612370488 | 5515121845123 | -261276623451526 | 11187678438452042 |
| 5 | 0 | 0 | 462 | -4208 | 308575 | -27242952 | 2075005637 | -134967532118 | 7642082210213 | -384898656237490 | 17575381337394925 |
| 6 | 0 | 0 | 9757 | -34468 | 914095 | -50492366 | 3204053300 | -195554839650 | 10983999127237 | -564354092549442 | 26674496909199694 |
| 7 | 0 | 0 | 95115 | -363608 | 4096989 | -134471610 | 6429657554 | -338709281142 | 17714458393384 | -885389528147694 | 41780584233048109 |
| 8 | 0 | 0 | 602395 | -3442542 | 24297636 | -473005618 | 16339619628 | -709931460036 | 33124639634829 | -1550574847245638 | 70711217658486539 |
| 9 | 0 | 0 | 2872686 | -27531174 | 159003934 | -2069571994 | 50880368591 | -1767456288994 | 71569396226022 | -3060796330213490 | 131835980366899491 |
| 10 | 0 | 0 | 11166111 | -182163160 | 1053629099 | -10307057780 | 187616361586 | -5139288077960 | 176685724431350 | -6773022816685924 | 271069267170147567 |

Table 10: GV invariants at genus (1,2)

| $d_2$ \ $d_1$ | 0 | 1 | 2 | 3 | 4 | 5 | 6 | 7 | 8 | 9 | 10 |
|---|---|---|---|---|---|---|---|---|---|---|---|
| 0 | | 0 | 36 | -3552 | 199129 | -8477200 | 305202621 | -9831079874 | 292451827680 | -8193427901442 | 219053227327973 |
| 1 | 0 | 0 | 37 | -5132 | 391895 | -21456544 | 951816717 | -36559985102 | 1265167969027 | -40459733831784 | 1216486396522567 |
| 2 | 0 | 0 | 37 | -5332 | 482505 | -31872458 | 1680167027 | -75235065398 | 2982734988243 | -107682958186546 | 3610552800188045 |
| 3 | 0 | -8 | 37 | -5342 | 512892 | -37960032 | 2277288722 | -115697208032 | 5159487585801 | -207532502054646 | 7683566995242455 |
| 4 | 0 | -120 | 403 | -9694 | 650547 | -45492860 | 2845195737 | -155997880696 | 7584685164224 | -332849815705118 | 13405043007333180 |
| 5 | 0 | -792 | 4032 | -44232 | 1456988 | -72718424 | 4003499489 | -214623624454 | 10718340308348 | -493969134826666 | 21087866983228178 |
| 6 | 0 | -3432 | 40032 | -227316 | 4852264 | -169805898 | 7373281571 | -346895883190 | 16401686036909 | -749687050494070 | 32594365559442838 |
| 7 | 0 | -11440 | 303420 | -1430960 | 18827088 | -497163382 | 17348476209 | -692265842390 | 29328965878560 | -1259736657853762 | 53323814192441639 |
| 8 | 0 | -31824 | 1731540 | -9791434 | 86102063 | -1681294978 | 48005169030 | -1640618797164 | 61576392616975 | -2420210604499130 | 96594746682174718 |
| 9 | 0 | -77520 | 7822110 | -64355702 | 441639078 | -6453703762 | 150166863631 | -4417828314310 | 147334819728793 | -5270128198705030 | 195620751168796920 |
| 10 | 0 | -170544 | 29481458 | -378136780 | 2411472333 | -27268753568 | 519116905858 | -13161855037508 | 390992623491806 | -12736589773157528 | 438155527249781100 |

Table 11: GV invariants at genus (0,3)

# E | Refined GV invariants for $\mathbb{C}^3/\mathbb{Z}_6$

GV invariants at genus (0,0):

| **0** | 0 | 1 | 2 | 3 | 4 | 5 | 6 | 7 | **1** | 0 | 1 | 2 | 3 | 4 | 5 | 6 | 7 |
|---|---|---|---|---|---|---|---|---|---|---|---|---|---|---|---|---|---|
| 0 | | 0 | 0 | 0 | 0 | 0 | 0 | 0 | 0 | -2 | 0 | 0 | 0 | 0 | 0 | 0 | 0 |
| 1 | -2 | -2 | 0 | 0 | 0 | 0 | 0 | 0 | 1 | -2 | -2 | 0 | 0 | 0 | 0 | 0 | 0 |
| 2 | 0 | -4 | 0 | 0 | 0 | 0 | 0 | 0 | 2 | 0 | -6 | 0 | 0 | 0 | 0 | 0 | 0 |
| 3 | 0 | -6 | -6 | 0 | 0 | 0 | 0 | 0 | 3 | 0 | -10 | -10 | 0 | 0 | 0 | 0 | 0 |
| 4 | 0 | -8 | -32 | -8 | 0 | 0 | 0 | 0 | 4 | 0 | -14 | -70 | -14 | 0 | 0 | 0 | 0 |
| 5 | 0 | -10 | -110 | -110 | -10 | 0 | 0 | 0 | 5 | 0 | -18 | -270 | -270 | -18 | 0 | 0 | 0 |
| 6 | 0 | -12 | -288 | -756 | -288 | -12 | 0 | 0 | 6 | 0 | -22 | -770 | -2200 | -770 | -22 | 0 | 0 |
| 7 | 0 | -14 | -644 | -3556 | -3556 | -644 | -14 | 0 | 7 | 0 | -26 | -1820 | -11544 | -11544 | -1820 | -26 | 0 |
| **2** | 0 | 1 | 2 | 3 | 4 | 5 | 6 | 7 | **3** | 0 | 1 | 2 | 3 | 4 | 5 | 6 | 7 |
| 0 | 0 | 0 | 0 | 0 | 0 | 0 | 0 | 0 | 0 | 0 | 0 | 0 | 0 | 0 | 0 | 0 | 0 |
| 1 | 0 | 0 | 0 | 0 | 0 | 0 | 0 | 0 | 1 | 0 | 0 | 0 | 0 | 0 | 0 | 0 | 0 |
| 2 | 0 | -6 | 0 | 0 | 0 | 0 | 0 | 0 | 2 | 0 | -4 | 0 | 0 | 0 | 0 | 0 | 0 |
| 3 | 0 | -12 | -12 | 0 | 0 | 0 | 0 | 0 | 3 | 0 | -12 | -12 | 0 | 0 | 0 | 0 | 0 |
| 4 | 0 | -18 | -96 | -18 | 0 | 0 | 0 | 0 | 4 | 0 | -20 | -110 | -20 | 0 | 0 | 0 | 0 |
| 5 | 0 | -24 | -416 | -416 | -24 | 0 | 0 | 0 | 5 | 0 | -28 | -518 | -518 | -28 | 0 | 0 | 0 |
| 6 | 0 | -30 | -1280 | -3850 | -1280 | -30 | 0 | 0 | 6 | 0 | -36 | -1710 | -5292 | -1710 | -36 | 0 | 0 |
| 7 | 0 | -36 | -3204 | -22176 | -22176 | -3204 | -36 | 0 | 7 | 0 | -44 | -4510 | -33000 | -33000 | -4510 | -44 | 0 |
| **4** | 0 | 1 | 2 | 3 | 4 | 5 | 6 | 7 | **5** | 0 | 1 | 2 | 3 | 4 | 5 | 6 | 7 |
| 0 | 0 | 0 | 0 | 0 | 0 | 0 | 0 | 0 | 0 | 0 | 0 | 0 | 0 | 0 | 0 | 0 | 0 |
| 1 | 0 | 0 | 0 | 0 | 0 | 0 | 0 | 0 | 1 | 0 | 0 | 0 | 0 | 0 | 0 | 0 | 0 |
| 2 | 0 | -6 | 0 | 0 | 0 | 0 | 0 | 0 | 2 | 0 | -8 | 0 | 0 | 0 | 0 | 0 | 0 |
| 3 | 0 | -10 | -10 | 0 | 0 | 0 | 0 | 0 | 3 | 0 | -14 | -14 | 0 | 0 | 0 | 0 | 0 |
| 4 | 0 | -20 | -112 | -20 | 0 | 0 | 0 | 0 | 4 | 0 | -18 | -126 | -18 | 0 | 0 | 0 | 0 |
| 5 | 0 | -30 | -576 | -576 | -30 | 0 | 0 | 0 | 5 | 0 | -30 | -630 | -630 | -30 | 0 | 0 | 0 |
| 6 | 0 | -40 | -2016 | -6368 | -2016 | -40 | 0 | 0 | 6 | 0 | -42 | -2254 | -7308 | -2254 | -42 | 0 | 0 |
| 7 | 0 | -50 | -5580 | -42400 | -42400 | -5580 | -50 | 0 | 7 | 0 | -54 | -6426 | -50652 | -50652 | -6426 | -54 | 0 |

GV invariants at genus (0,1):

| **0** | 0 | 1 | 2 | 3 | 4 | 5 | 6 | 7 | **1** | 0 | 1 | 2 | 3 | 4 | 5 | 6 | 7 |
|---|---|---|---|---|---|---|---|---|---|---|---|---|---|---|---|---|---|
| 0 | | 0 | 0 | 0 | 0 | 0 | 0 | 0 | 0 | 0 | 0 | 0 | 0 | 0 | 0 | 0 | 0 |
| 1 | 0 | 0 | 0 | 0 | 0 | 0 | 0 | 0 | 1 | 0 | 0 | 0 | 0 | 0 | 0 | 0 | 0 |
| 2 | 0 | 0 | 0 | 0 | 0 | 0 | 0 | 0 | 2 | 0 | 0 | 0 | 0 | 0 | 0 | 0 | 0 |
| 3 | 0 | 0 | 0 | 0 | 0 | 0 | 0 | 0 | 3 | 0 | 0 | 0 | 0 | 0 | 0 | 0 | 0 |
| 4 | 0 | 0 | 9 | 0 | 0 | 0 | 0 | 0 | 4 | 0 | 0 | 16 | 0 | 0 | 0 | 0 | 0 |
| 5 | 0 | 0 | 68 | 68 | 0 | 0 | 0 | 0 | 5 | 0 | 0 | 144 | 144 | 0 | 0 | 0 | 0 |
| 6 | 0 | 0 | 300 | 1016 | 300 | 0 | 0 | 0 | 6 | 0 | 0 | 704 | 2608 | 704 | 0 | 0 | 0 |
| 7 | 0 | 0 | 988 | 7792 | 7792 | 988 | 0 | 0 | 7 | 0 | 0 | 2496 | 22672 | 22672 | 2496 | 0 | 0 |
| **2** | 0 | 1 | 2 | 3 | 4 | 5 | 6 | 7 | **3** | 0 | 1 | 2 | 3 | 4 | 5 | 6 | 7 |
| 0 | 0 | 0 | 0 | 0 | 0 | 0 | 0 | 0 | 0 | 0 | 0 | 0 | 0 | 0 | 0 | 0 | 0 |
| 1 | 0 | 0 | 0 | 0 | 0 | 0 | 0 | 0 | 1 | 0 | 0 | 0 | 0 | 0 | 0 | 0 | 0 |
| 2 | 0 | 0 | 0 | 0 | 0 | 0 | 0 | 0 | 2 | 0 | 0 | 0 | 0 | 0 | 0 | 0 | 0 |
| 3 | 0 | 0 | 0 | 0 | 0 | 0 | 0 | 0 | 3 | 0 | 0 | 0 | 0 | 0 | 0 | 0 | 0 |
| 4 | 0 | 0 | 21 | 0 | 0 | 0 | 0 | 0 | 4 | 0 | 0 | 24 | 0 | 0 | 0 | 0 | 0 |
| 5 | 0 | 0 | 204 | 204 | 0 | 0 | 0 | 0 | 5 | 0 | 0 | 248 | 248 | 0 | 0 | 0 | 0 |
| 6 | 0 | 0 | 1073 | 4184 | 1073 | 0 | 0 | 0 | 6 | 0 | 0 | 1368 | 5492 | 1368 | 0 | 0 | 0 |
| 7 | 0 | 0 | 4032 | 40032 | 40032 | 4032 | 0 | 0 | 7 | 0 | 0 | 5368 | 56344 | 56344 | 5368 | 0 | 0 |
| **4** | 0 | 1 | 2 | 3 | 4 | 5 | 6 | 7 | **5** | 0 | 1 | 2 | 3 | 4 | 5 | 6 | 7 |
| 0 | 0 | 0 | 0 | 0 | 0 | 0 | 0 | 0 | 0 | 0 | 0 | 0 | 0 | 0 | 0 | 0 | 0 |
| 1 | 0 | 0 | 0 | 0 | 0 | 0 | 0 | 0 | 1 | 0 | 0 | 0 | 0 | 0 | 0 | 0 | 0 |
| 2 | 0 | 0 | 0 | 0 | 0 | 0 | 0 | 0 | 2 | 0 | 0 | 0 | 0 | 0 | 0 | 0 | 0 |
| 3 | 0 | 0 | 0 | 0 | 0 | 0 | 0 | 0 | 3 | 0 | 0 | 0 | 0 | 0 | 0 | 0 | 0 |
| 4 | 0 | 0 | 25 | 0 | 0 | 0 | 0 | 0 | 4 | 0 | 0 | 24 | 0 | 0 | 0 | 0 | 0 |
| 5 | 0 | 0 | 276 | 276 | 0 | 0 | 0 | 0 | 5 | 0 | 0 | 288 | 288 | 0 | 0 | 0 | 0 |
| 6 | 0 | 0 | 1589 | 6484 | 1589 | 0 | 0 | 0 | 6 | 0 | 0 | 1736 | 7224 | 1736 | 0 | 0 | 0 |
| 7 | 0 | 0 | 6448 | 70100 | 70100 | 6448 | 0 | 0 | 7 | 0 | 0 | 7272 | 81432 | 81432 | 7272 | 0 | 0 |

GV invariants at genus (1,0):

| **0** | 0 | 1 | 2 | 3 | 4 | 5 | 6 | 7 | **1** | 0 | 1 | 2 | 3 | 4 | 5 | 6 | 7 |
|---|---|---|---|---|---|---|---|---|---|---|---|---|---|---|---|---|---|
| 0 | | 0 | 0 | 0 | 0 | 0 | 0 | 0 | 0 | 1 | 0 | 0 | 0 | 0 | 0 | 0 | 0 |
| 1 | 1 | 1 | 0 | 0 | 0 | 0 | 0 | 0 | 1 | 1 | 1 | 0 | 0 | 0 | 0 | 0 | 0 |
| 2 | 0 | 10 | 0 | 0 | 0 | 0 | 0 | 0 | 2 | 0 | 11 | 0 | 0 | 0 | 0 | 0 | 0 |
| 3 | 0 | 35 | 35 | 0 | 0 | 0 | 0 | 0 | 3 | 0 | 45 | 45 | 0 | 0 | 0 | 0 | 0 |
| 4 | 0 | 84 | 359 | 84 | 0 | 0 | 0 | 0 | 4 | 0 | 119 | 635 | 119 | 0 | 0 | 0 | 0 |
| 5 | 0 | 165 | 1987 | 1987 | 165 | 0 | 0 | 0 | 5 | 0 | 249 | 4095 | 4095 | 249 | 0 | 0 | 0 |
| 6 | 0 | 286 | 7620 | 20554 | 7620 | 286 | 0 | 0 | 6 | 0 | 451 | 17545 | 51556 | 17545 | 451 | 0 | 0 |
| 7 | 0 | 455 | 23414 | 134882 | 134882 | 23414 | 455 | 0 | 7 | 0 | 741 | 58110 | 384852 | 384852 | 58110 | 741 | 0 |
| **2** | 0 | 1 | 2 | 3 | 4 | 5 | 6 | 7 | **3** | 0 | 1 | 2 | 3 | 4 | 5 | 6 | 7 |
| 0 | 0 | 0 | 0 | 0 | 0 | 0 | 0 | 0 | 0 | 0 | 0 | 0 | 0 | 0 | 0 | 0 | 0 |
| 1 | 0 | 0 | 0 | 0 | 0 | 0 | 0 | 0 | 1 | 0 | 0 | 0 | 0 | 0 | 0 | 0 | 0 |
| 2 | 0 | 11 | 0 | 0 | 0 | 0 | 0 | 0 | 2 | 0 | 10 | 0 | 0 | 0 | 0 | 0 | 0 |
| 3 | 0 | 46 | 46 | 0 | 0 | 0 | 0 | 0 | 3 | 0 | 46 | 46 | 0 | 0 | 0 | 0 | 0 |
| 4 | 0 | 129 | 731 | 129 | 0 | 0 | 0 | 0 | 4 | 0 | 130 | 751 | 130 | 0 | 0 | 0 | 0 |
| 5 | 0 | 284 | 5380 | 5380 | 284 | 0 | 0 | 0 | 5 | 0 | 294 | 5891 | 5891 | 294 | 0 | 0 | 0 |
| 6 | 0 | 535 | 25327 | 78341 | 25327 | 535 | 0 | 0 | 6 | 0 | 570 | 29847 | 94930 | 29847 | 570 | 0 | 0 |
| 7 | 0 | 906 | 90234 | 652272 | 652272 | 90234 | 906 | 0 | 7 | 0 | 990 | 113047 | 863476 | 863476 | 113047 | 990 | 0 |
| **4** | 0 | 1 | 2 | 3 | 4 | 5 | 6 | 7 | **5** | 0 | 1 | 2 | 3 | 4 | 5 | 6 | 7 |
| 0 | 0 | 0 | 0 | 0 | 0 | 0 | 0 | 0 | 0 | 0 | 0 | 0 | 0 | 0 | 0 | 0 | 0 |
| 1 | 0 | 0 | 0 | 0 | 0 | 0 | 0 | 0 | 1 | 0 | 0 | 0 | 0 | 0 | 0 | 0 | 0 |
| 2 | 0 | 35 | 0 | 0 | 0 | 0 | 0 | 0 | 2 | 0 | 84 | 0 | 0 | 0 | 0 | 0 | 0 |
| 3 | 0 | 45 | 45 | 0 | 0 | 0 | 0 | 0 | 3 | 0 | 119 | 119 | 0 | 0 | 0 | 0 | 0 |
| 4 | 0 | 130 | 751 | 130 | 0 | 0 | 0 | 0 | 4 | 0 | 129 | 1023 | 129 | 0 | 0 | 0 | 0 |
| 5 | 0 | 295 | 6036 | 6036 | 295 | 0 | 0 | 0 | 5 | 0 | 295 | 6627 | 6627 | 295 | 0 | 0 | 0 |
| 6 | 0 | 580 | 31835 | 103132 | 31835 | 580 | 0 | 0 | 6 | 0 | 581 | 33551 | 111566 | 33551 | 581 | 0 | 0 |
| 7 | 0 | 1025 | 126302 | 1000100 | 1000100 | 126302 | 1025 | 0 | 7 | 0 | 1035 | 134325 | 1101510 | 1101510 | 134325 | 1035 | 0 |

Table 12: Refined GV invariants for $\mathbb{C}^3/\mathbb{Z}_6$. Degree $d_1$ goes from left to right, $d_3$ from top to bottom and $d_2$ is indexed in the upper-left corner in each table.

# F | Fourier coefficients of Siegel modular forms

| | | | | | |
|---|---|---|---|---|---|
| $(n,r,m)$ | $(0,0,0)$ | $(0,0,1)$ | $(0,0,2)$ | $(0,0,3)$ | $(0,0,4)$ |
| | $(1,1,1)$ | $(1,0,1)$ | $(1,1,2)$ | $(1,0,2)$ | |
| | $(1,1,3)$ | $(1,0,3)$ | $(1,1,4)$ | $(1,0,4)$ | |
| | $(2,2,2)$ | $(2,1,2)$ | $(2,0,2)$ | | |
| | $(2,2,3)$ | $(2,1,3)$ | $(2,0,3)$ | | |
| | $(2,2,4)$ | $(2,1,4)$ | $(2,0,4)$ | | |
| | $(3,3,3)$ | $(3,2,3)$ | $(3,1,3)$ | $(3,0,3)$ | |
| | $(3,3,4)$ | $(3,2,4)$ | $(3,1,4)$ | $(3,0,4)$ | |
| | $(4,4,4)$ | $(4,3,4)$ | $(4,2,4)$ | $(4,1,4)$ | $(4,0,4)$ |
| $E_4^{(2)}$ | 1 | 240 | 2160 | 6720 | 17520 |
| | 13440 | 30240 | 138240 | 181440 | |
| | 362880 | 497280 | 967680 | 997920 | |
| | 604800 | 967680 | 1239840 | | |
| | 1814400 | 2903040 | 2782080 | | |
| | 5114880 | 5806080 | 7439040 | | |
| | 3642240 | 5987520 | 6531840 | 8467200 | |
| | 10644480 | 13426560 | 17418240 | 15980160 | |
| | 20818560 | 24192000 | 35804160 | 34974720 | 41882400 |
| $E_6^{(2)}$ | 1 | -504 | -16632 | -122976 | -532728 |
| | 44352 | 166320 | 2128896 | 3792096 | |
| | 15422400 | 23462208 | 65995776 | 85322160 | |
| | 24881472 | 65995776 | 90644400 | | |
| | 234311616 | 453454848 | 530228160 | | |
| | 1126185984 | 1724405760 | 2066692320 | | |
| | 883802304 | 1945345248 | 2818924416 | 3327730560 | |
| | 4864527360 | 8158449600 | 11304437760 | 12013404480 | |
| | 12809611584 | 22751511552 | 34911765504 | 42077629440 | 46585733040 |
| $\chi_{10}$ | 0 | 0 | 0 | 0 | 0 |
| | 1 | -2 | -16 | 36 | |
| | 99 | -272 | -240 | 1056 | |
| | 240 | -240 | 32 | | |
| | -1800 | 2736 | -1464 | | |
| | 4352 | -6816 | -576 | | |
| | 15399 | -19008 | 27270 | -43920 | |
| | -6864 | -26928 | 44064 | 12544 | |
| | 135424 | -22000 | 65280 | -36432 | -279040 |
| $E_{12}^{(2)}$ | 53678953 | 5089790160 | 10428980037840 | 901646147263680 | 21358556207286480 |
| | 22266840960 | 456798756960 | 162868282536960 | 661522702800960 | |
| | 18728849326561920 | 46719773564929920 | 486707206711864320 | 957976883554934880 | |
| | 46765376055216000 | 486707206711864320 | 958912407409188960 | | |
| | 9975044773603728000 | 43296648289119912960 | 67657429630096225920 | | |
| | 341726484448551121920 | 994500257327275345920 | 1388669119242478032960 | | |
| | 232922965682969976960 | 1387314320747141666880 | 3553058504433847368960 | 4778384234988952224000 | |
| | 11078677991333226193920 | 39296440893457709619840 | 78584973005595527976960 | 97978466594063423571840 | |
| | 98074242084224505939840 | 409369370741219790720000 | 1021197866857079343344640 | 1703669378274682309370880 | 2010986989911829611208800 |
| $E_\Delta^{(2)}$ | 0 | 12096 | -290304 | 3048192 | -17805312 |
| | 158976 | 2146176 | 377726976 | 1558123776 | |
| | 44644753152 | 111108644352 | 1155817838592 | 2276348012928 | |
| | -3910809600 | -25315338240 | -54993613824 | | |
| | -210642882048 | -1097120692224 | -1622158990848 | | |
| | -10065668456448 | -28301296693248 | -39925363636224 | | |
| | 2356270891776 | 8269812764928 | 24778417466880 | 31936386585600 | |
| | 18898645911552 | 108976387336704 | 198828131659776 | 245736461090304 | |
| | -245453236125696 | -584737669785600 | -1744357612732416 | -2891211375900672 | -3257305227141120 |
| $\phi_{12}^{(2)}$ | 0 | 0 | 0 | 0 | 0 |
| | 1 | 10 | -88 | -132 | |
| | 1275 | 736 | -8040 | -2880 | |
| | 2784 | -8040 | 17600 | | |
| | 13080 | -14136 | -54120 | | |
| | -64768 | 389520 | -232320 | | |
| | 48303 | 38016 | -256410 | 1073520 | |
| | -806520 | 938400 | -1227600 | -2309120 | |
| | 3392512 | 2311640 | -5917440 | 6141960 | 15902720 |

Table 13: Fourier coefficients $c\big(f, \left(\begin{smallmatrix} n & r/2 \\ r/2 & m \end{smallmatrix}\right)\big)$ of some genus 2 Siegel modular forms in [LMF15].

| $(n,r,m)$ | $u_1^2 v_1^2$ | $u_1 u_2 v_1^2$ | $u_2^2 v_1^2$ | $u_1^2 v_1 v_2$ | $u_1 u_2 v_1 v_2$ | $u_2^2 v_1 v_2$ | $u_1^2 v_2^2$ | $u_1 u_2 v_2^2$ | $u_2^2 v_2^2$ |
|---|---|---|---|---|---|---|---|---|---|
| $(1,1,1)$ | $-\frac{1}{100}$ | $-\frac{1}{100}$ | $-\frac{1}{400}$ | $-\frac{1}{100}$ | $-\frac{1}{40}$ | $-\frac{1}{100}$ | $-\frac{1}{400}$ | $-\frac{1}{100}$ | $-\frac{1}{100}$ |
| $(1,0,1)$ | $\frac{1}{50}$ | $0$ | $-\frac{1}{200}$ | $0$ | $-\frac{17}{100}$ | $0$ | $-\frac{1}{200}$ | $0$ | $\frac{1}{50}$ |
| $(1,1,2)$ | $\frac{4}{25}$ | $\frac{4}{25}$ | $\frac{3}{50}$ | $\frac{4}{25}$ | $\frac{39}{25}$ | $\frac{19}{25}$ | $\frac{3}{50}$ | $\frac{19}{25}$ | $-\frac{28}{25}$ |
| $(1,0,2)$ | $-\frac{9}{25}$ | $0$ | $\frac{9}{100}$ | $0$ | $\frac{117}{50}$ | $0$ | $\frac{9}{100}$ | $0$ | $\frac{63}{25}$ |
| $(1,1,3)$ | $-\frac{99}{100}$ | $-\frac{99}{100}$ | $-\frac{243}{400}$ | $-\frac{99}{100}$ | $-\frac{4311}{200}$ | $-\frac{1089}{100}$ | $-\frac{243}{400}$ | $-\frac{1089}{100}$ | $\frac{693}{100}$ |
| $(1,0,3)$ | $\frac{68}{25}$ | $0$ | $-\frac{18}{25}$ | $0$ | $-\frac{348}{25}$ | $0$ | $-\frac{18}{25}$ | $0$ | $-\frac{576}{25}$ |
| $(1,1,4)$ | $\frac{12}{5}$ | $\frac{12}{5}$ | $\frac{33}{10}$ | $\frac{12}{5}$ | $\frac{669}{5}$ | $69$ | $\frac{33}{10}$ | $69$ | $\frac{192}{5}$ |
| $(1,0,4)$ | $-\frac{264}{25}$ | $0$ | $\frac{84}{25}$ | $0$ | $\frac{1416}{25}$ | $0$ | $\frac{84}{25}$ | $0$ | $-\frac{264}{25}$ |
| $(2,2,2)$ | $\frac{84}{5}$ | $\frac{84}{5}$ | $\frac{384}{25}$ | $\frac{84}{5}$ | $\frac{492}{25}$ | $\frac{84}{5}$ | $\frac{384}{25}$ | $\frac{84}{5}$ | $\frac{84}{5}$ |
| $(2,1,2)$ | $-\frac{84}{5}$ | $-\frac{903}{25}$ | $-\frac{2541}{50}$ | $-\frac{903}{25}$ | $-\frac{2067}{25}$ | $-\frac{903}{25}$ | $-\frac{2541}{50}$ | $-\frac{903}{25}$ | $-\frac{84}{5}$ |
| $(2,0,2)$ | $\frac{56}{25}$ | $0$ | $\frac{344}{5}$ | $0$ | $\frac{56}{5}$ | $0$ | $\frac{344}{5}$ | $0$ | $\frac{56}{25}$ |
| $(2,2,3)$ | $-126$ | $-126$ | $\frac{3573}{50}$ | $-126$ | $\frac{2241}{25}$ | $\frac{4482}{25}$ | $\frac{3573}{50}$ | $\frac{4482}{25}$ | $\frac{2466}{25}$ |
| $(2,1,3)$ | $\frac{4788}{25}$ | $\frac{8631}{25}$ | $\frac{2331}{50}$ | $\frac{8631}{25}$ | $\frac{9963}{25}$ | $\frac{324}{25}$ | $\frac{2331}{50}$ | $\frac{324}{25}$ | $-\frac{10116}{25}$ |
| $(2,0,3)$ | $-\frac{2562}{25}$ | $0$ | $-\frac{1179}{10}$ | $0$ | $\frac{1701}{5}$ | $0$ | $-\frac{1179}{10}$ | $0$ | $\frac{18738}{25}$ |
| $(2,2,4)$ | $\frac{7616}{25}$ | $\frac{7616}{25}$ | $-\frac{13248}{25}$ | $\frac{7616}{25}$ | $-\frac{28704}{25}$ | $-\frac{31408}{25}$ | $-\frac{13248}{25}$ | $-\frac{31408}{25}$ | $\frac{14272}{25}$ |
| $(2,1,4)$ | $-\frac{11928}{25}$ | $-\frac{35994}{25}$ | $\frac{31197}{25}$ | $-\frac{35994}{25}$ | $-\frac{21282}{25}$ | $\frac{24966}{25}$ | $\frac{31197}{25}$ | $\frac{24966}{25}$ | $\frac{27264}{25}$ |
| $(2,0,4)$ | $-\frac{1008}{25}$ | $0$ | $-\frac{42912}{25}$ | $0$ | $-\frac{84528}{25}$ | $0$ | $-\frac{42912}{25}$ | $0$ | $-\frac{68976}{25}$ |
| $(3,3,3)$ | $\frac{3969}{100}$ | $\frac{3969}{100}$ | $\frac{56241}{400}$ | $\frac{3969}{100}$ | $-\frac{32427}{200}$ | $\frac{3969}{100}$ | $\frac{56241}{400}$ | $\frac{3969}{100}$ | $\frac{3969}{100}$ |
| $(3,2,3)$ | $\frac{40392}{25}$ | $-\frac{9504}{25}$ | $-\frac{16632}{25}$ | $-\frac{9504}{25}$ | $-\frac{80784}{25}$ | $-\frac{9504}{25}$ | $-\frac{16632}{25}$ | $-\frac{9504}{25}$ | $\frac{40392}{25}$ |
| $(3,1,3)$ | $-\frac{115587}{50}$ | $-\frac{72981}{50}$ | $-\frac{12663}{40}$ | $-\frac{72981}{50}$ | $\frac{51597}{20}$ | $-\frac{72981}{50}$ | $-\frac{12663}{40}$ | $-\frac{72981}{50}$ | $-\frac{115587}{50}$ |
| $(3,0,3)$ | $\frac{5508}{5}$ | $0$ | $\frac{5157}{5}$ | $0$ | $-\frac{59886}{5}$ | $0$ | $\frac{5157}{5}$ | $0$ | $\frac{5508}{5}$ |
| $(3,3,4)$ | $-\frac{200772}{25}$ | $-\frac{200772}{25}$ | $-\frac{79101}{50}$ | $-\frac{200772}{25}$ | $\frac{144243}{25}$ | $\frac{132957}{25}$ | $-\frac{79101}{50}$ | $\frac{132957}{25}$ | $\frac{27456}{25}$ |
| $(3,2,4)$ | $\frac{125136}{25}$ | $\frac{278784}{25}$ | $\frac{70794}{25}$ | $\frac{278784}{25}$ | $-\frac{68724}{25}$ | $\frac{6336}{25}$ | $\frac{70794}{25}$ | $\frac{6336}{25}$ | $-\frac{88308}{25}$ |
| $(3,1,4)$ | $\frac{19656}{25}$ | $\frac{71496}{25}$ | $\frac{72009}{25}$ | $\frac{71496}{25}$ | $\frac{748386}{25}$ | $\frac{37314}{25}$ | $\frac{72009}{25}$ | $\frac{37314}{25}$ | $-\frac{176256}{25}$ |
| $(3,0,4)$ | $\frac{10512}{5}$ | $0$ | $-\frac{32112}{5}$ | $0$ | $\frac{44736}{5}$ | $0$ | $-\frac{32112}{5}$ | $0$ | $\frac{488384}{25}$ |
| $(4,4,4)$ | $-\frac{3136}{25}$ | $-\frac{3136}{25}$ | $\frac{210464}{25}$ | $-\frac{3136}{25}$ | $-\frac{430336}{25}$ | $-\frac{3136}{25}$ | $\frac{210464}{25}$ | $-\frac{3136}{25}$ | $-\frac{3136}{25}$ |
| $(4,3,4)$ | $3520$ | $-\frac{127941}{5}$ | $-\frac{60217}{10}$ | $-\frac{127941}{5}$ | $-\frac{293807}{5}$ | $-\frac{127941}{5}$ | $-\frac{60217}{10}$ | $-\frac{127941}{5}$ | $3520$ |
| $(4,2,4)$ | $\frac{42816}{5}$ | $\frac{102192}{5}$ | $-\frac{55536}{5}$ | $\frac{102192}{5}$ | $\frac{221664}{5}$ | $\frac{102192}{5}$ | $-\frac{55536}{5}$ | $\frac{102192}{5}$ | $\frac{42816}{5}$ |
| $(4,1,4)$ | $\frac{145728}{25}$ | $-\frac{1227897}{25}$ | $-\frac{230571}{10}$ | $-\frac{1227897}{25}$ | $-\frac{921789}{5}$ | $-\frac{1227897}{25}$ | $-\frac{230571}{10}$ | $-\frac{1227897}{25}$ | $\frac{145728}{25}$ |
| $(4,0,4)$ | $-\frac{194944}{5}$ | $0$ | $\frac{254656}{5}$ | $0$ | $-34816$ | $0$ | $\frac{254656}{5}$ | $0$ | $-\frac{194944}{5}$ |

Table 14: Fourier coefficients $c\big(\phi_{10}^{(2)} f_{\mathrm{RS}}, \big(\begin{smallmatrix} n & r/2 \\ r/2 & m \end{smallmatrix}\big)\big)$ of $\phi_{10}^{(2)} f_{\mathrm{RS}}$ with notation as in Section 9.2.

| | | | | | | | | | |
|---|---|---|---|---|---|---|---|---|---|
| | $(-1,1,-1)$ | $(-1,2,-1)$ | $(0,0,-1)$ | $(0,1,-1)$ | $(0,2,-1)$ | $(1,-1,-1)$ | $(1,0,-1)$ | | |
| | $(1,1,-1)$ | $(1,2,-1)$ | $(2,-2,-1)$ | $(2,-1,-1)$ | $(2,0,-1)$ | $(2,1,-1)$ | $(2,2,-1)$ | | |
| | $(-1,0,0)$ | $(-1,1,0)$ | $(-1,2,0)$ | $(0,-1,0)$ | $(0,1,0)$ | $(0,2,0)$ | $(1,-2,0)$ | $(1,-1,0)$ | | |
| $(n,r,m)$ | $(1,0,0)$ | $(1,1,0)$ | $(1,2,0)$ | $(2,-2,0)$ | $(2,-1,0)$ | $(2,0,0)$ | $(2,1,0)$ | $(2,2,0)$ | | |
| | $(-1,-1,1)$ | $(-1,0,1)$ | $(-1,1,1)$ | $(-1,2,1)$ | $(0,-2,1)$ | $(0,-1,1)$ | $(0,0,1)$ | $(0,1,1)$ | $(0,2,1)$ | $(1,-2,1)$ |
| | $(1,-1,1)$ | $(1,0,1)$ | $(1,1,1)$ | $(1,2,1)$ | $(2,-2,1)$ | $(2,-1,1)$ | $(2,0,1)$ | $(2,1,1)$ | $(2,2,1)$ | |
| | $(-1,-2,2)$ | $(-1,-1,2)$ | $(-1,0,2)$ | $(-1,1,2)$ | $(-1,2,2)$ | $(0,-2,2)$ | $(0,-1,2)$ | $(0,0,2)$ | $(0,1,2)$ | $(0,2,2)$ |
| | $(1,-2,2)$ | $(1,-1,2)$ | $(1,0,2)$ | $(1,1,2)$ | $(1,2,2)$ | $(2,-2,2)$ | $(2,-1,2)$ | $(2,0,2)$ | $(2,1,2)$ | $(2,2,2)$ |
| | 1 | 2 | 2 | 24 | 48 | 3 | 48 | | | |
| | 327 | 648 | 4 | 72 | 648 | 3272 | 6404 | | | |
| | 2 | 24 | 48 | 24 | 600 | 1152 | 48 | 600 | | |
| | −648 | 8376 | 15600 | 1152 | 8376 | −12800 | 85176 | 154752 | | |
| | 3 | 48 | 327 | 648 | 48 | 600 | −648 | 8376 | 15600 | −648 |
| | 25353 | −50064 | 130329 | 209304 | −50064 | 561576 | −1127472 | 1598376 | 2023536 | |
| | 4 | 72 | 648 | 3272 | 6404 | 1152 | 8376 | −12800 | 85176 | 154752 |
| | −50064 | 561576 | −1127472 | 1598376 | 2023536 | −3859456 | 18458000 | −32861184 | 28698000 | 16620544 |

Table 15: Fourier coefficients $c\big(\phi_{10}^{(2)\,-1}, \big(\begin{smallmatrix} n & r/2 \\ r/2 & m \end{smallmatrix}\big)\big)$ of $\phi_{10}^{(2)\,-1}$ in the Weyl chamber described in Section 9.2.

| $(n,r,m)$ | $u_1^2 v_1^2$ | $u_1 u_2 v_1^2$ | $u_2^2 v_1^2$ | $u_1^2 v_1 v_2$ | $u_1 u_2 v_1 v_2$ | $u_2^2 v_1 v_2$ | $u_1^2 v_2^2$ | $u_1 u_2 v_2^2$ | $u_2^2 v_2^2$ |
|---|---|---|---|---|---|---|---|---|---|
| $(0,0,0)$ | $-\frac{1}{100}$ | $\frac{1}{100}$ | $-\frac{1}{400}$ | $\frac{1}{100}$ | $-\frac{1}{40}$ | $\frac{1}{100}$ | $-\frac{1}{400}$ | $\frac{1}{100}$ | $-\frac{1}{100}$ |
| $(0,1,0)$ | $0$ | $\frac{1}{50}$ | $-\frac{1}{100}$ | $\frac{1}{50}$ | $-\frac{11}{50}$ | $\frac{1}{50}$ | $-\frac{1}{100}$ | $\frac{1}{50}$ | $0$ |
| $(0,2,0)$ | $0$ | $\frac{1}{50}$ | $-\frac{1}{50}$ | $\frac{1}{50}$ | $-\frac{11}{25}$ | $\frac{1}{50}$ | $-\frac{1}{50}$ | $\frac{1}{50}$ | $0$ |
| $(1,-1,0)$ | $-\frac{4}{25}$ | $\frac{4}{25}$ | $\frac{1}{100}$ | $\frac{4}{25}$ | $-\frac{7}{50}$ | $-\frac{1}{50}$ | $\frac{1}{100}$ | $-\frac{1}{50}$ | $0$ |
| $(1,0,0)$ | $-\frac{8}{5}$ | $-\frac{6}{25}$ | $\frac{1}{50}$ | $-\frac{6}{25}$ | $\frac{11}{25}$ | $0$ | $\frac{1}{50}$ | $0$ | $0$ |
| $(1,1,0)$ | $-\frac{4}{25}$ | $-\frac{16}{25}$ | $\frac{1}{100}$ | $-\frac{16}{25}$ | $-\frac{7}{50}$ | $\frac{1}{50}$ | $\frac{1}{100}$ | $\frac{1}{50}$ | $0$ |
| $(1,2,0)$ | $0$ | $-\frac{12}{25}$ | $0$ | $-\frac{12}{25}$ | $0$ | $0$ | $0$ | $0$ | $0$ |
| $(2,-2,0)$ | $-\frac{2}{5}$ | $\frac{2}{5}$ | $0$ | $\frac{2}{5}$ | $-\frac{9}{25}$ | $-\frac{1}{50}$ | $0$ | $-\frac{1}{50}$ | $0$ |
| $(2,-1,0)$ | $-\frac{38}{25}$ | $\frac{7}{10}$ | $\frac{3}{100}$ | $\frac{7}{10}$ | $-\frac{3}{50}$ | $-\frac{1}{50}$ | $\frac{3}{100}$ | $-\frac{1}{50}$ | $0$ |
| $(2,0,0)$ | $-\frac{372}{25}$ | $-\frac{18}{25}$ | $\frac{3}{50}$ | $-\frac{18}{25}$ | $\frac{33}{25}$ | $0$ | $\frac{3}{50}$ | $0$ | $0$ |
| $(2,1,0)$ | $-\frac{38}{25}$ | $-\frac{107}{50}$ | $\frac{3}{100}$ | $-\frac{107}{50}$ | $-\frac{3}{50}$ | $\frac{1}{50}$ | $\frac{3}{100}$ | $\frac{1}{50}$ | $0$ |
| $(2,2,0)$ | $-\frac{2}{5}$ | $-\frac{46}{25}$ | $0$ | $-\frac{46}{25}$ | $-\frac{9}{25}$ | $\frac{1}{50}$ | $0$ | $\frac{1}{50}$ | $0$ |
| $(0,-1,1)$ | $0$ | $-\frac{1}{50}$ | $\frac{1}{100}$ | $-\frac{1}{50}$ | $-\frac{7}{50}$ | $\frac{4}{25}$ | $\frac{1}{100}$ | $\frac{4}{25}$ | $-\frac{4}{25}$ |
| $(0,0,1)$ | $0$ | $0$ | $\frac{1}{50}$ | $0$ | $\frac{11}{25}$ | $-\frac{6}{25}$ | $\frac{1}{50}$ | $-\frac{6}{25}$ | $-\frac{8}{5}$ |
| $(0,1,1)$ | $0$ | $\frac{1}{50}$ | $\frac{1}{100}$ | $\frac{1}{50}$ | $-\frac{7}{50}$ | $-\frac{16}{25}$ | $\frac{1}{100}$ | $-\frac{16}{25}$ | $-\frac{4}{25}$ |
| $(0,2,1)$ | $0$ | $0$ | $0$ | $0$ | $0$ | $-\frac{12}{25}$ | $0$ | $-\frac{12}{25}$ | $0$ |
| $(1,-2,1)$ | $-\frac{8}{5}$ | $\frac{86}{25}$ | $-\frac{91}{50}$ | $\frac{86}{25}$ | $-\frac{173}{25}$ | $\frac{86}{25}$ | $-\frac{91}{50}$ | $\frac{86}{25}$ | $-\frac{8}{5}$ |
| $(1,-1,1)$ | $\frac{256}{25}$ | $-\frac{256}{25}$ | $\frac{314}{25}$ | $-\frac{256}{25}$ | $\frac{28}{5}$ | $-\frac{256}{25}$ | $\frac{314}{25}$ | $-\frac{256}{25}$ | $\frac{256}{25}$ |
| $(1,0,1)$ | $-\frac{432}{25}$ | $0$ | $-\frac{561}{25}$ | $0$ | $-\frac{6}{5}$ | $0$ | $-\frac{561}{25}$ | $0$ | $-\frac{432}{25}$ |
| $(1,1,1)$ | $\frac{256}{25}$ | $\frac{256}{25}$ | $\frac{314}{25}$ | $\frac{256}{25}$ | $\frac{28}{5}$ | $\frac{256}{25}$ | $\frac{314}{25}$ | $\frac{256}{25}$ | $\frac{256}{25}$ |
| $(1,2,1)$ | $-\frac{8}{5}$ | $-\frac{86}{25}$ | $-\frac{91}{50}$ | $-\frac{86}{25}$ | $-\frac{173}{25}$ | $-\frac{86}{25}$ | $-\frac{91}{50}$ | $-\frac{86}{25}$ | $-\frac{8}{5}$ |
| $(2,-2,1)$ | $-\frac{2016}{25}$ | $\frac{2016}{25}$ | $-\frac{993}{25}$ | $\frac{2016}{25}$ | $-\frac{1758}{25}$ | $\frac{864}{25}$ | $-\frac{993}{25}$ | $\frac{864}{25}$ | $-\frac{432}{25}$ |
| $(2,-1,1)$ | $\frac{10278}{25}$ | $-\frac{10143}{50}$ | $\frac{20487}{100}$ | $-\frac{10143}{50}$ | $\frac{3903}{50}$ | $-\frac{2052}{25}$ | $\frac{20487}{100}$ | $-\frac{2052}{25}$ | $\frac{2052}{25}$ |
| $(2,0,1)$ | $-\frac{16448}{25}$ | $0$ | $-\frac{1658}{5}$ | $0$ | $-\frac{508}{25}$ | $0$ | $-\frac{1658}{5}$ | $0$ | $-\frac{3232}{25}$ |
| $(2,1,1)$ | $\frac{10278}{25}$ | $\frac{10143}{50}$ | $\frac{20487}{100}$ | $\frac{10143}{50}$ | $\frac{3903}{50}$ | $\frac{2052}{25}$ | $\frac{20487}{100}$ | $\frac{2052}{25}$ | $\frac{2052}{25}$ |
| $(2,2,1)$ | $-\frac{2016}{25}$ | $-\frac{2016}{25}$ | $-\frac{993}{25}$ | $-\frac{2016}{25}$ | $-\frac{1758}{25}$ | $-\frac{864}{25}$ | $-\frac{993}{25}$ | $-\frac{864}{25}$ | $-\frac{432}{25}$ |
| $(0,-2,2)$ | $0$ | $-\frac{1}{50}$ | $0$ | $-\frac{1}{50}$ | $-\frac{9}{25}$ | $\frac{2}{5}$ | $0$ | $\frac{2}{5}$ | $-\frac{2}{5}$ |
| $(0,-1,2)$ | $0$ | $-\frac{1}{50}$ | $\frac{3}{100}$ | $-\frac{1}{50}$ | $-\frac{3}{50}$ | $\frac{7}{10}$ | $\frac{3}{100}$ | $\frac{7}{10}$ | $-\frac{38}{25}$ |
| $(0,0,2)$ | $0$ | $0$ | $\frac{3}{50}$ | $0$ | $\frac{33}{25}$ | $-\frac{18}{25}$ | $\frac{3}{50}$ | $-\frac{18}{25}$ | $-\frac{372}{25}$ |
| $(0,1,2)$ | $0$ | $\frac{1}{50}$ | $\frac{3}{100}$ | $\frac{1}{50}$ | $-\frac{3}{50}$ | $-\frac{107}{50}$ | $\frac{3}{100}$ | $-\frac{107}{50}$ | $-\frac{38}{25}$ |
| $(0,2,2)$ | $0$ | $\frac{1}{50}$ | $0$ | $\frac{1}{50}$ | $-\frac{9}{25}$ | $-\frac{46}{25}$ | $0$ | $-\frac{46}{25}$ | $-\frac{2}{5}$ |
| $(1,-2,2)$ | $-\frac{432}{25}$ | $\frac{864}{25}$ | $-\frac{993}{25}$ | $\frac{864}{25}$ | $-\frac{1758}{25}$ | $\frac{2016}{25}$ | $-\frac{993}{25}$ | $\frac{2016}{25}$ | $-\frac{2016}{25}$ |
| $(1,-1,2)$ | $\frac{2052}{25}$ | $-\frac{2052}{25}$ | $\frac{20487}{100}$ | $-\frac{2052}{25}$ | $\frac{3903}{50}$ | $-\frac{10143}{50}$ | $\frac{20487}{100}$ | $-\frac{10143}{50}$ | $\frac{10278}{25}$ |
| $(1,0,2)$ | $-\frac{3232}{25}$ | $0$ | $-\frac{1658}{5}$ | $0$ | $-\frac{508}{25}$ | $0$ | $-\frac{1658}{5}$ | $0$ | $-\frac{16448}{25}$ |
| $(1,1,2)$ | $\frac{2052}{25}$ | $\frac{2052}{25}$ | $\frac{20487}{100}$ | $\frac{2052}{25}$ | $\frac{3903}{50}$ | $\frac{10143}{50}$ | $\frac{20487}{100}$ | $\frac{10143}{50}$ | $\frac{10278}{25}$ |
| $(1,2,2)$ | $-\frac{432}{25}$ | $-\frac{864}{25}$ | $-\frac{993}{25}$ | $-\frac{864}{25}$ | $-\frac{1758}{25}$ | $-\frac{2016}{25}$ | $-\frac{993}{25}$ | $-\frac{2016}{25}$ | $-\frac{2016}{25}$ |
| $(2,-2,2)$ | $-\frac{79104}{25}$ | $\frac{79104}{25}$ | $-\frac{78936}{25}$ | $\frac{79104}{25}$ | $-\frac{15888}{5}$ | $\frac{79104}{25}$ | $-\frac{78936}{25}$ | $\frac{79104}{25}$ | $-\frac{79104}{25}$ |
| $(2,-1,2)$ | $\frac{235482}{25}$ | $-\frac{234789}{50}$ | $\frac{941511}{100}$ | $-\frac{234789}{50}$ | $\frac{30081}{10}$ | $-\frac{234789}{50}$ | $\frac{941511}{100}$ | $-\frac{234789}{50}$ | $\frac{235482}{25}$ |
| $(2,0,2)$ | $-\frac{332568}{25}$ | $0$ | $-\frac{332979}{25}$ | $0$ | $-\frac{6978}{5}$ | $0$ | $-\frac{332979}{25}$ | $0$ | $-\frac{332568}{25}$ |
| $(2,1,2)$ | $\frac{235482}{25}$ | $\frac{234789}{50}$ | $\frac{941511}{100}$ | $\frac{234789}{50}$ | $\frac{30081}{10}$ | $\frac{234789}{50}$ | $\frac{941511}{100}$ | $\frac{234789}{50}$ | $\frac{235482}{25}$ |
| $(2,2,2)$ | $-\frac{79104}{25}$ | $-\frac{79104}{25}$ | $-\frac{78936}{25}$ | $-\frac{79104}{25}$ | $-\frac{15888}{5}$ | $-\frac{79104}{25}$ | $-\frac{78936}{25}$ | $-\frac{79104}{25}$ | $-\frac{79104}{25}$ |

Table 16: Fourier coefficients $c\big(f_{RS}, \big(\begin{smallmatrix} n & r/2 \\ r/2 & m \end{smallmatrix}\big)\big)$ of $f_{RS}$ in the Weyl chamber described in Section 9.2.

Bethe Center for Theoretical Physics and
Hausdorff Center for Mathematics
Universität Bonn
D-53115 Bonn, Germany
E-mail: aklemm@th.physik.uni-bonn.de

Rittenhouse Laboratory
University of Pennsylvania
19102 Philadelphia, USA
E-mail: mporet@sas.upenn.edu

Bethe Center for Theoretical Physics
Universität Bonn
D-53115 Bonn, Germany
E-mail: schimann@th.physik.uni-bonn.de

Max Planck Institute for Mathematics, Vivatsgasse 7, D-53111, Bonn, Germany
E-mail: martin@raum-brothers.eu
Homepage: http://raum-brothers.eu/martin